\documentclass[aps,article,author-year,notitlepage]{revtex4-1}
\usepackage{subeqn}
\usepackage{graphicx}
\usepackage{float}
\usepackage[T1]{fontenc}
\usepackage[latin1]{inputenc}
\usepackage{amssymb}
\include{epsf}
\epsfverbosetrue

\newcommand{\ts}{\vspace*{.2cm}}

\newcommand{\bea}{\begin{eqnarray}}
\newcommand{\eea}{\end{eqnarray}}

\newcommand{\eg}{{\rm EGG}}
\newcommand{\bb}{{\rm BBALL}}
\newcommand{\mer}{{\rm MERCEDES}}
\newcommand{\ha}{{\rm HAIR}}
\newcommand{\tw}{{\rm TWISTED}}
\newcommand{\lo}{{\rm LOOPY}}
\newcommand{\ey}{{\rm EYEBALL}}
\newcommand{\ta}{{\rm TARGET}}

\usepackage{latexsym} % Gets \Box etc
\usepackage{amssymb}  % \gtrsim, \geqslant, etc etc:
\usepackage{epsfig}       % For PostScript figures
\newcommand{\beq}{\begin{equation}}
\newcommand{\eeq}{\end{equation}}
\newcommand{\ba}{\begin{array}}
\newcommand{\ea}{\end{array}}

\newcommand\comment[1]{ \hbox{[{\it Comment suppressed here.}\/]} }
\newcommand\hide[1]

\newcommand{\Tr}{\hbox{Tr}}

\newcommand{\skipover}[1]{}

\makeatletter %\catcode`\@=11

\def\appendix{\par                              % Have \appendix say
    \setcounter{section}{0}                     % `Appendix A', not just `A'
    \setcounter{subsection}{0}
    \renewcommand{\theequation}{\Alph{section}.\arabic{equation}}
    \renewcommand{\thesection}{Appendix \Alph{section}}
}

\def\applabel#1{\@bsphack
  \protected@write\@auxout{}%
         {\string\newlabel{#1}{{\Alph{section}}{\thepage}}}%
  \@esphack}
\def\section{
\setcounter{equation}{0}        % Reset eqn numbers at start of section
\@startsection {section}{1}{\z@}{-3.5ex plus -1ex minus
 -.2ex}{2.3ex plus .2ex}{\large\bf}}
\renewcommand{\theequation}{\arabic{section}.\arabic{equation}}

\def\subsection{\@startsection{subsection}{2}{\z@}{-3.25ex plus -1ex minus
 -.2ex}{1.5ex plus .2ex}{\normalsize\bf}}

\def\subsubsection{\@startsection{subsubsection}{3}{\z@}{-3.25ex plus
 -1ex minus -.2ex}{1.5ex plus .2ex}{\normalsize}}

\makeatother   %\catcode`\@=12

\newsavebox{\eqlabel}
\makeatletter  %\catcode`\@=11
\newlength{\numblen}
\newsavebox{\eqnumb}
\def\@eqnnum{\savebox{\eqnumb}{\rm (\theequation)}%
\settowidth{\numblen}{\usebox{\eqnumb}}%
\makebox[\numblen][l]{\usebox{\eqnumb}~~~\usebox{\eqlabel}}}
\makeatother   %\catcode`\@=12

\newenvironment{equationwithlabel}[1]{ %
  \savebox{\eqlabel}{#1}
  \begin{equation}\label{#1} }{\end{equation}} %\savebox{\eqlabel}{~}}
\newcommand{\beql}[1]{\begin{equationwithlabel}{#1}}
\newcommand{\eeql}{\end{equationwithlabel}}
\begin{document}

\large

\title{Techniques for $n$-Particle Irreducible Effective Theories}

\author{M.E. Carrington}
\email{carrington@brandonu.ca}
\affiliation{Department of Physics, Brandon University, Brandon, Manitoba, R7A 6A9 Canada\\ and \\  Winnipeg Institute for Theoretical Physics, Winnipeg, Manitoba }
\author{Yun Guo}
\email{guoyun@brandonu.ca}
\affiliation{Department of Physics, Brandon University, Brandon, Manitoba, R7A 6A9 Canada\\ and \\  Winnipeg Institute for Theoretical Physics, Winnipeg, Manitoba }

\begin{abstract}
In this paper we show that the skeleton diagrams in the $m$-Loop $n$PI effective action correspond to an infinite resummation of perturbative diagrams which is void of double counting at the $m$-Loop level. We also show that the variational equations of motion produced by the $n$-Loop $n$PI effective theory are equivalent to the Schwinger-Dyson equations, up to the order at which they are consistent with the underlying symmetries of the original theory. We use a diagrammatic technique to obtain the 5-Loop 5PI effective action  for a scalar theory with cubic and quartic interactions, and verify that the result satisfies these two statements.
\end{abstract}
\maketitle

\section{Introduction}
\label{introSection}

The $n$-particle irreducible ($n$PI) effective action is the set of skeleton diagrams produced by the $n$th Legendre transform. It is a functional of the $n$-point functions of the theory, which are treated as variational parameters.
The variational equations of motion (eom's) are determined by functionally differentiating the action with respect to its arguments (and setting all sources to zero). In this paper we  show that: (1) the skeleton diagrams in the $m$-Loop $n$PI effective action correspond to an infinite resummation of perturbative diagrams which is void of double counting at the $m$-Loop level; and (2) the eom's produced by the $n$-Loop $n$PI effective theory are equivalent to the Schwinger-Dyson (sd) equations, up to the order at which they are consistent with the underlying symmetries of the original theory.

We comment that although both the $n$PI eom's and the sd  equations  are sets of coupled nonlinear integral equations that contain nonperturbative physics, there are
significant differences between them.
For an $n$PI effective theory, the effective action is truncated, and the resulting eom's form a closed set. In contrast, the sd equations form an infinite hierarchy of coupled equations which must be truncated in order to do calculations.
In addition, there are fundamental differences in the basic structure of the two sets of equations. In the sd equation, all graphs contain one bare vertex and are not symmetric with respect to permutations of external legs. The $n$PI eom's are symmetric and (for $n>2$) some graphs contain no bare vertices.
In light of these remarks, the statement (2) above seems unlikely to be true. In fact, as we will show in a fairly simple way, statement (2) is a direct consequence of statement (1).

The derivation of the sd equations is tedious but straightforward. In Refs. \cite{cvitanovic,kajantie}, an analytic method is used to produce the sd equations up to the level of the 4-point function. In Ref. \cite{alkoferALG}, an algorithm is implemented in a downloadable mathematica package which produces the sd equations to arbitrary order. In Figs. \ref{sdPIeqnLABEL} and \ref{sdUeqnLABEL} we give the results for the sd equations that we will use.

In principle, the calculation of a set of Legendre transforms is also a well defined problem, but the computation becomes extremely complicated beyond the lowest levels.
The 4PI effective action was introduced in Refs. \cite{deDom1,deDom2}. It was first discussed in the context of relativistic field theories in  Ref.  \cite{norton}. The  3-Loop 4PI effective action was calculated in Refs. \cite{bergesGauge,deDom1,deDom2,mec4PI}, and the 4-Loop 4PI effective action was calculated in \cite{ek4PI}. The effective action has not previously been calculated beyond the level of the fourth Legendre transform.  In this paper we use a diagrammatic technique to obtain the 5-Loop 5PI effective action for the theory defined in Eq. (\ref{scl}). We verify that the result satisfies both statements (1) and (2) from the first paragraph of this introduction. We also find that the 5PI effective action is not 5-particle irreducible: it contains diagrams that can be divided into two pieces, each of which contains at least one loop, by cutting five or fewer lines. The result suggests that the $n$-Loop $n$PI effective action is $n$-particle irreducible for $n\le 4$ only. Throughout this paper we will use ``$n$PI effective action'' to mean the effective action produced by taking $n$ Legendre transforms. We stress that the important point is that Legendre transforms produce an effective action that does not double count at the level of the truncation.

In this paper, consider a scalar theory that has both cubic and quartic  couplings.
We study this theory because both the sd equations and the $n$PI eom's have the same structure as the corresponding equations for QCD.
All of the results in this paper can be generalized to other theories in a straightforward way.

This paper is organized as follows: In Sec. \ref{notationSection} we define our notation and review some results for the 4-Loop 4PI effective action. In Sec. \ref{fifthLTsection}, we present our calculation of the 5-Loop 5PI effective action.
In Sec. \ref{sdnPISection} we show that the $n$PI effective action is void of double counting, and that the eom's are equivalent to the sd equations, at the truncation order. In Sec. \ref{PiUeqnSection} we verify that the eom's for the 2- and 3-point vertex functions satisfy the statements made in Sec. \ref{sdnPISection}, up to the level of the 5-Loop 5PI effective action. In Sec. \ref{concSection} we present our conclusions. Some details are left to the appendixes.

\section{Preliminaries}
\label{notationSection}
\subsection{Notation}

Throughout this paper we use $L$ to indicate the loop order in the skeleton expansion. We also use ``$n$-Loop'' to mean terms in the skeleton expansion with $L\le n$ loops, and ``$n$-loop'' to mean terms in the skeleton expansion with $L =  n$ loops.

We denote connected and proper vertices by ${\cal V}^c$ and ${\cal V}$, respectively. In addition, for the correlation functions with up to 5 external legs we use
\bea
{\rm 1-point~function}&&=: \phi\nonumber \\
{\rm 2-point~function}&&~ \{{\rm connected/self-energy/bare/effective~bare}\} =: \{D,~\Pi,~D^{oo},~D^0\}\nonumber\\
{\rm 3-point~function}&&~ \{{\rm connected/proper/bare/effective~bare}\} =: \{U^c,~U,~U^{oo},~U^0\}\nonumber\\
{\rm 4-point~function}&&~ \{{\rm connected/proper/bare}\}=: \{V^c,~V,~V^{0}\}\nonumber\\
{\rm 5-point~function}&& ~\{{\rm connected/proper}\}=: \{W^c,~W\}\nonumber
\eea
For example, ${\cal V}_3=U$, ${\cal V}^c_5=W^c$, etc.

In coordinate space, each function has arguments that correspond to the space-time coordinates of its legs.
%When we write equations that refer to \npoint functions with an arbitrary number of legs, we simply suppress these arguments (and write the vertices $V_j$, where $j$ indicates the number of legs).
We use a compactified notation in which the space-time coordinates for a given leg are represented by a single numerical subscript. For example, the field expectation value is written $\phi_i:=\phi(x_i)$,
the propagator is written $D_{ij}:=D(x_i,x_j)$, the bare 4-point vertex is written $V^0_{ijkl}:=V^0(x_i,x_j,x_k,x_l)$, etc.  We also use an Einstein convention in which a repeated index implies an integration over space-time variables.
%In some equations we use the generic notation ${\cal V}^c_i$ for a connected correlation function and ${\cal V}_i$ for a proper vertex function, where the index $i$ refers to the number of legs.

We define all propagators and vertices with factors of $i$ so that figures look as simple as possible:
lines, and intersections of lines, correspond directly to propagators and vertices, with no additional factors of plus or minus $i$.

Using this notation we write the classical action
\bea
\label{scl}
S_{cl}[\varphi]=\frac{1}{2}\varphi_i\big[i\,(D^{oo}_{ij})^{-1}\big]\varphi_j-\frac{i}{\;3!} U_{ijk}^{oo}\varphi_i\varphi_j\varphi_k-\frac{i}{\;4!}V_{ijkl}^0\varphi_i\varphi_j\varphi_k\varphi_l\,.
\eea

The $n$PI effective action is obtained by taking the $n$th Legendre transform of the generating functional which is constructed by coupling the field to $n$ source terms:
\bea
\label{genericGamma}
&& Z[J,R,R^{(3)},R^{(4)},R^{(5)}\dots]=\int d\varphi  \;{\rm Exp}[i\,{\cal X}]\,,\\[2mm] &&{\cal X}=S_{cl}[\varphi]+J_i\varphi_i+\frac{1}{2}R_{ij}\varphi_i\varphi_j + \frac{1}{3!}R^{(3)}_{ijk}\varphi_i\varphi_j\varphi_k + \frac{1}{4!} R^{(4)}_{ijkl}\varphi_i\varphi_j\varphi_k\varphi_l + \frac{1}{5!} R^{(5)}_{ijkln}\varphi_i\varphi_j\varphi_k\varphi_l\varphi_n+\cdots\,,\nonumber\\[4mm]
&&{\cal W}[J,R,R^{(3)},R^{(4)},R^{(5)}\dots]=-i \,{\rm Ln} Z[J,R,R^{(3)},R^{(4)},R^{(5)}\dots]\,,\nonumber\\[4mm]
&&\Gamma[\phi,D,U,V,W\dots] = {\cal W} - J_i\frac{\delta {\cal W}}{\delta J_i} - R_{ij}\frac{\delta {\cal W}}{\delta R_{ij}} - R^{(3)}_{ijk}\frac{\delta {\cal W}}{\delta R^{(3)}_{ijk}} - R^{(4)}_{ijkl}\frac{\delta {\cal W}}{\delta R^{(4)}_{ijkl}}  - R^{(5)}_{ijkln}\frac{\delta {\cal W}}{\delta R^{(5)}_{ijkln}}-\cdots\nonumber
\eea
We define connected green functions
\bea
\label{Wders}
{\cal V}^c_{t_1,t_2, \cdots t_k} = \langle\varphi_{t_1}\varphi_{t_2}\varphi_{t_3}\dots\varphi_{t_k}\rangle_c = -(-i)^{k+1}\frac{ \delta^k W}{\delta J_{t_k}\dots \delta J_{t_3}\delta J_{t_2}\delta J_{t_1}}\,,
\eea
which allows us to write
\bea\label{defcon}
\frac{\delta {\cal W}}{\delta J_i} &&= \langle\varphi_i\rangle = \phi_i \,,\\
2\frac{\delta {\cal W}}{\delta R_{ij}} &&= \langle\varphi_i \varphi_j\rangle = D_{ij}+\phi_i \phi_j \,,\nonumber\\
3!\frac{\delta {\cal W}}{\delta R^{(3)}_{ijk}} &&= \langle\varphi_i \varphi_j \varphi_k\rangle  =U^c_{ijk}+D_{jk} \phi_i+D_{ik} \phi_j+D_{ij} \phi_k+\phi_i \phi_j \phi_k\,,\nonumber\\
4!\frac{\delta \mathcal{W}}{\delta R^{(4)}_{ijkl}}&& = \langle\varphi_i \varphi_j \varphi_k \varphi_l\rangle
 = V^c_{ijkl} + (U^c_{ijk}\, \phi_l + 3\, {\rm perms}) + (D_{ij}
D_{kl} + 2\, {\rm perms})\nonumber \\ && ~~~~~~~~~~~~~~~~~~+(D_{ij}\, \phi_k
\phi_l + 5\, {\rm perms}) + \phi_i\phi_j\phi_k\phi_l \, ,\nonumber
\eea
where the notation ``perms'' indicates terms obtained by permuting the indices of the previous term, without regard for order. For example, $(U^c_{ijk}\, \phi_l + 3\, {\rm perms})=U^c_{ijk}\, \phi_l + U^c_{ijl}\, \phi_k + U^c_{ikl}\, \phi_j + U^c_{jkl}\, \phi_i$.

We define the effective bare propagator and 3-point vertex\footnote{There is no vertex $V^{oo}$ in Eq. (\ref{scl}) since the effective bare 4-point vertex is identical to the bare 4-point vertex. We use $U^0$ for the effective bare vertex since this vertex appears most often in equations.}
\bea
\label{free}
&&(D^0_{ij}(\phi))^{-1}=-i \frac{\delta^2 S_{cl}[\phi]}{\delta \phi_j\delta \phi_i} = (D_{ij}^{oo})^{-1}-U_{ijk}^{oo}\phi_k-\frac{1}{2}V_{ijkl}^0\phi_k\phi_l\,,\\
&&
U^0_{ijk}(\phi)=i\frac{\delta^3 S_{cl}[\phi]}{\delta \phi_k\delta \phi_j\delta \phi_i}= -\frac{\delta (D^{0}_{ij}(\phi))^{-1}}{\delta \phi_k} = U_{ijk}^{oo}+\phi_l V_{ijkl}^0\,.\nonumber
\eea
The $n$PI effective action  depends on $\phi$ only through the effective bare propagator and effective bare 3-vertex. We will suppress the argument and write $D^0_{ij}(\phi)$ as $D^0_{ij}$ and $U^0_{ijk}(\phi)$ as $U^0_{ijk}$.

We define proper vertices as derivatives of the 1PI effective action:
\bea
\label{properDefn}
\Gamma[\phi]&&=\mathcal{W}[J]-J_i \phi_i \,, \\[2mm]
{\cal V}_{t_1,t_2, \cdots t_k} &&= i \frac{\delta^k \Gamma[\phi]}{\delta \phi_{t_k}\dots \delta \phi_{t_3}\delta \phi_{t_2}\delta \phi_{t_1}}\,.\nonumber
\eea
The 3- and 4-point connected and proper vertices satisfy
\bea
\label{proper}
U^c_{ijk}&& =D_{i t_1} D_{j t_2} D_{k t_3}
   U_{t_1 t_2 t_3}\,,\\
V^c_{ijkl}&& =D_{i t_1} D_{j t_2}
   D_{k t_3} D_{l t_4}
   V_{t_1 t_2 t_3 t_4}+D_{it_1}D_{j t_2} D_{k t_3}
   D_{l t_4}D_{t_5 t_6}
   U_{t_1 t_6 t_3}
   U_{t_2 t_5 t_4}\nonumber\\
&&+D_{i t_1} D_{j t_2}
   D_{k t_3} D_{l t_4} D_{t_6 t_5}
   U_{t_1 t_2 t_6}
   U_{t_3 t_5 t_4}+D_{i t_1} D_{j t_2}
   D_{k t_3} D_{l t_4} D_{t_6 t_5}
   U_{t_1 t_5 t_4}
   U_{t_6 t_2 t_3}\,.\nonumber
\eea

It is straightforward to obtain equations analogous to Eqs. (\ref{defcon}) and (\ref{proper}) at arbitrarily higher orders,  but the resulting expressions are tedious to write.
To present these equations in a more compact form, we introduce a simplified notation in which we suppress space-time indices: we write ${\cal V}^c_{t_1,t_2, \cdots t_i}$ as ${\cal V}_i^c$ and ${\cal V}_{t_1,t_2, \cdots t_i}$ as ${\cal V}_i$.
We give an example of this notation in Eq. (\ref{notExample}):
\bea
\label{notExample}
 &&(3)\,D^5 U^2 := D_{it_1}D_{j t_2} D_{k t_3}
   D_{l t_4}D_{t_5 t_6}
   U_{t_1 t_6 t_3}
   U_{t_2 t_5 t_4}\nonumber\\
&&~~+D_{i t_1} D_{j t_2}
   D_{k t_3} D_{l t_4} D_{t_6 t_5}
   U_{t_1 t_2 t_6}
   U_{t_3 t_5 t_4}+D_{i t_1} D_{j t_2}
   D_{k t_3} D_{l t_4} D_{t_6 t_5}
   U_{t_1 t_5 t_4}
   U_{t_6 t_2 t_3}\,.
\eea
Note that when space-time indices are included, the three terms that correspond to the three different 2 $\leftrightarrow$ 2 channels are all written separately, but when we suppress indices, the distinction between these three channels is lost. We indicate that all three channels are included in one term by writing the factor (3) in front of the term on the left-hand side (lhs) of Eq. (\ref{notExample}).

\begin{figure}[H]
\begin{center}
\epsfig{file=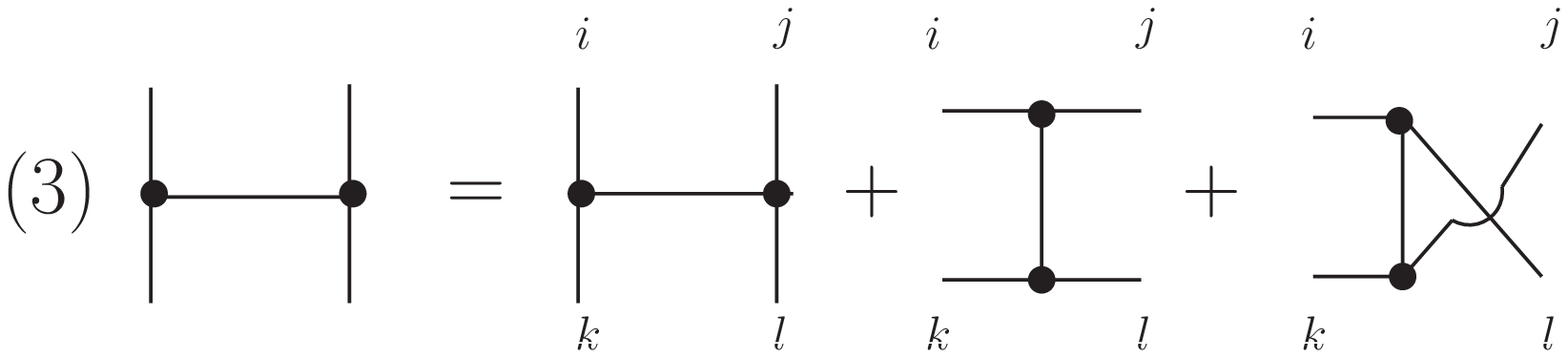,width=10.5cm}
\end{center}
\vspace*{-0.5cm} \caption{\small Diagrammatic representation of Eq. (\ref{notExample}).} \label{fig:fig1}
\end{figure}

Many of the equations we will write in this paper are easier to understand as diagrams. In some cases, we will give only the diagrammatic form of an equation. As an illustration, in Fig. \ref{fig:fig1} we show the diagram\footnote{Figures in this paper are drawn using Jaxodraw \cite{jaxo}. } that corresponds to Eq. (\ref{notExample}).

Using the notation that suppresses space-time indices, Eq. (\ref{defcon}) can be rewritten in a compact way. Including the result for the derivative with respect to  $R^{(5)}$ we have
\bea
\label{connect2}
\frac{\delta {\cal W}}{\delta J} &&= \phi\,,\\
2\frac{\delta {\cal W}}{\delta R} &&= D+ \phi^2 \,,\nonumber\\
3!\frac{\delta {\cal W}}{\delta R^{(3)}} &&= U^c + (3) D \phi + \phi^3\,,\nonumber\\
4!\frac{\delta \mathcal{W}}{\delta R^{(4)}}&&
 = V^c + (4) U^c \phi + (3) D^2 + (6) D \phi^2 + \phi^4 \, ,\nonumber\\
5!\frac{\delta \mathcal{W}}{\delta R^{(5)}}&&
 = W^c + (5) V^c \phi+(10) U^c \phi^2 +(10) U^c D + (10) D \phi^3 +(15) D^2 \phi +\phi^5 \, .\nonumber
 \eea
Similarly, we can rewrite Eq. (\ref{proper}). Including the result for the connected 5-point vertex we have
\bea
\label{proper2}
U^c &=&  D^3 U
\, ,   \\
V^c &=& D^4 V + (3) \, D^5 U^2  \nonumber
\, , \\
W^c &=& \, D^5 W + (10)\, \, D^6 U V + (15)\, D^7 U^3 \, .\nonumber
 \eea

\begin{figure}[H]
\begin{center}
\includegraphics[width=5cm]{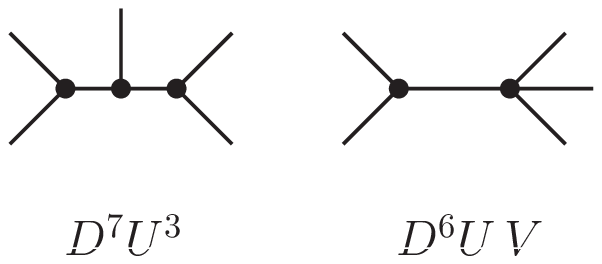}
\end{center}
\caption{\label{D7U3LABEL}Two terms from the third line in Eq. (\ref{proper2}).}
\end{figure}

In Eqs. (\ref{notExample}),  (\ref{connect2}),  (\ref{proper2}) and Fig. \ref{fig:fig1}, the bracketed numerical coefficients indicate the number of
permutations of the external legs which have been combined by the notation that suppresses space-time indices. We will use this notation throughout this paper.
These numerical coefficients are easy to understand when they are written as products of combinatoric factors.
We give two examples below. We use $C^n_m:= m!/(n!(m-n)!)$.
The terms $D^7 U^3$ and $D^6 U V$ are shown in Fig. \ref{D7U3LABEL}.
For $D^7 U^3$ there are $15$ different permutations of the external legs. The factor $15$ is determined as follows. We need to assign 5 indices to the 5 external legs. First, we assign one index to the external leg in the middle, which gives $C_5^1$ different choices; then we assign 2 of the remaining 4 indices to the 2 external legs on the left side, which gives $C_4^2$ different choices; finally, we assign the remaining 2 indices to the 2 legs on the right side, which gives $C_2^2$ different choices. Note that since the 2 pairs of external legs on the left and right sides are symmetric, we need to introduce a factor of $1/2$ to account for this symmetry. Combining we obtain a factor $C_5^1 C_4^2 C_2^2/2 = 15 $ for this diagram. Using the same method, the factor for the term $D^6 U V$ is $C_5^2 C_3^3 = 10$ (note that $C_5^2 C_3^3=C_5^3 C_2^2$, which means that one can work from either side of the diagram and obtain the same result).
\par

\subsection{$n$PI Effective Action}
\label{nPIsection}

The $n$PI effective action is obtained from the last line of Eq. (\ref{genericGamma}). The result can be written:
\bea
\label{gammaGen}
&&\Gamma[\phi,D,U,V,W\dots] \\
&&~~=S_{cl}[\phi]+
    \frac{i}{2} {\rm Tr} \,{\rm Ln}D^{-1}  +
\frac{i}{2} {\rm Tr}\left[ \left(D^0\right)^{-1} D\right] -i\Phi^0[\phi,D,U,V]-i\Phi^{int}[D,U,V,W\dots]~~+~~{\rm const} \,.\nonumber
\eea
The terms $\Phi^0[\phi,D,U,V]$ and $\Phi^{int}[D,U,V,W\dots]$ contain all contributions to the effective action which have two or more loops.
%We define:
%\bea
%\label{phiDefn}
%\Phi^0 = i\,\Gamma^0\,,~~\Phi^{int} = i\,\Gamma^{int}\,,~~\Phi=\Phi^{0}+\Phi^{int}
%\eea
% which allows us to represent $\Phi^0$ and $\Phi^{int}$ as series of diagrams with all factors of $i$ absorbed into the definitions of the propagators and vertices.
For $n\ge 4$ the $\Phi^0$ piece includes all terms that contain bare vertices, and therefore $\Phi^{int}$ is $\phi$-independent. We also define $\Phi=\Phi^{0}+\Phi^{int}$.

In general, we use $n$ to indicate the order of the Legendre transform, or equivalently, the highest variational correlation function in the effective action, and $m$ to denote the order of the skeleton loop expansion. The $n$PI effective action at $m$-Loop order is written\footnote{Note that from Eq. (\ref{properDefn}) we have ${\cal V}_2=D^{-1}$ and thus $\Gamma_n^{(m)}[{\cal V}_i]\,,i\in\{1,2,\dots n\}$ really means $\Gamma_n^{(m)}[\phi,D^{-1},U,V,\dots]$ and not $\Gamma_n^{(m)}[\phi,D,U,V,\dots]$. We ignore this point to avoid introducing unnecessary notation.} $\Gamma_n^{(m)}[{\cal V}_i]\,,i\in\{1,2,\dots n\}$.
In Ref. \cite{bergesGauge} it is argued that at $n$-Loop order the $n$PI effective action provides a complete self-consistent description. In \ref{hierSection} we verify that this result holds at the 4-Loop level. The conclusion is that we only need to consider $m\ge n$. When $m=n$ we drop the superscript on the effective action that indicates the Loop order. In some cases, we write the functional arguments explicitly, and drop the redundant subscript. For example, the $m$-Loop 3PI effective action is written $\Gamma^{(m)}[\phi,D,U]$, and the 5-Loop 5PI effective action is written $\Gamma[\phi,D,U,V,W]$.

\ts

The eom's are obtained by functionally differentiating the effective action with respect to the arguments (and setting all sources to zero). For the $n$PI effective action there are $n$ eom's given by:
\bea
\label{generaleom}
\frac{\delta \Gamma^{(m)}_n[{\cal V}_i]}{\delta {\cal V}_j}=0\,,~~~~~\{i,j\}\in\{1,\dots,n\}\,.
\eea

\subsection{4PI Effective Action}
\label{4PIsection}

\begin{figure}[H]
\begin{center}
\includegraphics[width=10.5cm]{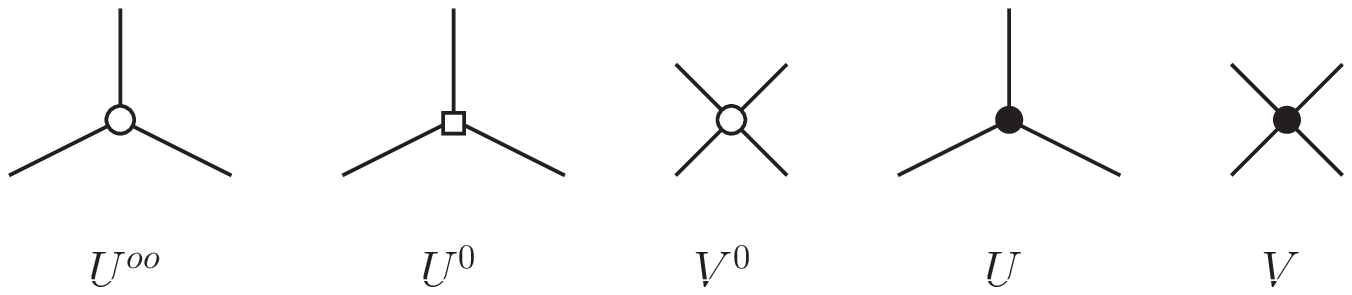}
\end{center}
\caption{\label{vertFirst}Diagrammatic notation for some vertices.}
\end{figure}

We give some results for the 4-Loop 4PI effective theory \cite{ek4PI}.  We introduce the diagrammatic notation shown in Fig. \ref{vertFirst} for bare, effective bare, and proper vertices [see Eqs. (\ref{scl}), (\ref{properDefn}) and (\ref{free})].
\par
\begin{figure}[H]
\begin{center}
\includegraphics[width=11.5cm]{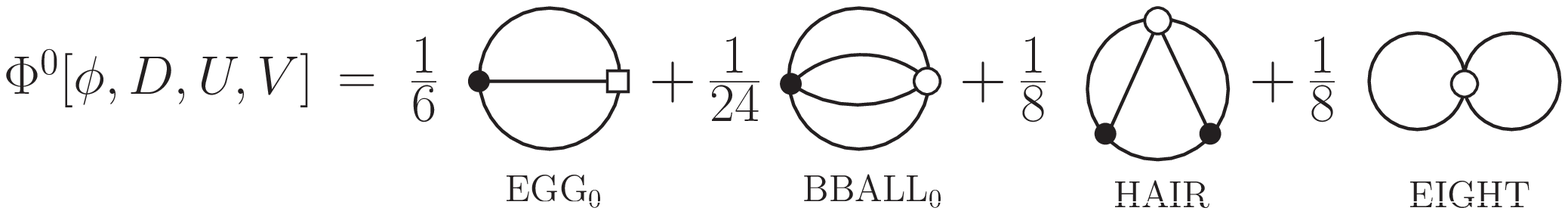}
\end{center}
\caption{\label{Phi0Fig}$\Phi^0$ for $n\ge 4$. }
\end{figure}
\begin{figure}[H]
\begin{center}
\includegraphics[width=16cm]{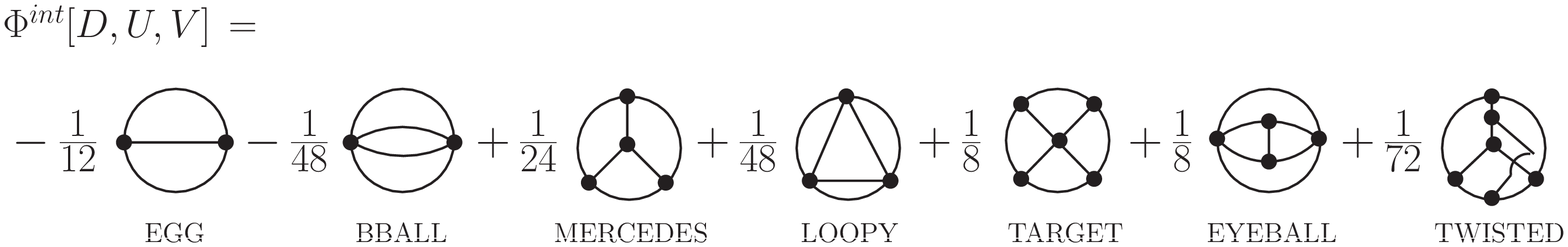}
\end{center}
\caption{\label{Phi4Fig}$\Phi^{int}$ for 4-Loop 4PI.}
\end{figure}

In Figs. \ref{Phi0Fig} and \ref{Phi4Fig} we show the results for $\Phi^0$ and $\Phi^{int}$ respectively. Each contribution is given a  name so that we can refer to the diagrams individually.
The result for $\Phi^0$  is complete for $n\ge 4$.
\par
\par

\begin{figure}[H]
\begin{center}
\includegraphics[width=18cm]{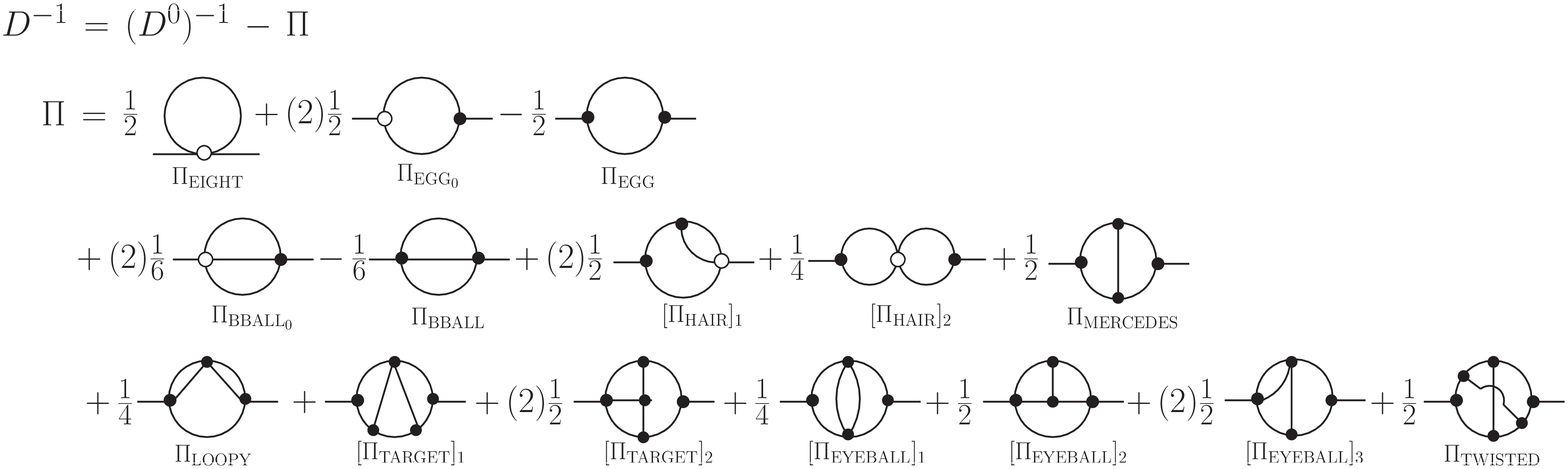}
\end{center}
\caption{\label{PIintLABEL}Integral equation for the 2-point vertex from the 4-Loop 4PI effective action.}
\end{figure}

\begin{figure}[H]
\begin{center}
\includegraphics[width=16cm]{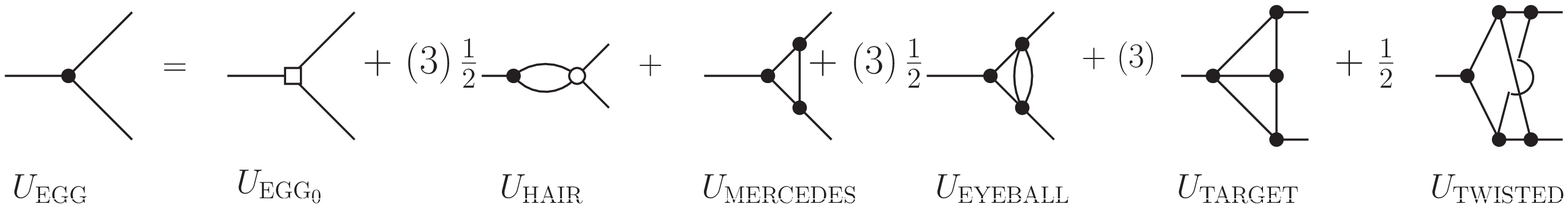}
\end{center}
\caption{\label{UintLABEL}Integral equation for the 3-point vertex from the 4-Loop 4PI effective action.}
\end{figure}

\begin{figure}[H]
\begin{center}
\includegraphics[width=12cm]{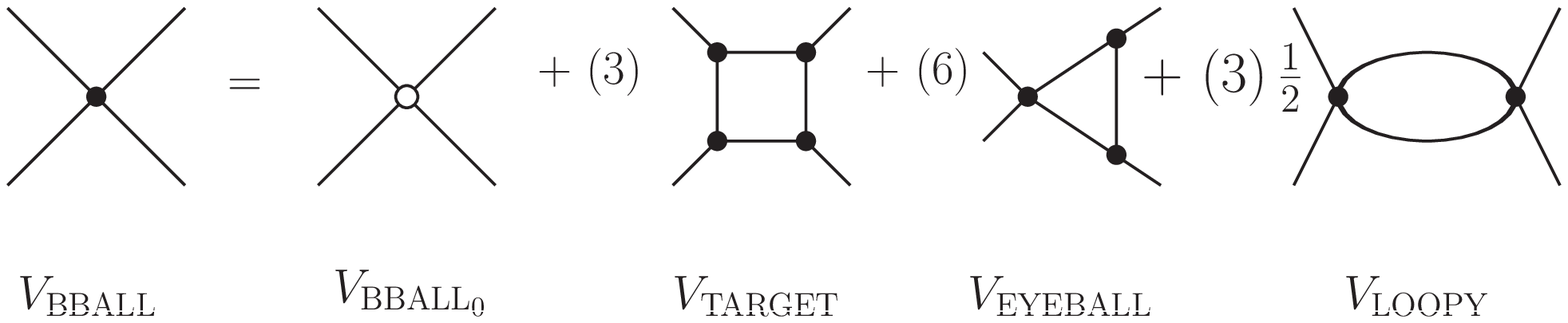}
\end{center}
\caption{\label{VintLABEL}Integral equation for the 4-point vertex from the 4-Loop 4PI effective action.}
\end{figure}

The equations of motion are obtained from Eqs. (\ref{gammaGen}) and (\ref{generaleom}), and Figs. \ref{Phi0Fig} and \ref{Phi4Fig}. The equations corresponding to $\delta \Gamma[\phi,D,U,V]/\delta D=0$, $\delta \Gamma[\phi,D,U,V]/\delta U=0$ and $\delta \Gamma[\phi,D,U,V]/\delta V=0$ are shown respectively in Figs. \ref{PIintLABEL},  \ref{UintLABEL}, and \ref{VintLABEL}. In each of these figures, the labels indicate the diagram in the effective action that produced each graph. In cases where one diagram in the effective action produces two distinct topologies in an equation of motion, we use subscripts 1,2 $\dots$ to distinguish them. For example, the HAIR graph in the effective action produces the two graphs labeled $[\Pi_\ha]_1$ and $[\Pi_\ha]_2$ in Fig. \ref{PIintLABEL}.
\par
\par
\par

\section{The 5-Loop 5PI effective action}
\label{fifthLTsection}

In this section, we derive the 5-Loop 5PI effective action. We use the method of successive Legendre transformations \cite{deDom1,mec4PI,bergesGauge}. In order to use this method at the 5-Loop 5PI level, we  introduce a bare 5-point vertex ($W^0$), for organizational purposes only. This point will be explained in the first subsection below. We use a diagrammatic technique that allows us to identify the classes of graphs that cancel to remove reducible diagrams from the final result.

\subsection{Tilded Effective Action}
\label{tildeSection}

\begin{figure}[H]
\begin{center}
\includegraphics[width=8cm]{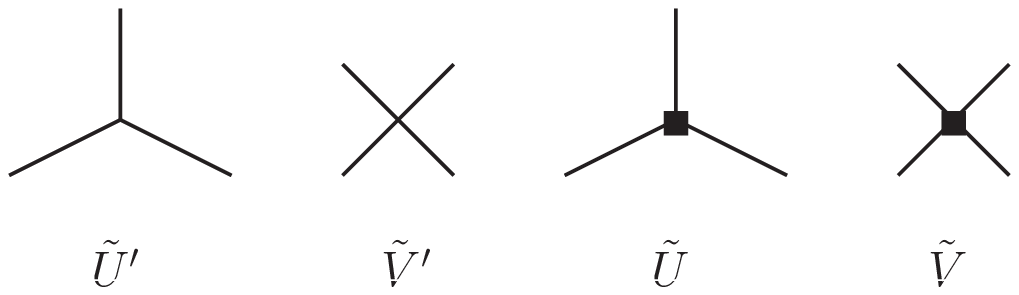}
\end{center}
\caption{\label{vertSecond}Diagrammatic notation for the tilde vertices and effective tilde vertices defined in Eqs. (\ref{tildeVert}) and  (\ref{tildePrime}), respectively.}
\end{figure}

We start by doing only the first two Legendre transforms, and absorbing the last three terms in the second line of Eq. (\ref{genericGamma}) into a modified interaction by defining the vertices
\beq
\label{tildeVert}
i\tilde{U}:= iU^{oo} - R^{(3)},~~
i\tilde{V} :=iV^{0} - R^{(4)},~~
i\tilde{W} :=i W^0 - R^{(5)}  \,.\label{eq:replace}
\eeq
We will refer to these vertices as ``tilde vertices''
and we write them collectively using the notation $\tilde {\cal V}_i\in\{\tilde U,\tilde V,\tilde W\}$.
The diagrammatic notation we will use for the tilde vertices [and for the effective tilde vertices which will be defined in Eq. (\ref{tildePrime})] is shown in Fig. \ref{vertSecond}.
\par

The 5PI effective action is defined in the last line of Eq. (\ref{genericGamma}). We rewrite this expression using tilde vertices as
\bea
\label{dormant}
\Gamma[\phi,D,U,V,W]&& = {\cal W} - J\frac{\delta {\cal W}}{\delta J} - R\frac{\delta {\cal W}}{\delta R} - R^{(3)}\frac{\delta {\cal W}}{\delta R^{(3)}} - R^{(4)}\frac{\delta {\cal W}}{\delta R^{(4)}}  - R^{(5)}\frac{\delta {\cal W}}{\delta R^{(5)}}\,,\\
&&= \tilde\Gamma[\phi,D] - R^{(3)}\frac{\delta {\cal W}}{\delta R^{(3)}} - R^{(4)}\frac{\delta {\cal W}}{\delta R^{(4)}}  - R^{(5)}\frac{\delta {\cal W}}{\delta R^{(5)}}\,,\nonumber
\eea
where we define:
\bea
\label{tildeGamma}
\tilde\Gamma[\phi,D]
=
\mathcal{W} -\frac{\delta
\mathcal{W}}{\delta J} J - \frac{\delta \mathcal{W}}{\delta R} R\, .
\eea
In order to construct $\tilde\Gamma[\phi,D]$, we start with the set of 5-Loop 2PI diagrams\footnote{We include some $L\ge 6$ loop  tadpole diagrams that do not change the result of the calculation. They are included for organisational purposes only. This point is explained in more detail in the discussion under Eq. (\ref{tildePrime}).} that correspond to a theory with a bare 3-point, 4-point and 5-point interaction, and then we replace the bare vertices by the tilde vertices. We will refer to $\tilde\Gamma[\phi,D]$ as the tilded 2PI effective action. In the rest of this section, a tilde over a functional always indicates that all bare vertices are replaced by tilde vertices\footnote{Note that the bare 5-point vertex $W^0$ has disappeared at this point in the calculation. It's only role is to produce diagrams in $\tilde\Gamma[\phi,D]$ that contain the 5-point tilde vertex $\tilde W$.}.

\ts

We can rewrite the dormant Legendre transforms in Eq. (\ref{dormant}) as functional derivatives of the tilded 2PI effective action
using the relations
\beq
\frac{\delta \tilde{\Gamma}}{\delta R^{(3)}}
= \frac{\delta \mathcal{W}}{\delta R^{(3)}}\, ,\qquad \frac{\delta
\tilde{\Gamma}}{\delta R^{(4)}} = \frac{\delta \mathcal{W}}{\delta
R^{(4)}}\, ,\qquad \frac{\delta \tilde{\Gamma}}{\delta R^{(5)}} =
\frac{\delta \mathcal{W}}{\delta R^{(5)}} \, . \label{relate}
\eeq
Using Eqs. (\ref{tildeVert}) and (\ref{relate}), Eq. (\ref{dormant}) becomes
\bea
\label{dormant2}
&&\Gamma[\phi,D,U,V,W] \\
&&~~=  \tilde{\Gamma}[\phi,D] - \frac{\delta
\tilde{\Gamma}[\phi,D]}{\delta \tilde{U}} (\tilde{U} - U^{oo}) -
\frac{\delta \tilde{\Gamma}[\phi,D]}{\delta \tilde{V}} (\tilde{V}
- V^{0})- \frac{\delta \tilde{\Gamma}[\phi,D]}{\delta
\tilde{W}}\tilde{W}~~+~~{\rm const}\, .\nonumber
\eea

The next step is to rewrite $\tilde\Gamma[\phi,D]$. From Eq. (\ref{gammaGen}) the (untilded) 2PI effective action is
\bea
\label{eq:2PIdef}
\Gamma[\phi,D] = \Gamma_1[\phi,D] - i \Phi[\phi,D] ~+~{\rm const}\,,
\eea
where we have combined the tree and 1-loop terms by introducing the notation
\bea
\label{Gamma1def}
\Gamma_1[\phi,D] := S_{cl}[\phi] + \frac{i}{2} \Tr \ln D^{-1} +
\frac{i}{2} \Tr\,\left[\left(D^{0}(\phi)\right)^{-1} D\right]\,.
\eea
Using Eq. (\ref{eq:2PIdef})  we rewrite the tilded 2PI effective action as
\bea
\label{Gammatilde}
\tilde\Gamma[\phi,D]&& = \tilde\Gamma_1[\phi,D]- i \tilde{\Phi}[\phi,D]~~+~~{\rm const}\,,\\[2mm]
&&  =  \Gamma_1[\phi,D] + \Delta \Gamma_1[\phi,D]- i \tilde{\Phi}[\phi,D]~~+~~{\rm const}\,,\nonumber
\eea
where we have defined
\bea
\label{delGamma1def}
\Delta \Gamma_1[\phi,D]:=\tilde \Gamma_1[\phi,D]-\Gamma_1[\phi,D]\,.
\eea

\begin{figure}[H]
\begin{center}
\includegraphics[width=7cm]{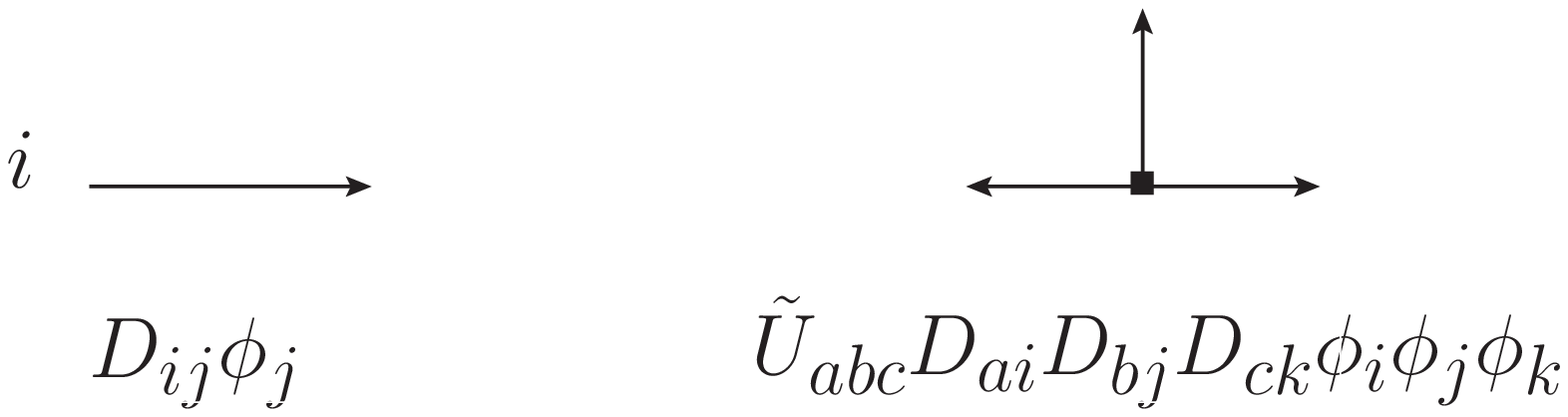}
\end{center}
\caption{\label{phiArrow}Notation used to draw $\phi$ fields.}
\end{figure}

\begin{figure}[H]
\begin{center}
\epsfig{file=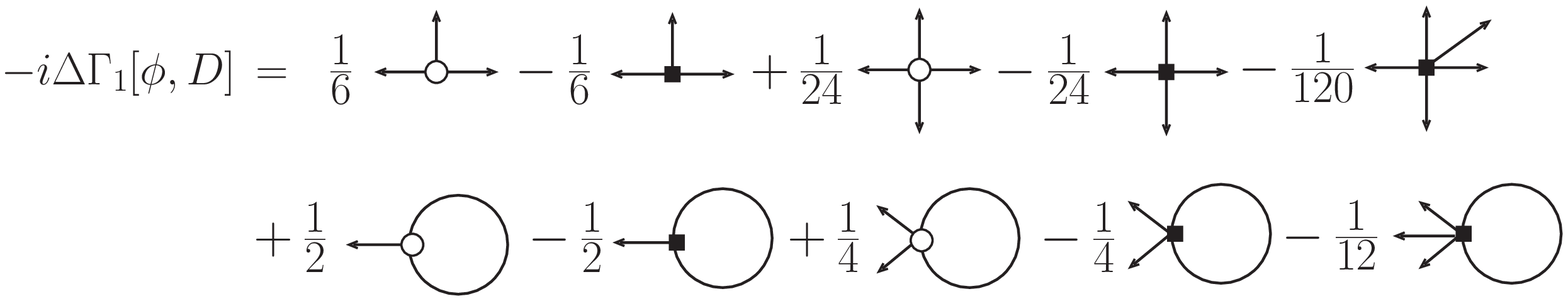,width=16cm}
\end{center}
\caption{\small Diagrammatic representation of Eq. (\ref{Gamma1final}).}\label{fig:fig2}
\end{figure}

From Eqs. (\ref{scl}), (\ref{tildeVert}) and (\ref{Gamma1def}) we obtain an explicit expression for $\Delta \Gamma_1[\phi,D]$:
\bea
\label{Gamma1final}
\Delta \Gamma_1[\phi,D] &=& - \frac{ i}{6} \phi^3 (U^{oo}- \tilde{U}) - \frac{ i}{24} \phi^4 (V^{0}- \tilde{V})
+ \frac{ i}{120} \phi^5  \tilde{W} \nonumber \\
&-& \frac{ i}{2} \Tr\,\left[\phi (U^{oo}- \tilde{U})+ \frac{1}{2} \phi^2 (V^{0}- \tilde{V}) - \frac{1}{6} \phi^3  \tilde{W}
\right ] D\,.
\eea
In order to represent this equation diagrammatically, we draw the $\phi$ fields as arrows. This notation is illustrated with two examples in Fig. \ref{phiArrow}. Equation (\ref{Gamma1final}) is shown in Fig. \ref{fig:fig2}.
\par

\begin{figure}[H]
\begin{center}
\epsfig{file=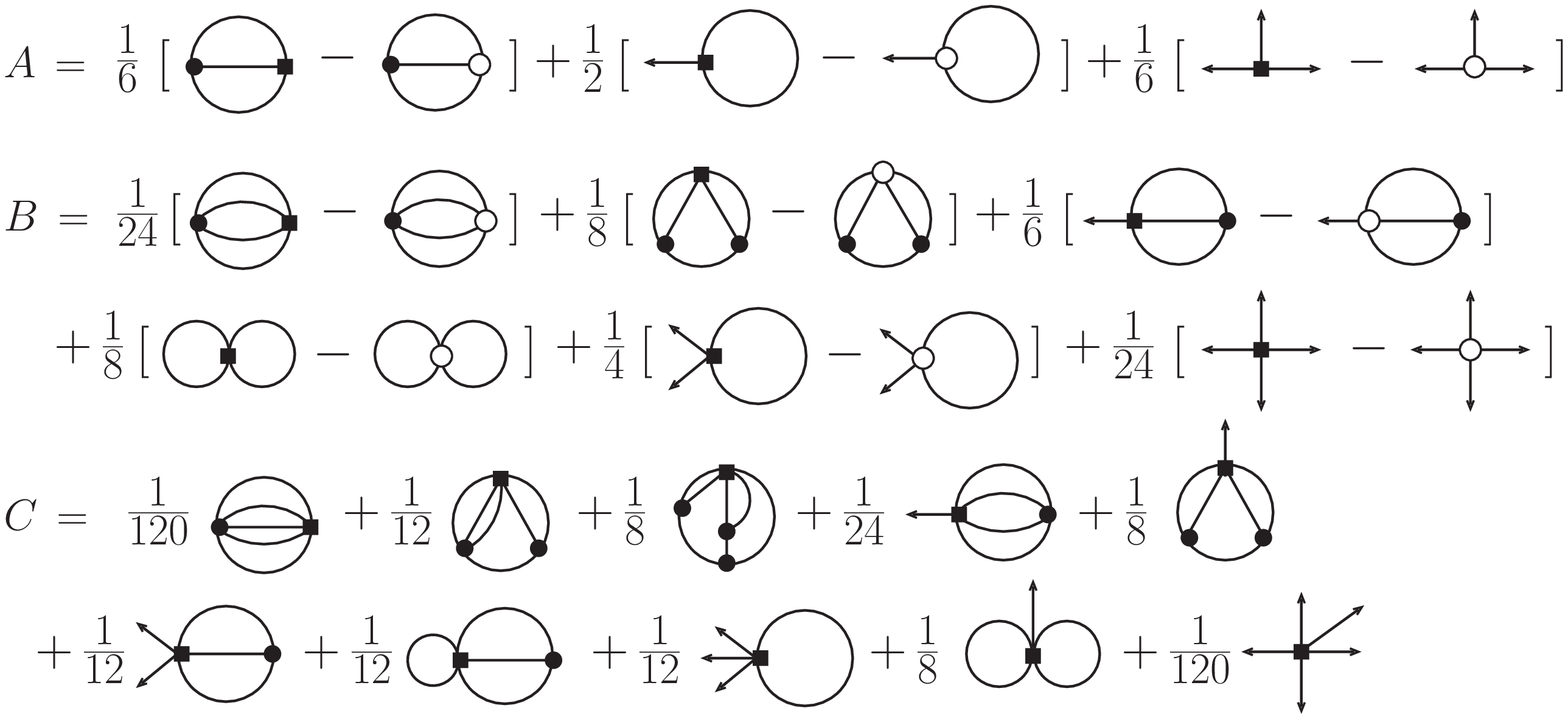,width=17cm}
\end{center}
\caption{\small Diagrammatic representation of Eqs. (\ref{ABCdef}) and (\ref{eq:DGV}).} \label{fig:fig3}
\end{figure}

Now we calculate the last three terms in Eq. (\ref{dormant2}). We define
\bea
\label{ABCdef}
A:= i
\frac{\delta\tilde{\Gamma}[\phi,D]}{\delta\tilde{U}}(\tilde{U}-U^{oo})
\,,~~
B:= i \frac{\delta\tilde{\Gamma}[\phi,D]}{\delta\tilde{V}}(\tilde{V}-V^0)\,,~~
C:= i \frac{\delta\tilde{\Gamma}[\phi,D]}{\delta\tilde{W}}\tilde{W}\,.
\eea
We can write the derivatives that appear in these terms using Eqs. (\ref{connect2}), (\ref{proper2}), (\ref{tildeVert}) and (\ref{relate}):
\bea
\label{eq:DGV}
\frac{\delta
\tilde{\Gamma}[\phi,D]}{\delta \tilde{U}} = &&- \frac{i}{6}
\left(D^3 U + (3) D \phi + \phi^3 \right)\, ,
 \\
\frac{\delta \tilde{\Gamma}[\phi,D]}{\delta \tilde{V}} =&& -
\frac{i}{24} \left(D^4 V + (3) D^5 U^2 + (4) D^3 U \phi + (3) D^2 + (6) D
\phi^2 + \phi^4 \right) \, , \nonumber \\
\frac{\delta \tilde{\Gamma}[\phi,D]}{\delta \tilde{W}} =&& -
\frac{i}{120} (D^5 W + (10) D^6 V U + (15) D^7 U^3 + (5) D^4 V \phi + (15)
D^5 U^2 \phi \nonumber \\ [2mm]
&& + (10) D^3 U \phi^2 + (10) D^4 U + (10) D
\phi^3 + (15) D^2 \phi + \phi^5 ) \, .\nonumber  \eea
Substituting (\ref{eq:DGV}) into Eqs. (\ref{ABCdef}) we obtain the result shown in Fig. \ref{fig:fig3}.

Using  Eqs. (\ref{Gammatilde}) and (\ref{ABCdef}) to rewrite Eq. (\ref{dormant2}) we obtain
\bea
\label{gammaSoFarx}
\Gamma[\phi,D,U,V,W] = \Gamma_1[\phi,D] + \underbrace{\Delta \Gamma_1[\phi,D]}_{{\rm Fig.}~\ref{fig:fig2}}- i \tilde{\Phi}[\phi,D]+ i \underbrace{(A+B+C)}_{{\rm Fig.}~\ref{fig:fig3}}~~+~~{\rm const}\,.
\eea
The diagrams in Fig. \ref{fig:fig2} exactly cancel with the tree and 1-loop diagrams in Fig. \ref{fig:fig3} [this is the reason that we chose to write the tilded 1-Loop effective action immediately in terms of the variational vertices in Eq. (\ref{Gammatilde})]. We obtain
\bea
\label{gammaSoFar}
\Gamma[\phi,D,U,V,W] = \Gamma_1[\phi,D] - i \tilde{\Phi}[\phi,D]+ i \underbrace{(A^\prime+B^\prime +C^\prime)}_{{\rm Fig.}~\ref{fig:fig3}~~L\ge 2}~~+~~{\rm const}\,,
\eea
where the primes indicate that the 1-Loop terms have been removed.

We compare Eq. (\ref{gammaSoFar}) with the expression for the 5PI effective action obtained from Eqs. (\ref{gammaGen}) and (\ref{Gamma1def}):
\beq
\label{gammaGoal}
\Gamma[\phi,D,U,V,W] = \Gamma_1(\phi,D)  -
i \Phi^0[\phi,D,U,V] - i \Phi^{\rm int}[D,U,V,W] ~~+~~{\rm const}\, .
\eeq
Equating the right sides of Eqs. (\ref{gammaSoFar}) and (\ref{gammaGoal})
we obtain
\bea
\label{match1}
\Phi^0[\phi,D,U,V] +\Phi^{\rm int}[D,U,V,W] = \tilde{\Phi}[\phi,D] - \underbrace{(A^\prime + B^\prime + C^\prime)}_{{\rm Fig.}~\ref{fig:fig3}~~L\ge 2}\,.
\eea
Note that Eq. (\ref{match1}) gives a formal result for  $\Phi^0[\phi,D,U,V]$ and $\Phi^{\rm int}[D,U,V,W]$ as a functional of the tilde vertices. In the next section, we show how to convert this expression to a functional of  variational proper vertices.

\ts

We can write the diagrams contained in the right-hand side (rhs) of Eq. (\ref{match1}) in a compact way by introducing ``effective tilde'' vertices.
The definitions we will use are
\bea
\label{tildePrime}
{\tilde{U}}^{\prime}:= {\tilde{U}}
+ \phi {\tilde{V}} +\frac{1}{2}(\phi^2 {\tilde{W}} + D
{\tilde{W}})\,,~~
{\tilde{V}}^{\prime}:={\tilde{V}} + \phi
{\tilde{W}}\,,~~
{\tilde{W}}^{\prime}:= {\tilde{W}}\,.
\eea
We write these vertices collectively using the notation $\tilde {\cal V}_i^\prime\in\{\tilde U^\prime,\tilde V^\prime,\tilde W^\prime\}$.
The diagrammatic notation we will use is shown in Fig. \ref{vertSecond}.

\begin{figure}[H]
\begin{center}
\includegraphics[width=11cm]{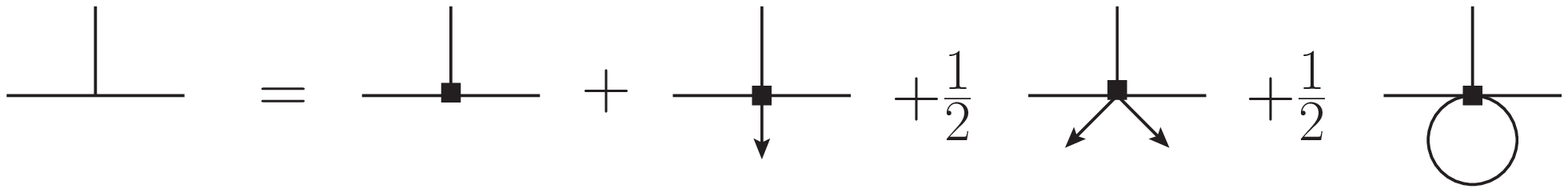}
\end{center}
\caption{\label{UtildePrime}The vertex ${\tilde{U}}^{\prime}$ defined in Eq. (\ref{tildePrime}).}
\end{figure}

We make some comments about the structure of the definitions of the effective tilde  vertices. Recall that the effective bare propagator and effective bare vertex [see Eq. (\ref{free})] are constructed so that the explicit $\phi$-dependence in the (untilded) $\Phi$ is absorbed into these definitions. Similarly, for the tilded functional $\tilde\Phi[\phi,D]$, we could remove all the explicit $\phi$-dependence with the definitions in Eq. (\ref{tildePrime}), if we did not include the last term in the definition of ${\tilde{U}}^{\prime}$, which produces the tadpole-type diagram in Fig. \ref{UtildePrime} (the last graph on the rhs).
\par
\begin{figure}[H]
\begin{center}
\includegraphics[width=9cm]{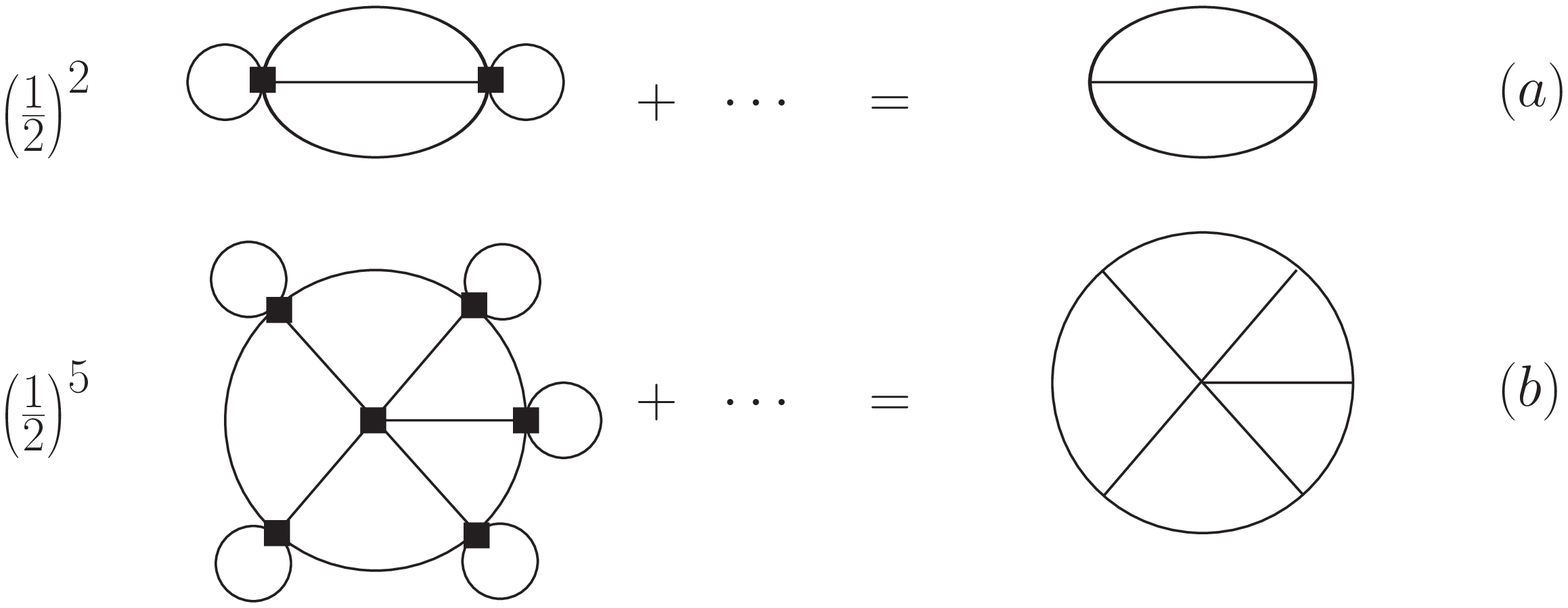}
\end{center}
\caption{\label{tadEx}Examples of tadpole graphs that are included using the definition of ${\tilde{U}}^{\prime}$ in Eq. (\ref{tildePrime}).}
\end{figure}

\noindent This tadpole-type term is included because it absorbs all of the tadpole graphs in the 5-Loop $\tilde\Gamma[\phi,D]$. An example of this is shown in part (a) of Fig. \ref{tadEx}.
In order to understand why all tadpole graphs are absorbed by this definition, we note that, in the absence of the 5-point interaction, there are no tadpole graphs in the 2PI effective action, at any loop order [because they are 2-particle reducible (2PR)], and therefore all possible tadpole graphs have the form of bubbles attached to 5-point vertices, as in the last graph in Fig. \ref{UtildePrime}. We include some $L \ge 6$ loop tadpoles in $\tilde\Gamma[\phi,D]$ so that the tadpole term in the definition of ${\tilde{U}}^{\prime}$ absorbs all tadpole graphs (see footnote 4). An example is shown in part (b) of Fig. \ref{tadEx}. These higher loop tadpoles are included for organizational purposes only, and allow us to present the calculation in a more compact way. We have checked that we get the same result for the 5PI effective action if we drop the tadpole term in the definition of the effective tilde $U$-vertex, and include only 5-Loop 2PI tadpole graphs in $\tilde\Phi[\phi,D]$ (which are then not absorbed into the effective vertices).
\par

\begin{figure}[H]
\begin{center}
\epsfig{file=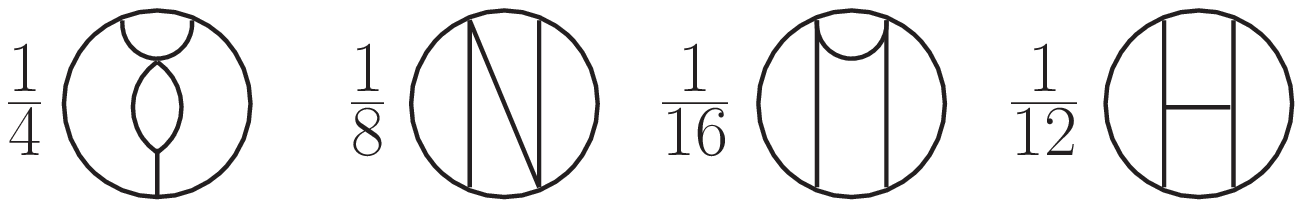,width=8cm}
\end{center}
\vspace*{-0.5cm} \caption{\small 4-loop diagrams in $\tilde\Phi^\prime[D]$ that are 2PI and 4PR. }
\label{add4LABEL}
\end{figure}

\begin{figure}[H]
\begin{center}
\epsfig{file=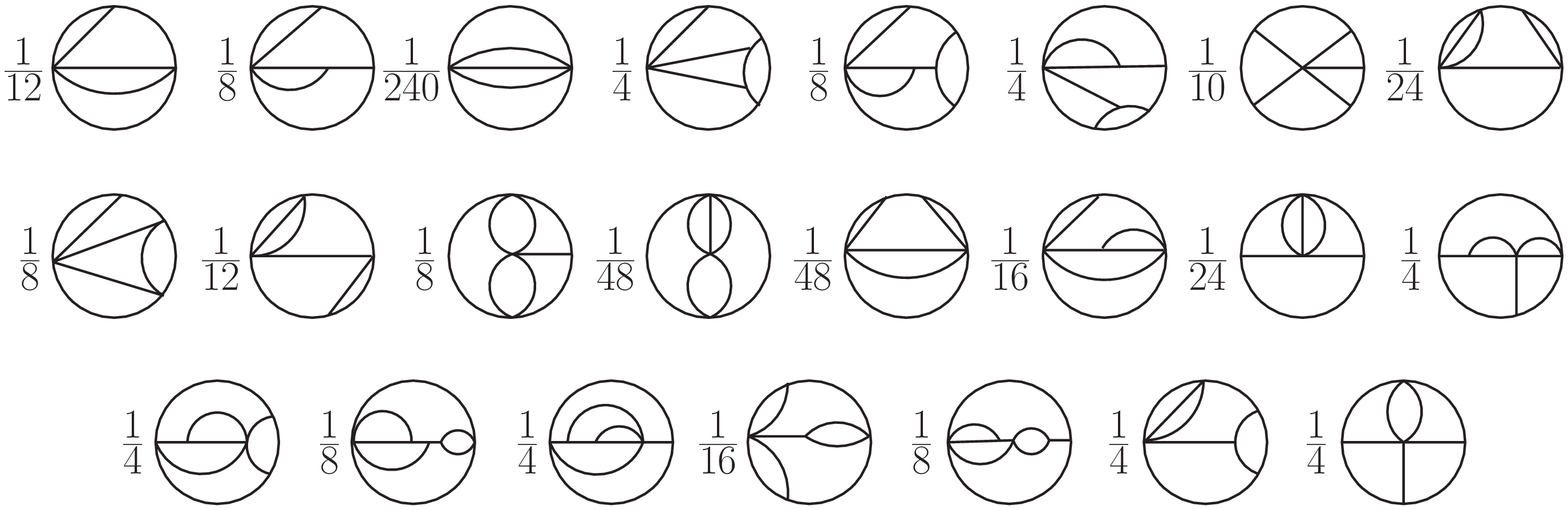,width=15cm}
\end{center}
\vspace*{-0.5cm} \caption{\small 4- and 5-loop diagrams in $\tilde\Phi^\prime[D]$ that contain the vertex $\tilde{W}^{\prime}$.}
\label{fig:fig5}
\end{figure}

We can use the vertices defined in Eq. (\ref{tildePrime}) to rewrite the two terms on the rhs of Eq. (\ref{match1}) in a compact way:

In the first term, all of the $\phi$-dependence is absorbed into the effective tilde vertices and we write: $\tilde\Phi[\phi,D]:=\tilde \Phi^\prime[D]$. The quantity $\tilde \Phi^\prime[D]$ contains only effective tilde vertices ($\tilde {\cal V}^\prime_i$) and no explicit $\phi$-dependence.
There are 71 diagrams in the 5-Loop result for  $\tilde{\Phi}^\prime[D]$:
\begin{itemize}
\item There are 9 diagrams with $L\le 4$ loops which are 4PI. We get these diagrams from Figs. \ref{Phi0Fig} and \ref{Phi4Fig} by replacing all vertices with tilde effective vertices $\tilde {\cal V}^\prime$ [note that the two figures contain 11 diagrams, but the (EGG, EGG$_0$) and (BBALL, BBALL$_0$) diagrams combine, which leaves 9 diagrams].
\item There are 4 additional 4-loop diagrams that are 2PI and 4PR, and  depend on $\tilde U^\prime$ and $\tilde V^\prime$ but not the effective 5-vertex $\tilde{W}^{\prime}$. These diagrams are shown in Fig. \ref{add4LABEL}.
\item There are 35 5-loop 2PI diagrams that do not contain the effective 5-vertex. These diagrams are given in Ref. \cite{kajantie}, in Eq. (59).
%survivors in redKeep.eps
\item There are 23 4- and 5-loop diagrams that contain the vertex
$\tilde{W}^{\prime}$ which are given in Fig. \ref{fig:fig5}.
\end{itemize}

\begin{figure}[H]
\begin{center}
\epsfig{file=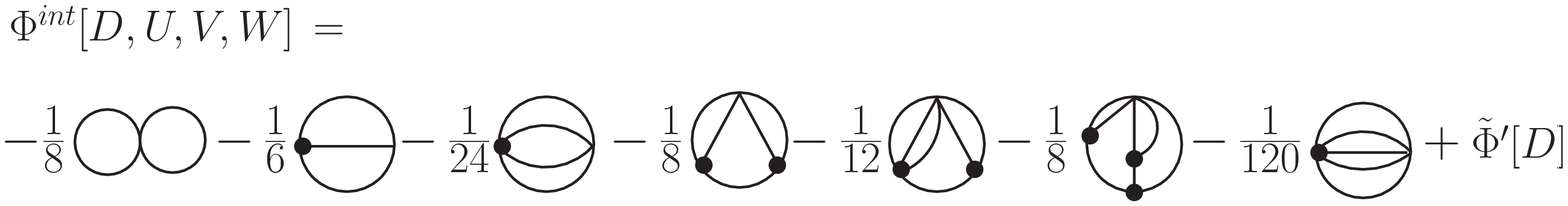,width=17cm}
\end{center}
\caption{\small Diagrammatic representation of $\Phi^{int}[D, U, V, W]$.}
\label{fig:fig4}
\end{figure}

The second term on the rhs of Eq. (\ref{match1}) is shown in Fig.  \ref{fig:fig3}. It is straightforward to rewrite these diagrams in terms of effective tilde vertices using Eq. (\ref{tildePrime}). As an example, we consider the six diagrams with the EGG topology\footnote{The six different graphs with the EGG topology differ from each other by having different vertices and sometimes different numerical coefficients. Throughout this section, all graphs with the EGG topology will be referred to generically as EGG diagrams, and similarly for all other topologies. The vertices for a specific diagram are shown in the corresponding figure.}. The 2nd diagram in $A$ and the 6th diagram in $B$ combine to give: $-\frac{1}{6}U\,D^3 \,U^0$. Including the minus sign in Eq. (\ref{match1}), this is the EGG$_0$ graph in $\Phi^0[\phi,D,U,V]$ (see Fig. \ref{Phi0Fig}). The 1st diagram in $A$, the 5th diagram in $B$, and the 6th and 7th diagrams in $C$ give: $\frac{1}{6}U\,D^3\,\tilde U^\prime$. Including the minus sign in Eq. (\ref{match1}), this is the 2nd graph in Fig. \ref{fig:fig4}. Combining all terms, we find that the $L\ge 2$ loop terms in Fig. \ref{fig:fig3} produce $\Phi^0[\phi,D,U,V]$ and seven additional diagrams.

\begin{figure}[H]
\begin{center}
\epsfig{file=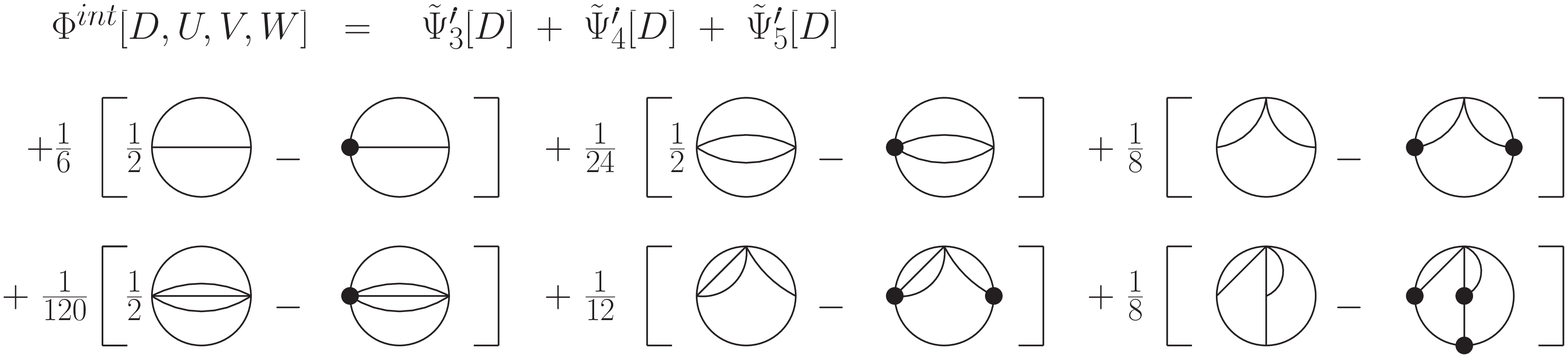,width=16.5cm}
\end{center}
\vspace*{-0.5cm} \caption{\small Rearranged version of Fig. \ref{fig:fig4}.}
\label{fig4NEW}
\end{figure}

Substituting these results into Eq. (\ref{match1}),  $\Phi^0[\phi,D,U,V]$ cancels and we obtain the result shown in Fig. \ref{fig:fig4}. The EIGHT diagram in Fig. \ref{fig:fig4} cancels identically with the EIGHT diagram in $\tilde\Phi^\prime[D]$. The remaining six diagrams in Fig. \ref{fig:fig4} have partners in $\tilde\Phi^\prime[D]$ which have the same topology, but all vertices are effective tilde vertices. We extract these diagrams and group the remaining terms into a set of functionals which we call $\tilde\Psi_i^\prime[D]$, where the subscript indicates loop order. These functionals include all of the  diagrams in the list under Fig. \ref{tadEx} except the EIGHT, EGG, EGG$_0$, BBALL, BBALL$_0$, and HAIR graphs, and the three 4-loop graphs at the beginning of the first line of Fig. \ref{fig:fig5}. This rearranged version of Eq. (\ref{fig:fig4}) is shown in Fig. \ref{fig4NEW}.

\subsection{Vertex Inversion}

In this subsection, the goal is to obtain expressions that will allow us to write the rhs of the equation represented in Fig. \ref{fig4NEW} in terms of proper vertices. The end result in this subsection is a set of expressions of the form
\bea
\label{results}
&& \tilde{U}^{\prime}= U + f_U^{(1)}[{\cal{V}}_i] + f_U^{(2)}[{\cal{V}}_i] + f_U^{(3)}[{\cal{V}}_i] +\cdots\\
&& \tilde{V}^{\prime}= V + f_V^{(1)}[{\cal{V}}_i] + f_V^{(2)}[{\cal{V}}_i] + f_V^{(3)}[{\cal{V}}_i] +\cdots\nonumber\\
&& \tilde{W}^{\prime}= W + f_W^{(1)}[{\cal{V}}_i] + f_W^{(2)}[{\cal{V}}_i] + f_W^{(3)}[{\cal{V}}_i] +\cdots\nonumber
\eea
where $f_U$, $f_V$, and $f_W$ are functionals of proper variational vertices, and the superscripts indicate their loop orders.  These expressions will be substituted into Fig. \ref{fig4NEW} to produce the final result for the 5-Loop 5PI effective action, as a functional of proper variational vertices.

The lowest order diagrams in Fig. \ref{fig4NEW} that contain the 3-vertex $\tilde{U}^{\prime}$ are the EGG-type diagrams in the first set of square brackets. Since these diagrams are themselves 2-loop, it appears that, in order to obtain the effective action to 5-Loop order,
 we need to keep the 3-loop term (of the form $f_U^{(3)}[{\cal{V}}_i]$) in Eq. (\ref{results}). The lowest order diagrams that contain the 4-vertex $\tilde{V}^{\prime}$ are the BBALL-type diagrams in the second set of square brackets, which are 3-loop, indicating that we need to keep the 2-loop term (of the form $f_V^{(2)}[{\cal{V}}_i]$) in the expansion of the 4-vertex $\tilde{V}^{\prime}$. The lowest order diagrams that contain the 5-vertex $\tilde{W}^{\prime}$ are the 4-loop graphs in the fourth set of square brackets, which indicates that we need to keep the  1-loop term (of the form $f_{W}^{(1)}[{\cal{V}}_i]$) in the expansion of the  5-vertex $\tilde{W}^{\prime}$.

From the argument above, we would conclude that in order to obtain the 5-Loop 5PI effective action, we need to keep terms ${\cal O} \{f_{U}^{(3)}[{\cal{V}}_i],f_{V}^{(2)}[{\cal{V}}_i],f_{W}^{(1)}[{\cal{V}}_i]\}$ in Eq. (\ref{results}).
However, it is easy to show that these highest order terms do not contribute. Consider the first set of square brackets in Fig. \ref{fig4NEW} which contains the EGG-type diagrams. Substituting the expansions in Eq. (\ref{results}) and keeping only the 5-loop contributions that contain $f_U^{(3)}$ we obtain
\bea
\frac{1}{2}\big(f_U^{(3)}\,D^3 \,U+U\,D^3 \,f_U^{(3)}\big)-U \,D^3\,f_U^{(3)} = 0\,.
\eea
It is easy to see that the 5-loop contributions which contain $f_{V}^{(2)}$ and $f_{W}^{(1)}$ from the second and fourth terms in square brackets  also cancel.
The result is that the highest terms that we need to keep have the form
$
\{f_{U}^{(2)}[{\cal{V}}_i],f_{V}^{(1)}[{\cal{V}}_i],f_{W}^{(0)}[{\cal{V}}_i]=W\}$,
which means that we can use Eq. (\ref{results}) in the form:
\bea
\label{resultsShort}
&& \tilde{U}^{\prime}= U + f_U^{(1)}[{\cal{V}}_i] + f_U^{(2)}[{\cal{V}}_i] \,,\\
&& \tilde{V}^{\prime}= V + f_V^{(1)}[{\cal{V}}_i] \, ,\nonumber\\
&& \tilde{W}^{\prime}= W \, .\nonumber
\eea

Now we describe how to calculate the $f$ functions in Eq. (\ref{resultsShort}). The first step is to rewrite the derivatives on the lhs of Eq. (\ref{eq:DGV}). We start by using Eqs. (\ref{Gammatilde}), (\ref{delGamma1def}) and (\ref{Gamma1final}) to separate the $L\le 1$ and $L\ge 2$ pieces of $\tilde\Gamma[\phi,D]$:
\bea
\label{eq:ex}
&& \frac{\delta
\tilde{\Gamma}[\phi,D]}{\delta \tilde{U}} = - \frac{i}{6}\left( \phi^3 +
 (3)  D \phi + 6
\frac{\delta \tilde{\Phi}[\phi,D]}{\delta \tilde{U}}\right)\, ,\\
&& \frac{\delta \tilde{\Gamma}[\phi,D]}{\delta \tilde{V}} = -
\frac{i}{24}\left( \phi^4 + (6) D \phi^2 + 24
\frac{\delta \tilde{\Phi}[\phi,D]}{\delta \tilde{V}}\right)\, ,\nonumber \\
&& \frac{\delta \tilde{\Gamma}[\phi,D]}{\delta \tilde{W}} = -
\frac{i}{120} \left( \phi^5 + (10) D \phi^3 +120 \frac{\delta
\tilde{\Phi}[\phi,D]}{\delta \tilde{W}}\right)\, .\nonumber \eea
Equating the right sides of Eqs. (\ref{eq:DGV}) and (\ref{eq:ex}) gives
\bea
\label{reuvwFirst}
&& D^3 U =  3! \frac{\delta \tilde{\Phi}[\phi,D]}{\delta \tilde{U}}
\, ,\\
&& D^4 V + (3) D^5 U^2
 + (4) D^3 U \phi + (3) D^2 =  4! \frac{\delta \tilde{\Phi}[\phi,D]}{\delta
\tilde{V}}  \, ,\nonumber\\
&& D^5 W +
(10)
D^6 V U + (15) D^7 U^3  + (5) D^4 V \phi + (15) D^5 U^2 \phi + (10) D^3 U \phi^2 \nonumber\\
&&\hspace*{7cm} + (10) D^4 U + (15) D^2 \phi  =  5!
\frac{\delta \tilde{\Phi}[\phi,D]}{\delta \tilde{W}} \, .\nonumber
\eea

As we have seen in the previous subsection, the quantity $\tilde{\Phi}[\phi,D]$ can be written in a compact way in terms of the effective tilde vertices as $\tilde\Phi^\prime[D]$. In order to make use of this result, we need to convert the functional derivatives with respect to tilde vertices in Eq. (\ref{reuvwFirst}) into derivatives with respect to effective tilde vertices.  Using Eq. (\ref{tildePrime}) to obtain $\delta\tilde W^\prime/\delta\tilde V=\delta\tilde W^\prime/\delta\tilde U = \delta\tilde V^\prime/\delta\tilde U=0$ we have
\bea
\label{chain}
&& \frac{\delta \tilde{\Phi}[\phi,D]}{\delta \tilde{U}}
%=\frac{\delta \tilde{\Phi}[\phi,D]}{\delta \tilde{U}^{\prime}}+\frac{\delta \tilde{\Phi}[\phi,D]}{\delta \tilde{V}^{\prime}}\frac{\delta \tilde{V}^{\prime}}{\delta \tilde{U}} + \frac{\delta \tilde{\Phi}[\phi,D]}{\delta \tilde{W}^{\prime}}\frac{\delta \tilde{W}^{\prime}}{\delta \tilde{U}}
=\frac{\delta \tilde{\Phi}^\prime[D]}{\delta \tilde{U}^{\prime}}\,,\\
&& \frac{\delta \tilde{\Phi}[\phi,D]}{\delta \tilde{V}}=\frac{\delta \tilde{\Phi}^\prime[D]}{\delta \tilde{U}^{\prime}}\frac{\delta \tilde{U}^{\prime}}{\delta \tilde{V}} + \frac{\delta \tilde{\Phi}^\prime[D]}{\delta \tilde{V}^{\prime}}\,,\nonumber\\
&& \frac{\delta \tilde{\Phi}[\phi,D]}{\delta \tilde{W}}=\frac{\delta \tilde{\Phi}^\prime[D]}{\delta \tilde{U}^{\prime}}\frac{\delta \tilde{U}^{\prime}}{\delta \tilde{W}}+\frac{\delta \tilde{\Phi}^\prime[D]}{\delta \tilde{V}^{\prime}}\frac{\delta \tilde{V}^{\prime}}{\delta \tilde{W}}  + \frac{\delta \tilde{\Phi}^\prime[D]}{\delta \tilde{W}^{\prime}}\,.\nonumber
\eea
Using Eq. (\ref{tildePrime}) again, we can obtain explicit expressions for $\delta\tilde U^\prime/\delta\tilde V$, $\delta\tilde U^\prime/\delta\tilde W$, $\delta\tilde V^\prime/\delta\tilde W$, and simultaneously solve the set of equations in Eq. (\ref{reuvwFirst}) and (\ref{chain}), eliminating the functional derivatives with respect to the tilde vertices. We obtain
\bea
\label{reuvw}
&&D^3 U =  6 \frac{\delta \tilde{\Phi}^\prime[D]}{\delta \tilde{U}^{\prime}}  \,,\\
&&  D^4 V + (3) D^5 U^2 + (3) D^2 = 24 \frac{\delta \tilde{\Phi}^\prime[D]}{\delta \tilde{V}^{\prime}} \,, \nonumber\\ && D^5 W + (10) D^6 V U + (15) D^7 U^3 = 120 \frac{\delta \tilde{\Phi}^\prime[D]}{\delta \tilde{W}^{\prime}} \,.\nonumber
\eea
We note that the EIGHT diagram in $\tilde{\Phi}^\prime[D]$ has the form $\frac{1}{8}D^2\,\tilde V^\prime$ and therefore the only  contribution it makes to Eq. (\ref{reuvw}) is to cancel the last term on the left side of the second equation. In the future we will drop this term, and remove the EIGHT diagram from $\tilde{\Phi}^\prime[D]$.

\ts

\begin{figure}[H]
\begin{center}
\epsfig{file=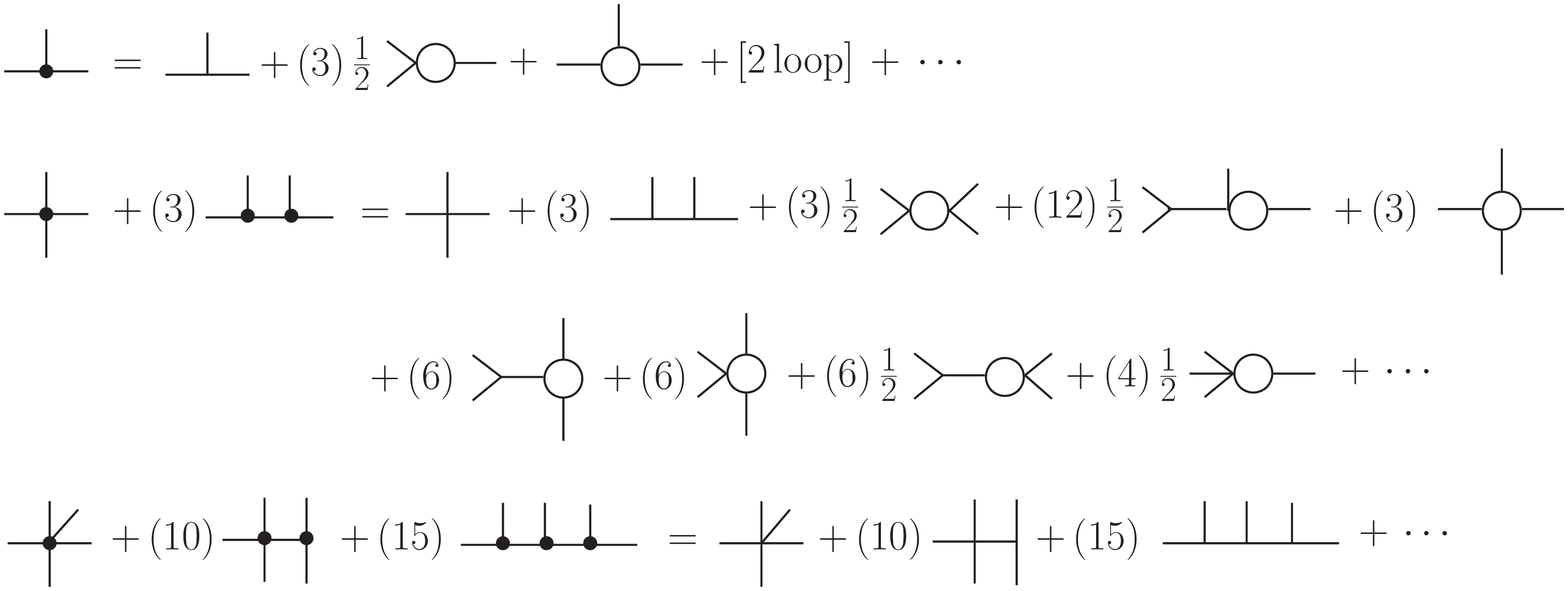,width=16cm}
\end{center}
\vspace*{-0.5cm} \caption{\small Diagrammatic representation of Eq.
(\ref{reuvw}).} \label{fig:fig6}
\end{figure}

\begin{figure}[H]
\begin{center}
\epsfig{file=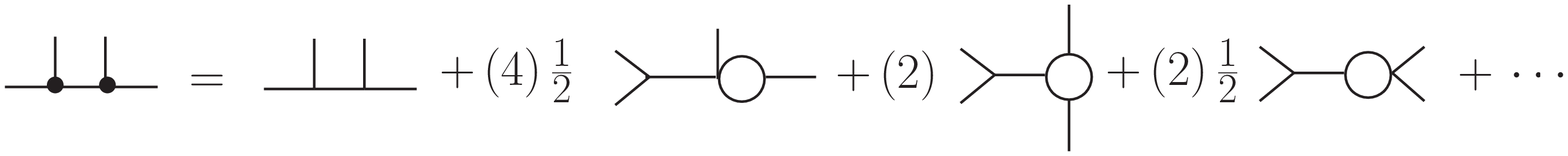,width=16cm}
\end{center}
\caption{\small Iteration of the first line in Fig. \ref{fig:fig6}. } \label{fig:fig8}
\end{figure}

\begin{figure}[H]
\begin{center}
\epsfig{file=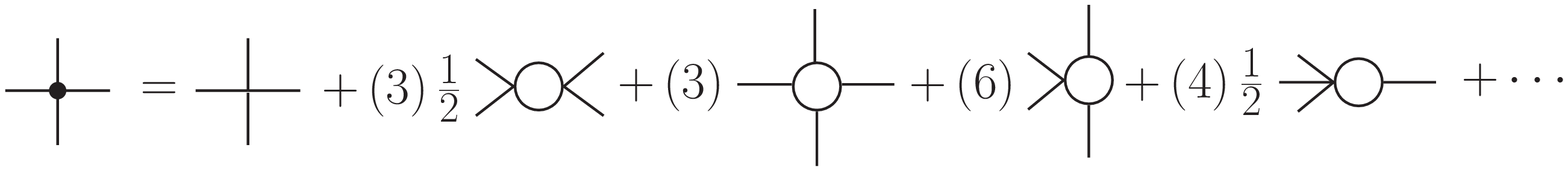,width=15cm}
\end{center}
\caption{\small Rearrangement of the second line in Fig. \ref{fig:fig6}.
} \label{fig:fig6b}
\end{figure}

The next step is to calculate the derivatives in Eq. (\ref{reuvw}). There are 71 diagrams in $\tilde\Phi^\prime[D]$, which are listed under Fig. \ref{tadEx}. However, for some terms, derivatives give contributions that are beyond the order to which we are working [see Eq. (\ref{resultsShort})].
  In Sec. \ref{accuracySection2} we discuss in general the relationship between the loop order of a diagram in the effective action and the loop order of the functional derivative of the diagram with respect to an $i$-point vertex. The result is given in Eq. (\ref{calL}). Using this result we find that, in Eq. (\ref{reuvw}), we only need to calculate derivatives of the $L\le 4$ loop terms in $\tilde\Phi^\prime[D]$. The results are shown in Fig. \ref{fig:fig6}. In accordance with Eq. (\ref{resultsShort}), we keep $L\le 2$ loop terms in the expansion of $U$, $L\le 1$ loop terms in the expansion of $V$, and tree graphs  in the expansion of $W$. The result for the 2-loop terms $f_U^{(2)}$ is not given because it cancels exactly, as we will show below.
We can simplify the results shown in Fig. \ref{fig:fig6}. Substituting the first and second equations into the third, it is clear that the equation for the 5-vertex can be rewritten $W=\tilde W^\prime + \dots$ [where the dots indicate terms with one or more loops which are beyond the order of Eq. (\ref{resultsShort})].  We can also rewrite the equation for the 4-vertex. We iterate the first line in the figure to obtain the result shown in Fig. \ref{fig:fig8} (dropping $L\ge 2$ loop terms). Multiplying this equation by three and subtracting from the second line in Fig. \ref{fig:fig6}, we can write the equation for the proper 4-vertex as shown in Fig.  \ref{fig:fig6b}.

\begin{figure}[H]
\begin{center}
\epsfig{file=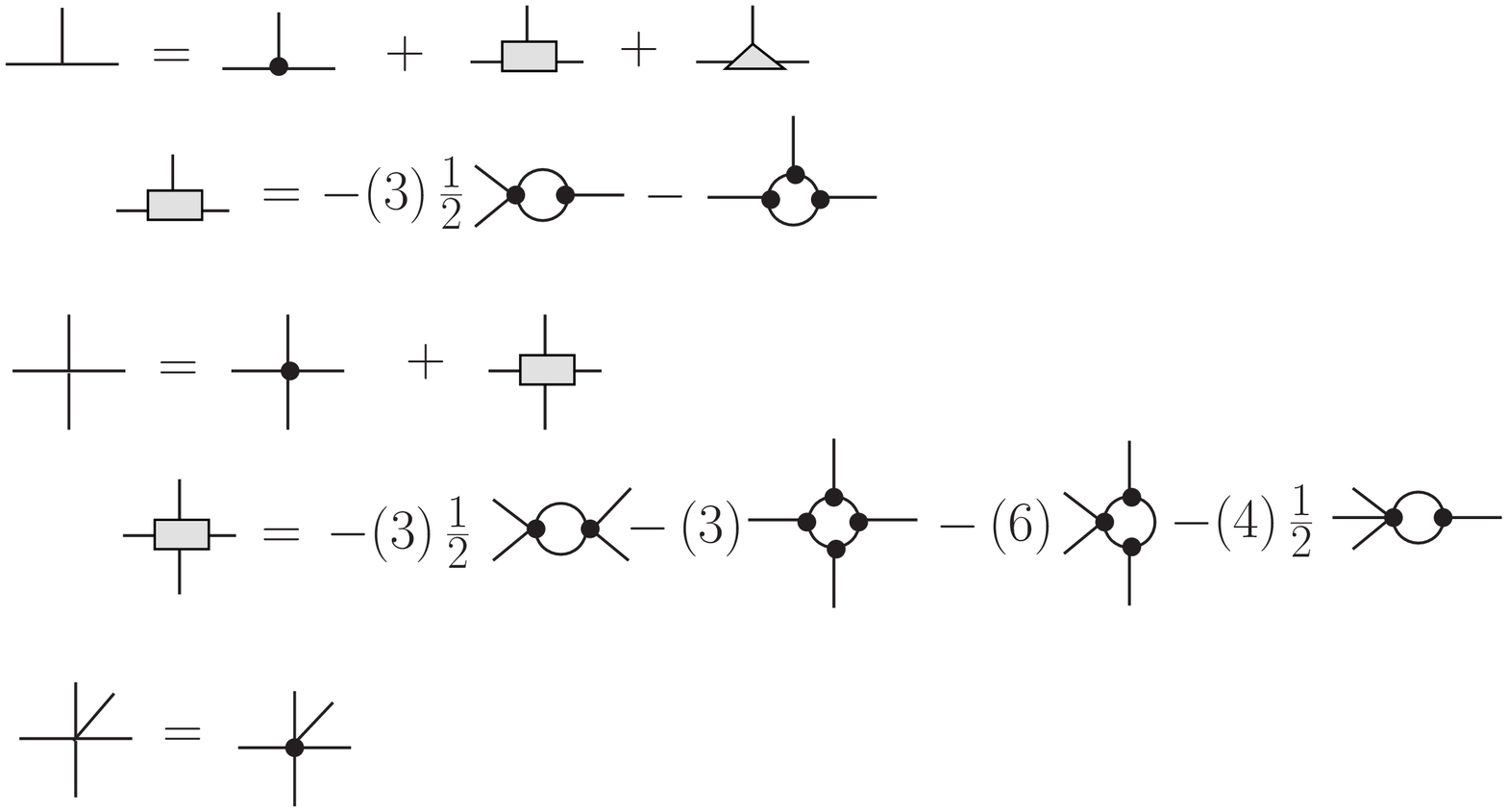,width=14.5cm}
\end{center}
\vspace*{-0.5cm} \caption{\small Diagrammatic representation of Eq. (\ref{resultsShort}).
The grey square blobs denote the contributions to the 3- and 4-point vertices at 1-loop order ($f_U^{(1)}$ and $f_V^{(1)}$), and the grey triangle blob denotes the  contribution to the 3-point vertex at 2-loop order ($f_U^{(2)}$). No result is given for the triangle blob, since it cancels in the final result [see below]. }
\label{fig:fig7}
\end{figure}

The final task is to invert these equations, which can be done using a straightforward iterative process. The complete set of inverted equations is shown in Fig. \ref{fig:fig7}, where we drop all terms that are beyond the order to which we are working (see Eq. (\ref{resultsShort})).

\subsection{Substitutions}
\label{subSection}

In this subsection we discuss the substitution of the results in Fig. \ref{fig:fig7} into the expression for the effective action in Fig. \ref{fig4NEW}.

\begin{figure}[H]
\begin{center}
\epsfig{file=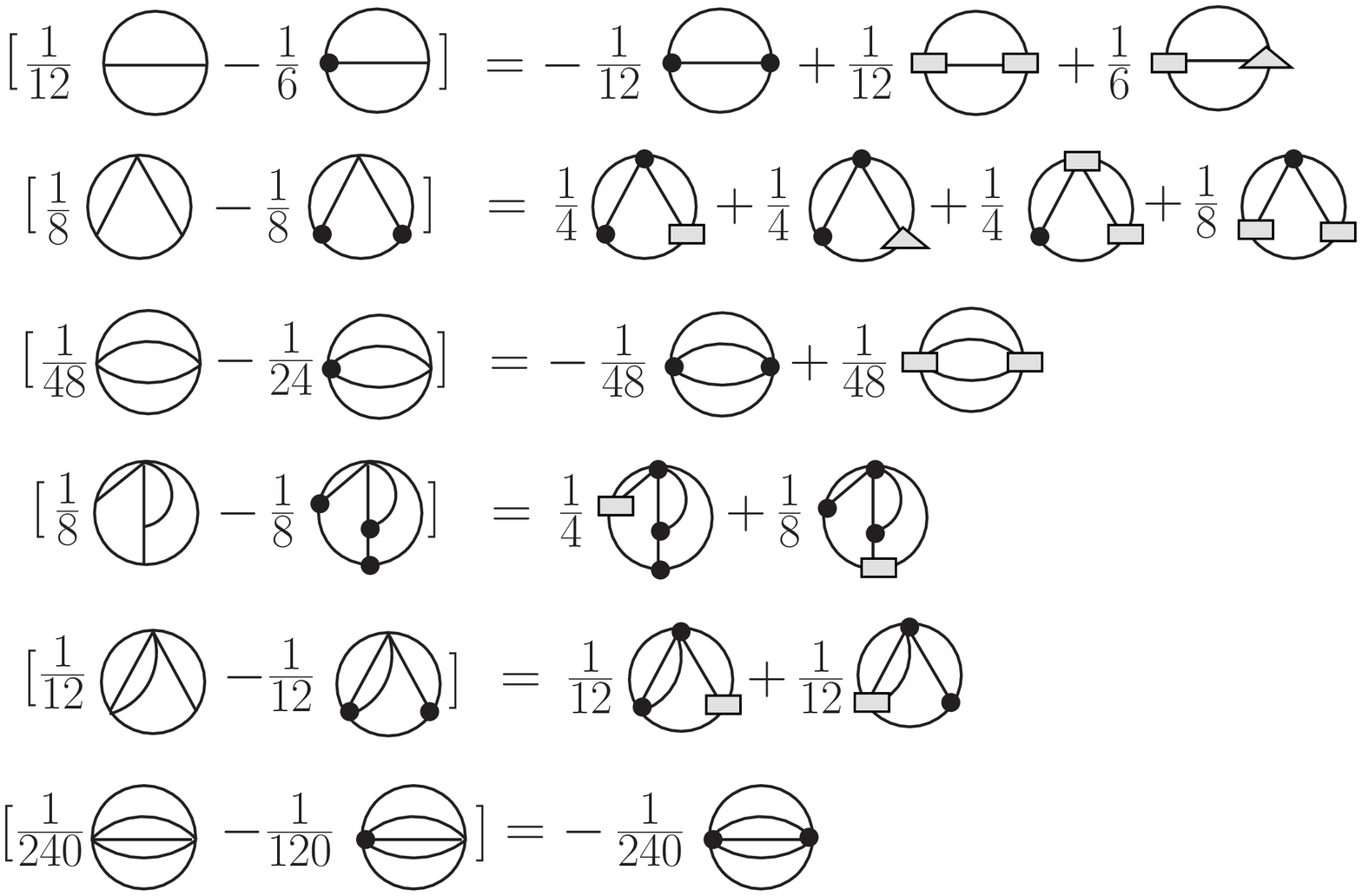,width=12.5cm}
\end{center}
\vspace*{-0.5cm} \caption{\small The result obtained from substituting Fig. \ref{fig:fig7} into Fig. \ref{fig4NEW}. }
\label{fig:fig9}
\end{figure}

First we consider the diagrams in square brackets in Fig. \ref{fig4NEW}.  The results are shown in Fig. \ref{fig:fig9}. We give an example to show how this figure is obtained. The two EGG-type graphs in the first set of square brackets in Fig. \ref{fig4NEW} can be written
\bea
\label{EGGex}
&&\frac{1}{12}D^3(\tilde{U}^{\prime})^2 - \frac{1}{6}D^3 U\tilde U^\prime =
\frac{1}{12}D^3 (U + f_U^{(1)} + f_U^{(2)})^2  - \frac{1}{6}D^3 U\tilde U^\prime\,, \\
&& ~~=\frac{1}{12}D^3 \bigg( 2  U \underbrace{(U + f_U^{(1)} + f_U^{(2)})}_{\tilde U^\prime} - U^2 + (f_U^{(1)})^2
+2f_U^{(1)} f_U^{(2)}\bigg) - \frac{1}{6}D^3 U\tilde U^\prime \, ,\nonumber
\\
&& ~~=\frac{1}{12}D^3 \bigg(  - U^2 + (f_U^{(1)})^2
+2f_U^{(1)} f_U^{(2)} \bigg)\, .\nonumber
\eea
Equation (\ref{EGGex}) corresponds to the first line in Fig. \ref{fig:fig9}. The last three terms in Eq. (\ref{EGGex}) are of loop order (2,4,5), respectively.

\begin{figure}[H]
\begin{center}
\epsfig{file=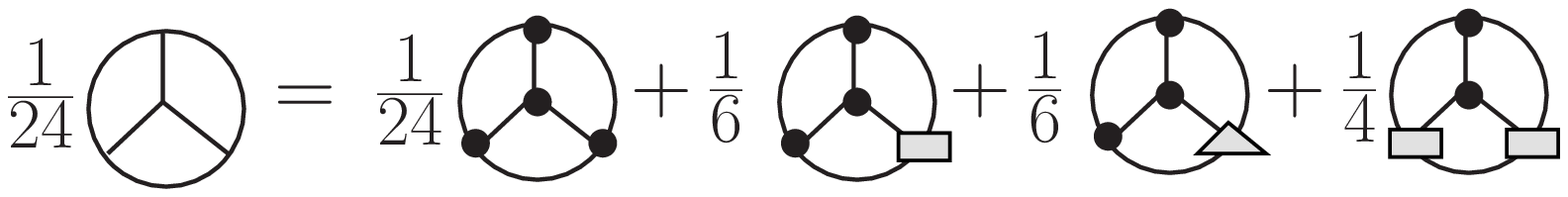,width=11cm}
\end{center}
\vspace*{-0.5cm} \caption{\small Result obtained from substituting Fig. \ref{fig:fig7} into the MERCEDES diagram. }
\label{mercLABEL}
\end{figure}

Next we consider the MERCEDES diagram, which is the only contribution to $\tilde\Psi^\prime_3[D]$. Using Fig. \ref{fig:fig7} we obtain the result shown in Fig. \ref{mercLABEL}. The graphs on the rhs of this figure are of loop order (3,4,5,5), respectively.

\ts

It is now straightforward to collect the 5-Loop terms in $\Phi^{int}[D,U,V,W]$.

For $L\le 4$ loops there are contributions from

\begin{figure}
\begin{center}
\epsfig{file=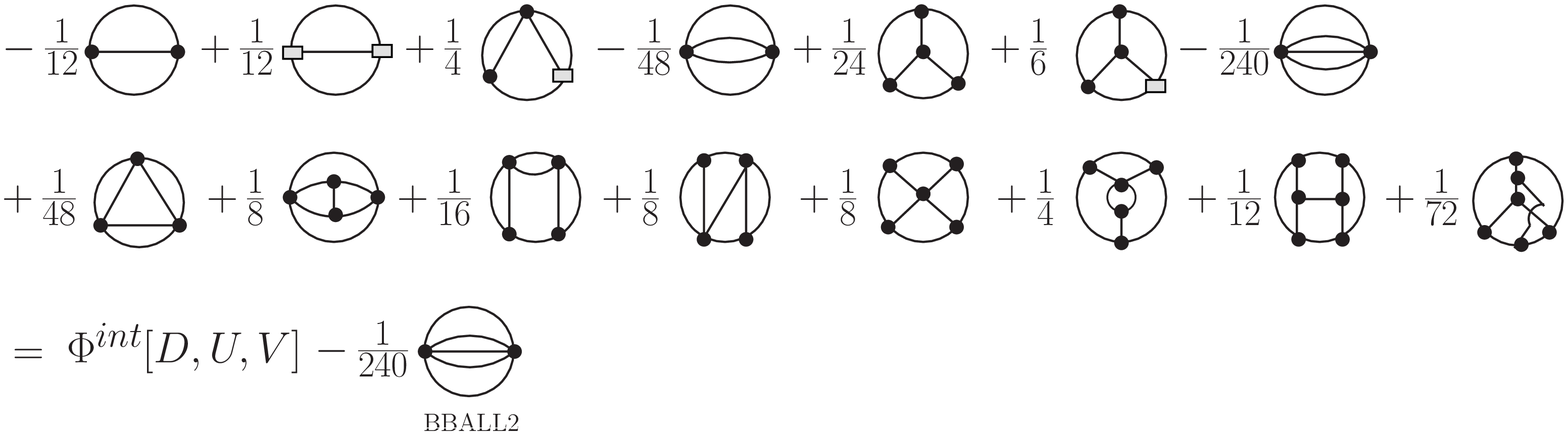,width=18cm}
\end{center}
\vspace*{-0.5cm} \caption{\small The $L\le 4$ loop diagrams contributing to
$\Phi^{int}[D,U,V,W]$. The diagrams in $\Phi^{int}[D,U,V]$ are shown in Fig. \ref{Phi4Fig}. } \label{fig:fig16}
\end{figure}

\begin{itemize}
\item  The $L\le 4$ loop diagrams in Fig. \ref{fig:fig9}.

\item The $L\le 4$ loop diagrams in Fig. \ref{mercLABEL}.

\item  The 4-loop diagrams in $\tilde{\Psi}_4^\prime[D]$ with the effective tilde vertices replaced by the proper ones.
\end{itemize}
Collecting these terms we obtain the set of 4-Loop graphs shown in the first two lines of Fig. \ref{fig:fig16}. After inserting the square blobs (see Fig. \ref{fig:fig7}), we find that the 4PR diagrams cancel and we are left with the 4PI terms in Fig. \ref{Phi4Fig}, and one additional graph that contains an effective 5-point vertex, which we call BBALL2. This result is shown in the third line of Fig. \ref{fig:fig16}.

For $L\le 5$ loops there are additional contributions from

\begin{itemize}
\item  The 5-loop diagrams in Fig. \ref{fig:fig9}.

\item The 5-loop diagrams in Fig. \ref{mercLABEL}.

\item The 4-loop diagrams in $\tilde{\Psi}^\prime_4[D]$ with one effective tilde vertex replaced by the corresponding grey square blob in Fig. \ref{fig:fig7}, and the other effective tilde vertices replaced by proper vertices.

\item The 5-loop diagrams in $\tilde{\Psi}^\prime_5[D]$ with all effective tilde vertices replaced by proper vertices.

\end{itemize}

\begin{figure}[H]
\begin{center}
\epsfig{file=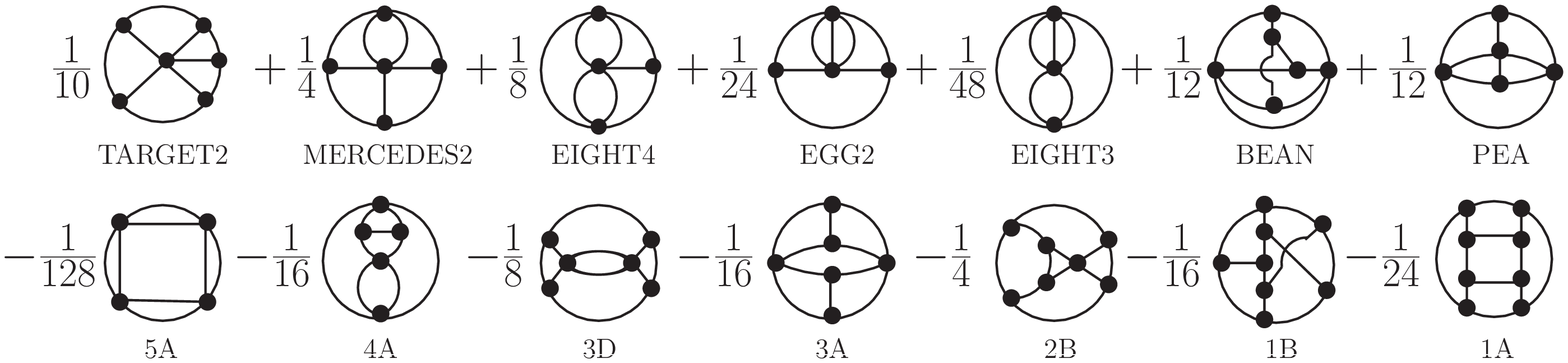,width=15cm}
\end{center}
\vspace*{-0.5cm} \caption{\small 5-loop diagrams contributing
to $\Phi^{int}[D,U,V,W]$.} \label{fig:fig17}
\end{figure}

There are three 5-loop diagrams that contain the
grey triangle blob ($f_U^{(2)}$). They are included in items 1 and 2 in the list above (see Figs. \ref{fig:fig9} and \ref{mercLABEL}). It is easy to show that the sum of these three graphs is zero, by substituting in the expression for the grey square blob in Fig. \ref{fig:fig7}. As a consequence of this cancellation, the explicit form of the 2-loop term $f_U^{(2)}$ is not needed.
%\begin{figure}[H]
%\begin{center}
%\epsfig{file=fig12.eps,width=8.5cm}
%\end{center}
%\vspace*{-0.5cm} \caption{\small An identity related to the grey
%triangle blob ($f_U^2$).} \label{fig:fig12}
%\end{figure}
Collecting all of the other terms in the list above produces, at the end of a long calculation, the diagrams shown in  Fig. \ref{fig:fig17}.

This result has several unexpected features. All of the diagrams in the second line of the figure have negative coefficients, and the coefficient for diagram 1A is twice the factor that would be produced by the normal combinatoric rules for calculating symmetry factors. These diagrams are also 5PR, which means that the ``5PI'' effective action is not 5-particle irreducible. This suggests that the $n$-Loop $n$PI effective action is $n$-particle irreducible for $n\le 4$ only. We discuss the significance of this in the next section.

\section{Structure of the $n$PI effective action and its eom's}
\label{sdnPISection}
In this section, we discuss the structure of the $n$PI effective action, and the relationship between the $n$PI eom's and the sd equations. From this point on we set $\phi$=0. We remind the reader that the effective bare propagator and effective bare  3-vertex are equal to their bare counterparts when $\phi$=0: $D^{oo}=\big.D^0\big|_{\phi=0}$ and $U^{oo}=\big.U^0\big|_{\phi=0}$ [see Eq. (\ref{free})].

\subsection{Skeleton Expansion of the Effective Action}
\label{accuracySection}

In this subsection, we make some general statements about the diagrams that can  appear in the skeleton expansion of the $m$-loop $n$PI effective action. We discuss which diagrams are topologically allowed in the 2PI effective action, and  which ones survive the Legendre transforms.  Using $I$ for the number of internal lines, $E$ for the number of external legs,  and $v_k$ for the number of $k$-point vertices, the standard topological  relations are
\bea
m=I-\sum_3^n v_k +1\,,~~~~
 2I + E=\sum_3^n k\,v_k\,.
\eea
Eliminating $I$ and setting $E = 0$ we get
\bea
\label{mprimeTopo}
m = 1+\sum_{k=3}\bigg(\frac{1}{2}k-1\bigg)v_k\,.
\eea

Before looking at general solutions to Eq. (\ref{mprimeTopo}), we consider a special class of graphs that contribute to the $m$-loop $n$PI effective action which we will call super-BBALL and super-BBALL$_0$ diagrams. In the next subsection, we will see that these diagrams play an important role in the eom's.
For $i\ge 2$ loops, nontadpole graphs with two ${\cal V}_{i+1}$ vertices, and no other vertices, are super-BBALL diagrams. Nontadpole graphs with one ${\cal V}_{i+1}$ vertex and one ${\cal V}^0_{i+1}$ vertex, and no other vertices, are super-BBALL$_0$ diagrams.
For example, the $i=\{2,3,4\}$ super-BBALL diagrams are  $\{\eg,~\bb, {\rm BBALL2}\}$, and the super-BBALL$_0$ diagrams are $\{\eg_0,~\bb_0\}$.
In addition, the terms $ i/2\, {\rm Tr} \,{\rm Ln}D^{-1}$ and $i/2 \,{\rm Tr}[(D^0)^{-1} D]$ in the 1-loop effective action will be called the 1-loop super-BBALL and super-BBALL$_0$, respectively.

For any loop number $m$, we define ${\cal V}_{k_{\rm max}}$ as the largest vertex (with $k_{\rm max}$ legs) that appears in the $L=m$ loop effective action.  We will argue below that for any $m$ the vertex ${\cal V}_{k_{\rm max}}$ appears only in the $m$-loop super-BBALL and super-BBALL$_0$ diagrams. From Eq. (\ref{mprimeTopo}), using $v_{k_{\rm max}}=2$ and $v_{k\ne k_{\rm max}}=0$, it follows immediately that $k_{\rm max}=m+1$.

\ts

Now we consider general 2PI solutions to Eq. (\ref{mprimeTopo}).

\ts

\noindent \underline{$m=1$}: Eq. (\ref{mprimeTopo}) gives $v_k=0$ for $k\in\{3,4,\dots n\}$, and therefore $k_{\rm max}=2$. %This corresponds to the fact that the 1-loop diagrams in the 5-loop 5PI effective action contain no vertices with 3 or more legs.
Since the 1-loop terms in the effective action are the 1-loop super-BBALL and super-BBALL$_0$ diagrams, we conclude that for $m=1$, the vertex ${\cal V}_{k_{\rm max}=(m+1)}$ appears only in the $m$-loop super-BBALL and super-BBALL$_0$ diagrams.

\ts

\noindent \underline{$m=2$}: Eq. (\ref{mprimeTopo}) gives $v_5=v_6=v_7=\dots=0$. The only 2-loop diagram that has a 4-point vertex is the EIGHT diagram, which has a bare 4-point vertex. The largest variational vertex that appears at the 2-loop level is the 3-vertex  in the EGG (and EGG$_0$) diagram, which is the 2-loop super-BBALL (and super-BBALL$_0$) diagram.
The conclusion is that for $m=2$ we have the same result as for the 1-loop case above:
the vertex ${\cal V}_{k_{\rm max}=(m+1)}$ appears only in the $m$-loop super-BBALL and super-BBALL$_0$ diagrams.

We note that if the EIGHT diagram contained a variational 4-vertex, it would produce a nonconnected contribution to the eom for the 4-point vertex.

\ts

\begin{figure}[H]
\begin{center}
\epsfig{file=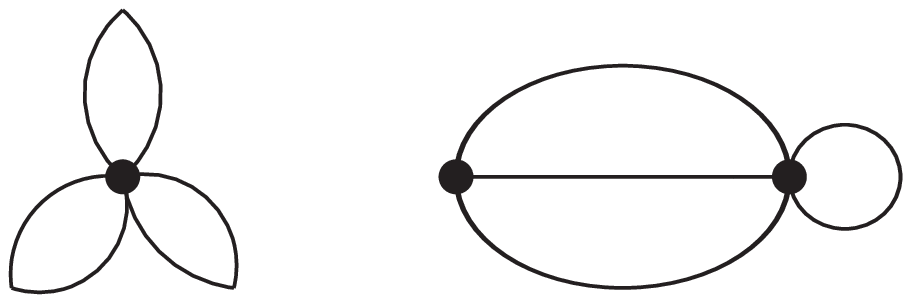,width=5cm}
\end{center}
\vspace*{-0.5cm} \caption{\small 3-loop diagrams that do not contribute to the effective action. }
\label{mprime3tad}
\end{figure}

\noindent \underline{$m=3$}: Eq. (\ref{mprimeTopo}) gives $v_i=0$ for $i\in\{7,8,\dots n\}$.
We look at the solutions to Eq. (\ref{mprimeTopo}) that have nonzero values of $v_k$ for $4\le k\le  6$.
The only possible solutions are
\bea
\label{solns3}
&& \{v_3=0,v_4=0,v_5=0,v_6=1\}~~\to~~ {\rm tadpole~1}\,,\\
&& \{v_3=1,v_4=0,v_5=1,v_6=0\}~~\to~~ {\rm tadpole~2}\,,\nonumber\\
&& \{v_3=2,v_4=1,v_5=0,v_6=0\}~~\to~~ {\rm \ha}\,,\nonumber\\
&& \{v_3=0,v_4=2,v_5=0,v_6=0\}~~\to~~ {\rm \bb~ and ~\bb_0}\,.\nonumber
\eea
The  first two solutions correspond to the tadpole graphs shown in Fig. \ref{mprime3tad}. It is easy to see that tadpole graphs with all lines joining at a single vertex must cancel (in the same way that the EIGHT graph that depends on the effective tilde vertex cancels in Fig. \ref{fig:fig4}). Therefore, the first diagram in Fig. \ref{mprime3tad} does not appear in the effective action. Our calculation of the 5-loop 5PI effective action shows that the second tadpole graph in Fig. \ref{mprime3tad} also cancels. The third solution in Eq. (\ref{solns3}) corresponds to the HAIR diagram, which has a bare 4-vertex. The fourth solution is the BBALL (and BBALL$_0$) diagram, which is the 3-loop super-BBALL (and super-BBALL$_0$) diagram.
The conclusion is that for $m=3$ we have the same result as for the $m\le 2$ cases above:
the vertex ${\cal V}_{k_{\rm max}= (m+1)}$ appears only in the $m$-loop super-BBALL and super-BBALL$_0$ diagrams.

We note that if the tadpole  diagrams did not cancel, they would give nonconnected contributions to the eom's for the 5- and 6-point vertices. Also, if the HAIR graph contained a variational 4-vertex, it would give a 1PR contribution to the eom for the 4-point vertex.

\ts

\begin{figure}[H]
\begin{center}
\epsfig{file=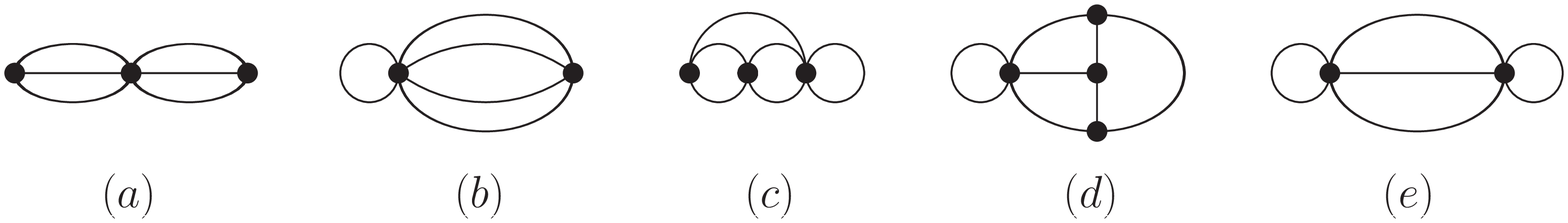,width=14cm}
\end{center}
\vspace*{-0.5cm} \caption{\small 4-loop diagrams that do not contribute to the effective action. }
\label{mprime4tad}
\end{figure}

\noindent \underline{$m=4$}: Eq. (\ref{mprimeTopo})  gives $v_i=0$ for $i\in\{9,10,\dots n\}$.  We look at the solutions to Eq. (\ref{mprimeTopo}) that have nonzero values of $v_k$ for $6\le k\le 8$. Two examples of graphs that contain a 6-point vertex are shown in parts (a) and (b) of Fig. \ref{mprime4tad}. These graphs give disconnected contributions to the eom for the vertex ${\cal V}_6$, and therefore they should cancel in the effective action.
We expect that all graphs with $v_k\ne 0$ for $6\le k\le 8$ would give  disconnected contributions to eom's, and  that they cancel from the effective action. This means that the largest vertex that will appear at the 4-loop level is the 5-vertex. The solutions to Eq. (\ref{mprimeTopo}) with $v_5 \ne 0$ and $v_k=0$ for $6\le k\le 8$ are
\bea
&& \{v_3=1,v_4=1,v_5=1,v_6=v_7 \dots = 0\}~~\to~~ {\rm tadpole~3~and ~first~ diagram~ in~ Fig. \ref{fig:fig5}}\,,\\
&& \{v_3=3,v_4=0,v_5=1,v_6=v_7 \dots = 0\}~~\to~~ {\rm tadpole~4~and ~second~ diagram~ in~ Fig. \ref{fig:fig5}}\,,\nonumber\\
&& \{v_3=0,v_4=0,v_5=2,v_6=v_7 \dots = 0\}~~\to~~ {\rm tadpole~5~and~ BBALL2}\,.\nonumber
\eea
The graphs labeled tadpole 3, tadpole 4, and tadpole 5 are shown in Fig. \ref{mprime4tad}, in parts (c), (d), and (e), respectively. Our calculation of the 5-Loop 5PI effective action proves that these graphs cancel.  In addition, the calculation shows that the first two diagrams in Fig. \ref{fig:fig5} cancel. The only surviving 4-loop diagram is  BBALL2.
The conclusion is that for $m=4$ we have the same result as for the $m\le 3$ cases above:
the vertex ${\cal V}_{k_{\rm max}= (m+1)}$ appears only in the $m$-loop super-BBALL diagram (there is no 4-loop super-BBALL$_0$ diagram because $W^0=0$).

We note that the first two graphs in Fig. \ref{fig:fig5} would produce 1PR contributions to the eom for the 5-point vertex.

We summarize below our results for $m\le 4$:
\begin{enumerate}

\item The only graph with ${\cal V}_{k_{\rm max}}$ that survives the Legendre transform is the $m$-loop super-BBALL (and for $m< 4$ the $m$-loop super-BBALL$_0$) diagram, and $k_{\rm max}=(m+1)$.

\item The $m$-loop graphs  that contain ${\cal V}_{k_{\rm max}}$ and are not super-BBALL (or super-BBALL$_0$) cancel. If they did not cancel, they would produce 1PR contributions to the eom for the vertex ${\cal V}_{k_{\rm max}}$.

\item Tadpole graphs would produce disconnected contributions to the eom of the vertex that joins the tadpole loop to the diagram.
In the calculation of the 5-Loop 5PI effective action, there is one 3-loop tadpole (the second graph in Fig. \ref{mprime3tad}), three 4-loop tadpoles (the last three graphs in Fig. \ref{mprime4tad}), and 14 5-loop tadpoles. We have checked that all of these graphs cancel. The EIGHT diagram depends only on the bare vertex.

\end{enumerate}

We expect that these results hold at higher orders. We assume below that for any $m$, the only graph that contains the vertex ${\cal V}_{k_{\rm max}}$ is the $m$-loop super-BBALL diagram (and for $m < 4$ the $m$-loop super-BBALL$_0$ diagram).
%It is equivalent to say that an arbitrary vertex ${\cal V}_i$ appears in the super-BBALL (or super-BBALL$_0$) diagram at $(i-1)$-loops, and all additional diagrams that contain the vertex ${\cal V}_i$ have $L>(i+1)$ loops.
In order to facilitate the discussion in the next section,  we separate the super-BBALL and super-BBALL$_0$ terms from the effective action by writing
\bea
\label{superSep}
\Gamma_n^{(m)} = \sum_i {\rm super\!\!-\!\!BBALL}^{(i-{\rm loop})} ~ + ~ \sum_i {\rm super\!\!-\!\!BBALL_0}^{(i-{\rm loop})} ~+~\hat \Gamma_n^{(m)}[{\cal V}_j]\,,
\eea
where $\hat \Gamma_n^{(m)}[{\cal V}_j] = -i\hat \Phi_n^{(m)}[{\cal V}_j]$ contains all terms in the effective action that are not super-BBALL or super-BBALL$_0$. The vertex ${\cal V}_i$ appears in the ($i-1)$ loop super-BBALL (and super-BBALL$_0$) diagram, and diagrams with $L>(i-1)$ loops in $\hat\Gamma_n^{(m)}[{\cal V}_j]$.

\subsection{Skeleton Expansion of the eom's}
\label{accuracySection2}

Now we consider the structure of the eom for the vertex ${\cal V}_i$ that will be produced from the effective action in Eq. (\ref{superSep}) using Eq. (\ref{generaleom}).

Taking the functional derivative of an $m$-loop graph in the effective action  with respect to the variational vertex ${\cal V}_i$ ($2\le i \le n$) opens $i-1$ loops. This means that an $m$-loop graph in the effective action produces a diagram with ${\cal L}[m,i]$ loops in the eom for the vertex ${\cal V}_i$, where we have defined
\bea
\label{calL}
{\cal L}[m,i] := m-i+1\,.
\eea
Note that the order of the original $m$-loop diagram in the effective action corresponds to $i=1$.

According to our definition (see Sec. \ref{accuracySection}), the super-BBALL and super-BBALL$_0$ diagrams have $(i-1)$ loops.  Equation (\ref{calL}) gives ${\cal L}[i-1,i]=0,$ which means that these graphs produce 0-loop (or tree) contributions to the eom for the vertex ${\cal V}_i$.
Derivatives of the super-BBALL and super-BBALL$_0$ diagrams give
$
-(1-2\delta_{i2})\;(1/i!)\;({\cal V}_i - {\cal V}^0_i).
$
The factor $(1-2\delta_{i2})$ gives $(1)$ for $i\ge 3$ and $(-1)$ for $i=2$. The sign difference for the 2-point function occurs because of the fact that it is conventional to write the effective action as a function of the propagator $D$ instead of the inverse propagator $D^{-1}$ (see footnote 3).
We give several examples: In the eom for the 2-vertex, the 1-loop contributions to the effective action produce $\frac{1}{2}(D^{-1}-(D^0)^{-1})$. In the eom for the 3-vertex, the EGG and EGG$_0$ graphs produce $-\frac{1}{6}U$ and $\frac{1}{6}U^0$, respectively. In the eom for the 4-vertex, the BBALL and BBALL$_0$ graphs produce $-\frac{1}{24}V$ and $\frac{1}{24}V^0$, respectively. In the eom for the 5-vertex, the BBALL2 graph produces $-\frac{1}{120}W$, and the absence of a bare 5-vertex corresponds to the absence of a BBALL2-type graph with a bare vertex, and the absence of a $W^0$ term in the eom.

Equation (\ref{calL}) also tells us that functionally differentiating the $L>(i-1)$ loop terms in $\hat\Gamma_n^{(m)}[{\cal V}_j]$ produces contributions with $L>0$ loops.
Combining pieces we can rewrite the eom:
\bea
\label{generaleom2}
{\cal V}_i && = {\cal V}_i^0 + {\rm fcn}_i[{\cal V}_j]\,,~~~~{\rm fcn}_i[{\cal V}_j] =   i! ~(1-2\delta_{i2})~\frac{\delta \hat \Phi_n^{(m)}[{\cal V}_j]}{\delta {\cal V}_i}\,,
\eea
where the functional ${\rm fcn}_i[{\cal V}_j]$ is $L\ge 1$ loops in the skeleton expansion.
To illustrate the notation we write out Eq. (\ref{generaleom2}) for $i=2$ and $i=3$:
\bea
\label{2ptEx}
&& \frac{\delta \Gamma_n^{(m)}[{\cal V}_j]}{\delta D}=0~~\to~~D^{-1} = (D^0)^{-1}-2!\frac{\delta \hat\Phi_n^{(m)}[{\cal V}_j]}{\delta D} = (D^0)^{-1}-\Pi[{\cal V}_j]\,,\\
&& \frac{\delta \Gamma_n^{(m)}[{\cal V}_j]}{\delta U}=0~~\to~~U = U^0+3!~\frac{\delta \hat\Phi_n^{(m)}[{\cal V}_j]}{\delta U} = U^0 + {\rm fcn_3}[{\cal V}_j]\,.\nonumber
\eea

\subsection{Perturbative Expansions}
\label{pertSection}

The $n$PI effective action represents a reorganization of perturbation theory that can be thought of as corresponding to an infinite resummation of some set of diagrams  with bare propagators and vertices.
The $n$PI formalism is of interest because it includes nonperturbative effects, and the goal is not to expand, but to solve the $n$PI eom's or the sd equations self-consistently. However, we will show in this section that we can obtain some interesting results about the structure of the nonperturbative variational eom's by considering their perturbative expansions. In order to see which diagrams are included in the perturbative expansion of the effective action, or in any of the vertices ${\cal V}_i$, we could expand the corresponding functional by substituting repeatedly the eom's. This procedure would replace variational propagators and vertices with bare ones, while generating higher and higher loop diagrams in the perturbative expansion that correspond to propagator- and vertex-corrected versions of the diagrams in the skeleton expansion.
In this subsection we consider the diagrams that would be produced by this procedure.
The goal is to show that the $m$-Loop effective action  produces {\it all} terms in the perturbative expansions of the effective action, and the vertices ${\cal V}_{ i}$, up to $L_{\rm pt}={\cal L}[m, i]$ loops. The significance of this result is explained in Sec. \ref{comparisonLABEL}.

Throughout this paper, we use $L$ to denote loop number in the skeleton expansion. In this section we introduce the notation $L_{\rm pt}$ to denote loop number in the perturbative expansion.

\ts

We consider adding $L=(m+1)$ loop terms to the skeleton expansion of $\hat\Gamma^{(m)}_n[{\cal V}_j]$. We write these terms $\hat\Gamma_n^{L=(m+1)}[{\cal V}_j]$ where the superscript indicates  only terms with $L=(m+1)$ loops are included (recall that $\hat\Gamma_n^{(m)}[{\cal V}_j]$ contains terms with $L\le m$ loops).  These $(m+1)$ loop terms will produce new contributions to the skeleton and perturbative expansions of the effective action, and each vertex ${\cal V}_i$.
We will show that all terms in $\hat\Gamma_n^{L=(m+1)}[{\cal V}_j]$ produce contributions to the perturbative expansion of the effective action and the vertex ${\cal V}_{i}$ at $L_{\rm pt}> {\cal L}[m, i]$ loops.

\ts

There are two types of contributions:

\noindent (1) As explained in the previous subsection, taking the functional derivative of $\hat\Gamma_n^{L=(m+1)}[{\cal V}_j]$ with respect to ${\cal V}_{ i}$ produces new terms in the skeleton expansion of ${\cal V}_{ i}$ of order ${\cal L}[m+1, i]$. It is clear that they  contribute at $L_{\rm pt} > {\cal L}[m, i]$ loops to the perturbative expansion, and give no contributions at $L_{\rm pt}\le {\cal L}[m, i]$ loops.

\noindent (2) We also need to consider lower order (old) terms in the skeleton expansion of order ${\cal L}[m^\prime, i]$, with $m^\prime\le m$, with an arbitrary internal variational vertex ${\cal V}_k$ replaced by a term in ${\rm fcn}_k[{\cal V}_j]$ [see Eq. (\ref{generaleom2})] which was produced by functional differentiation of $\hat\Gamma_n^{L=(m+1)}[{\cal V}_j]$. For any $k$, the new contributions to the vertex ${\cal V}_k$ are of order ${\cal L}[m+1,k]$. The substitution produces terms of order $L={\cal L}[m^\prime, i]+{\cal L}[m+1,k]$ in the skeleton expansion. From Eq. (\ref{calL}) we have  ${\cal L}[m+1,k]\ge {\cal L}[m+1,k_{\rm max}]$ which means $L\ge {\cal L}[m^\prime, i]+{\cal L}[m+1,k_{\rm max}]$. Using Eq. (\ref{calL}) with $k_{\rm max}=m^\prime +1$ (see Sec. \ref{accuracySection}) we obtain $L>{\cal L}[m, i]$, and therefore $L_{\rm pt}> {\cal L}[m, i]$ loops.

\ts

We have shown that both types of contributions from terms in $\hat\Gamma^{L=(m+1)}_n[{\cal V}_j]$ contribute to the perturbative expansion of the effective action and the vertices ${\cal V}_{ i}$ at $L_{\rm pt}> {\cal L}[m, i]$ loops.
We also know that without truncation the expanded effective action and vertices ${\cal V}_{ i}$  exactly match the perturbative expansion. These two statements together allow us to conclude that the $m$-Loop effective action must produce {\it all} terms in the perturbative expansions of the effective action and the vertices ${\cal V}_{ i}$ up to $L_{\rm pt}={\cal L}[m, i]$ loops.

\subsubsection{Structure of the Effective Action}

It is usually said that the fact that the 2PI effective action does not contain 2PR diagrams is evidence that double counting does not occur at this level. One argues that any 2PR diagram in the effective action would correspond to a propagator correction of a lower loop diagram, and thus would appear twice in the expanded series, or equivalently, it would appear with the wrong symmetry factor.

\begin{figure}[H]
\begin{center}
\includegraphics[width=8cm]{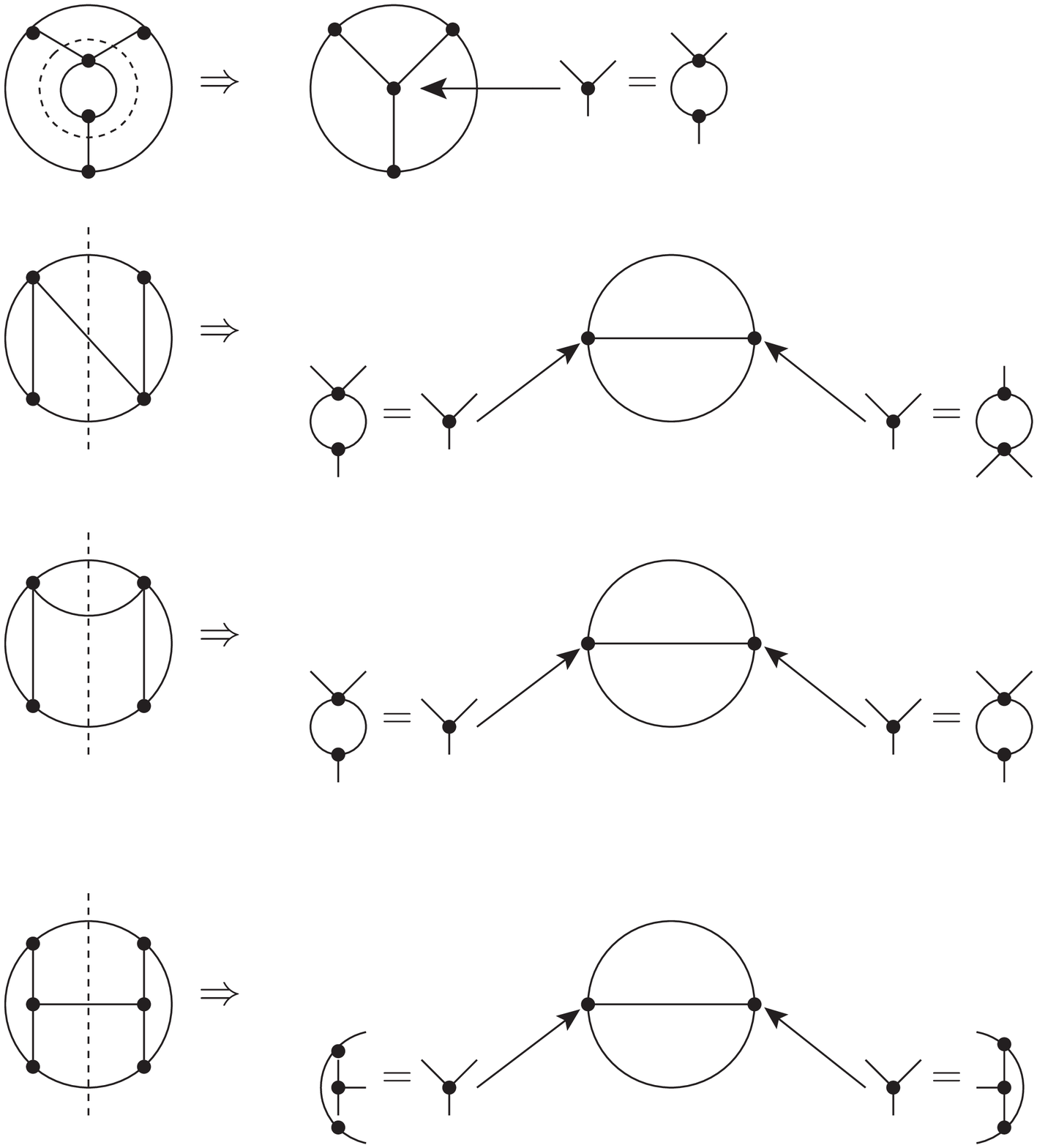}
\end{center}
\caption{\label{4PRfig}The left side of the figure shows 4-loop 4PR diagrams that do not contribute to the 4PI effective action. The dotted lines divide the diagrams into two pieces, each of which contains a closed loop. The right side of the figure shows the corresponding vertex-corrected lower loop diagram. }
\end{figure}

We can try to extend this argument beyond the 2PI level. An $n$PR diagram can be  defined to be a diagram that cannot be divided into two pieces by cutting $n$ or fewer lines such that each piece contains at least one closed loop. We expect that any diagram that could be cut in this way would correspond to a vertex correction of a lower loop diagram. To illustrate this point, we consider the 4-Loop 4PI effective action. The diagrams that survive the Legendre transform are not 4PR (see Figs. \ref{Phi0Fig} and \ref{Phi4Fig}). The 4-loop 4PR diagrams that are canceled by the Legendre transform  are shown in Fig. \ref{add4LABEL}. They can all  be written as vertex-corrected lower loop diagrams. This is shown in Fig. \ref{4PRfig}.
\par

The fact that all $n$PR diagrams are removed by the Legendre transforms does not guarantee that the expanded effective action will agree with the perturbative expansion to the truncation order, since it is not obvious that all graphs would be produced with the correct symmetry factors. In addition, the argument discussed above is not valid for the 5-Loop 5PI effective action, since all of the diagrams in the second line in Fig. \ref{fig:fig17} are 5PR. The fact that the ``5PI'' effective action is not 5-particle irreducible suggests that the  $n$-Loop $n$PI effective action is $n$-particle irreducible for $n\le 4$ only (as discussed in the introduction, we use ``$n$PI effective action'' to mean the action produced by taking $n$ Legendre transforms).

From these results, the naive conclusion is that the $n$-Loop $n$PI effective action does not make sense for $n\ge 5$.
However, we have shown in this section that the expanded $L\le m$ loop skeleton diagrams of the effective action must match the perturbative expansion at $L_{\rm pt}\le {\cal L}[m,1]=m$ loops. This is equivalent to proving that the $m$-Loop $n$PI effective action is void of double counting at $m$ Loops. We conclude that the effective action produced by the Legendre transforms does not double count at the level of the truncation, but it is not necessarily $n$-particle irreducible.

\subsubsection{Structure of the eom's}

Since the $m$-Loop effective action matches the perturbative expansions of the vertices ${\cal V}_{ i}$ up to $L_{\rm pt}={\cal L}[m, i]$ loops, we can rewrite Eq. (\ref{generaleom2}) as
\bea
\label{genericEOM}
{\cal V}_i = {\cal V}_i^0~~+ \underbrace{~~~~{\rm fcn}_i[{\cal V}_j]~~~~}_{L\le {\cal L}[m,i] ~\big|~L_{\rm pt}\le {\cal L}[m,i]_{\rm all} }\,,~~\{i,j\}\in\{2,3,\dots n\}\,.
\eea
The subscript $L\le {\cal L}[m, i] ~\big|~L_{\rm pt}\le {\cal L}[m, i]_{\rm all}$ indicates that ${\cal V}_{ i}$ contains only terms with $L\le {\cal L}[m, i]$ loops in the skeleton expansion, and all terms with $L_{\rm pt}\le {\cal L}[m, i]$ loops in the perturbative expansion. The terms that are included in the skeleton expansion depend on the result for $\Gamma_n^{(m)}$. The terms in the perturbative expansion can be obtained in the usual way from the $m$-Loop 1PI effective action [see Eq. (\ref{properDefn})], or by simply writing down all possible ${\cal L}[m, i]$ Loop 1PI diagrams and calculating the symmetry factors using a combinatoric formula.
%We discuss the significance of Eq. (\ref{genericEOM}) in sub-section \ref{comparisonLABEL}.

\subsubsection{Interpretation}

We have shown above that the $m$-Loop effective action must produce {\it all} terms in the perturbative expansions of the effective action and the vertices ${\cal V}_{ i}$ up to $L_{\rm pt}={\cal L}[m, i]$ loops. This result has important consequences when we consider the role of symmetries in nonperturbative calculations.

One of the very attractive features of $n$PI effective theories is that the eom's guarantee that linearly realized global symmetries of the original theory are respected, and that the conservation laws that follow from Noether's theorem are satisfied \cite{baym1,baym2}. However, if the original theory has a local symmetry, the variational $i$-point functions will not obey standard Ward-like identities (see \cite{bergesReview} and references therein).

The issue is particularly important in the context of gauge theories. When calculating physical observables, we expect that gauge independence is encoded in the Ward identities. For an $n$PI effective action, the vertex functions that are defined as derivatives of the variational extrema of  the action will satisfy a set of symmetry identities that have the same structure as 1PI Ward identities \cite{JSward}\footnote{In Ref. \cite{JSward} the effective action is defined as a function of connected vertices.}.
These kinds of symmetry identities have been used to prove the renormalizability of the 2PI effective action for QED \cite{reinosaRenorm2}. However, these symmetry constraints do not directly address the question of the gauge invariance of observables calculated from the $n$PI effective action.
For any diagram, the $i$-point functions that correspond to internal lines and vertices must be determined by a fully self-consistent variational procedure, and  will not satisfy standard Ward identities. As a consequence, quantities  calculated with $n$PI techniques that are supposed to correspond to physical observables can contain gauge dependent contributions. We need to be able to quantify and control this gauge dependence.

In this paper we are working with a scalar theory with cubic and quartic couplings. The $m$-Loop $n$PI effective action matches the perturbative expansion to $m$ loops.
The  vertex functions match the perturbative expansion to ${\cal L}[m,i]$ loops, which means that they have the correct crossing symmetry to this order.
Symmetry breaking contributions appear at order ${\cal L}[m,i]+1 = {\cal L}[m+1,i]$ loops, which is the same order as terms that would be produced by $L=(m+1)$ loop terms in the effective action which were dropped when we performed the  truncation. We conclude that the variational vertex functions respect crossing symmetry to the truncation order.

The toy model that we study in this paper has the same basic diagrammatic structure as QED or QCD. It should be straightforward to use the same method to show that the $m$-Loop $n$PI effective action for these theories (and their eom's) matches the corresponding perturbative results to order ${\cal L}[m,i]$. Since the perturbative loop expansion is gauge invariant at every loop order, we would conclude that the $m$-Loop $n$PI effective action is gauge invariant to $m$ loops (or to order $g^{2m-2}$), which is the truncation order.

%It has been shown previously that the gauge dependence of the effective action always appears at higher order than the truncation order. If the effective action is truncated at $n$-Loops ($\sim g^{2n-2}$) in the skeleton expansion, the non-perturbative result for the resummed effective action that is obtained by substituting the variational solutions can be gauge dependent only at order $g^{2n}$.
The same result has been obtained previously using a completely different approach. In Ref. \cite{smit}, the authors look at the  truncated effective action, evaluated at the solutions obtained from the self-consistent  eom's.  They consider the behavior of this resummed effective action under Becchi-Rouet-Stora-Tyutin transformations  and prove that it is gauge invariant to the truncation order. In Ref. \cite{HZ}, the authors obtain the same result using the Nielsen identities to study cancellations between the gauge dependence of the effective action and the variational solutions.

\subsection{Comparison of the $n$-Loop $n$PI eom's and the Schwinger-Dyson Equations}
\label{comparisonLABEL}

\begin{figure}[H]
\begin{center}
\includegraphics[width=12cm]{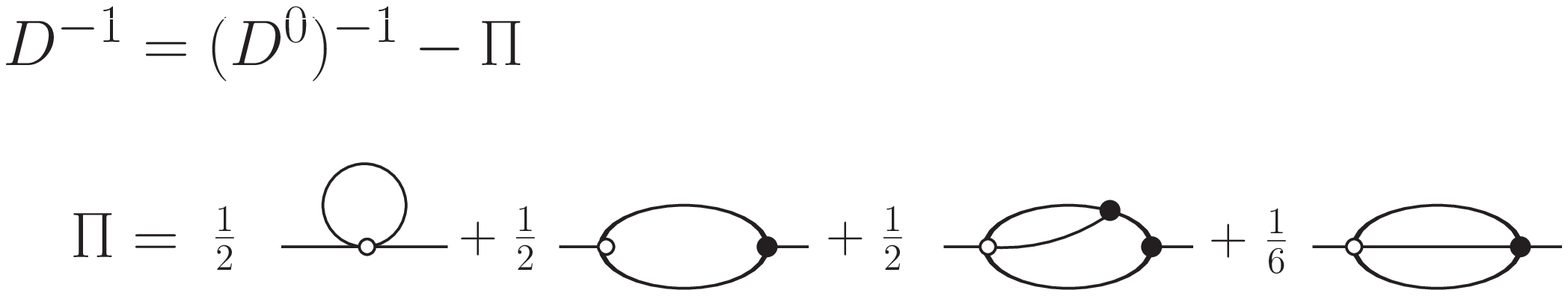}
\end{center}
\caption{\label{sdPIeqnLABEL}Schwinger-Dyson equation for the 2-point vertex. }
\end{figure}

\begin{figure}
\begin{center}
\includegraphics[width=14cm]{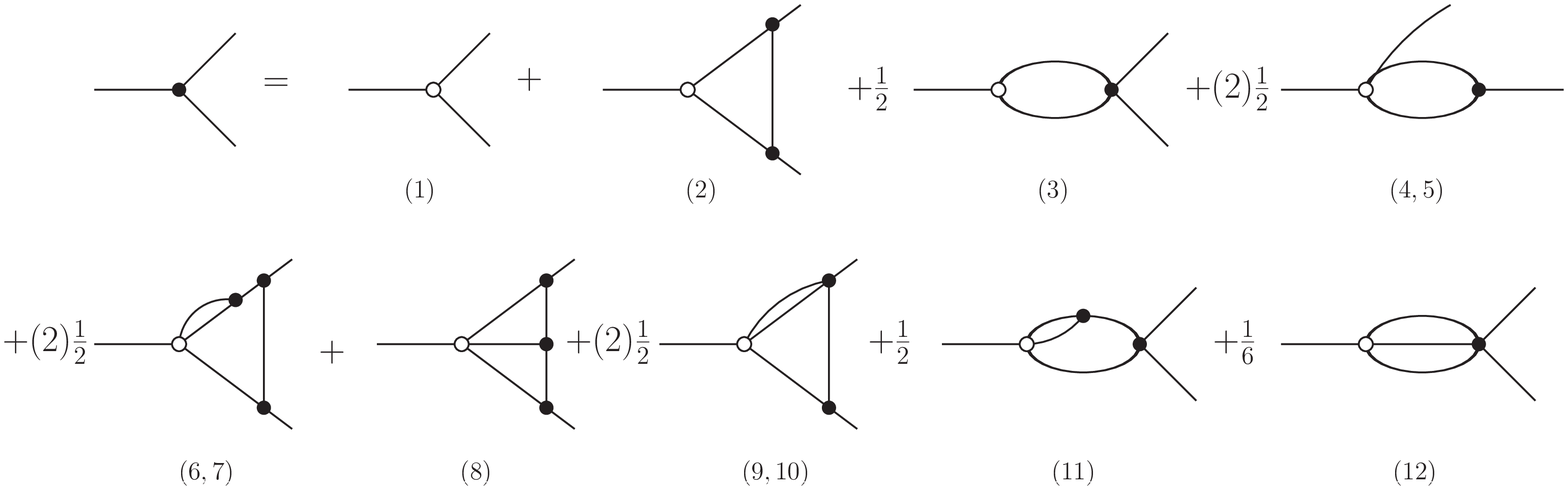}
\end{center}
\caption{\label{sdUeqnLABEL}Schwinger-Dyson equation for the 3-point vertex.}
\end{figure}

The Schwinger-Dyson equations form an infinite hierarchy of coupled nonlinear integral equations for all the $n$-point functions of the theory.
Their derivation is tedious but straightforward \cite{cvitanovic,kajantie,alkoferALG}. The $i$-point function, which we write ${\cal V}_i^{sd}$, satisfies an integral equation that depends on the vertices ${\cal V}_j^{sd}$ with $2 \le j \le i+2$ which has the form
\bea
\label{sdGeneral}
{\cal V}_i^{sd} = {\cal V}_i^0+f_i^{sd}[{\cal V}_j^{sd}]\,;~~2 \le j \le i+2\,.
\eea
The sd equations for the 2- and 3-point functions are shown in Figs. \ref{sdPIeqnLABEL} and \ref{sdUeqnLABEL}, respectively. We give the equation for the 4-point function in  \ref{sdAppendix}. In all diagrams that correspond to sd equations, the numerical factors in brackets refer to permutations of the legs on the rhs of the graph.
\par
\par
\noindent

\begin{figure}[H]
\begin{center}
\includegraphics[width=10cm]{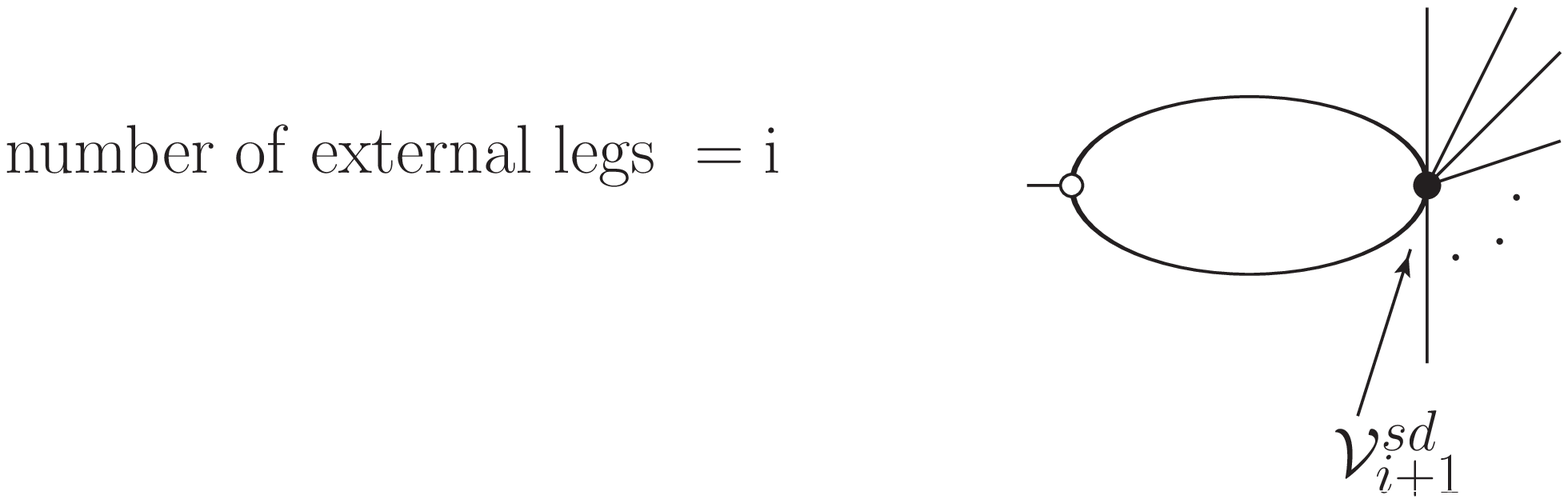}
\end{center}
\caption{\label{worstFIGlabel}A term from the sd equation for the vertex ${\cal V}^{sd}_{i}$ which produces a term that fails to match the perturbative expansion at the lowest loop order. The dots denote legs that are not drawn.  }
\end{figure}

We can truncate the sd equations by setting ${\cal V}^{sd}_{p+1}={\cal V}^0_{p+1}$ for some value of $p$. We consider iterating the resulting set of equations to obtain a series of perturbative diagrams. For each $i$-point function, we want to determine the order ${\bf L}[p,i]$ to which this series of diagrams matches the perturbative expansion. The diagram in the sd equation for the vertex $V^{sd}_{i}$ that produces a term that fails to match at the smallest $L_{\rm pt}$ is the 1-loop diagram that contains the vertex $V^{sd}_{j=i+1}$. The graph is shown in Fig. \ref{worstFIGlabel}. We start by considering $i=p$. Since we have set ${\cal V}^{sd}_{p+1}= {\cal V}^0_{p+1}$ in order to truncate the series, the graph that corresponds to  a 1-loop insertion in the place of the vertex on the rhs of this diagram will be missing. Since the diagram is itself 1-loop, the expanded equation for ${\cal V}^{sd}_{p}$ will be missing graphs with $L_{\rm pt}=2$, but will match the perturbative expansion to $L_{\rm pt}=1$. We write this ${\bf L}[p,p]=1$. Now we look at the same graph with $i=p-1$. We have just seen that the ${\cal V}^{sd}_{p}$ vertex on the rhs is missing 2-loop terms. This means that the expansion of the vertex ${\cal V}^{sd}_{p-1}$ will be missing graphs with $L_{\rm pt}=3$, but will match the perturbative expansion to $L_{\rm pt}=2$. We write this ${\bf L}[p,p-1]=2$. Continuing in the same way we get ${\bf L}[p,p-2]=3$, ${\bf L}[p,p-3]=4$ $\cdots$. The general expression corresponding to these results is
\bea
\label{boldL}
{\bf L}[p,i] = p-i+1\,,
\eea
where ${\bf L}[p,i]$ gives the order to which the vertex ${\cal V}^{sd}_i$ will match the perturbative expansion, if the sd hierarchy is truncated by setting ${\cal V}^{sd}_{p+1} = 0$.
\par

\ts

Now we can compare the eom's produced by the $n$PI effective action and the sd equations.
If neither the $n$PI effective action nor the sd hierarchy is truncated, the two sets of equations are equivalent.
Calculations involve truncation, so we should compare the truncated sets of equations. We consider the eom's produced by the $m$-Loop $n$PI effective action\footnote{From \ref{hierSection}, only $m\ge n$ is possible.}, and the sd hierarchy truncated by setting ${\cal V}^{sd}_{p+1}={\cal V}^{0}_{p+1}$.
At first glance, sd equations and $n$PI eom's do not appear to have the same structure at all. All contributions to the eom from diagrams in $\Phi^{int}$ will contain only corrected vertices (no bare vertices), and are missing from the sd equation. The contributions to the eom from $\Phi^0$ contain bare vertices, but they are symmetric in permutations of the external legs. In contrast, in the sd equation, all graphs contain a bare vertex on the lhs.
However, from Eqs. (\ref{calL}) and (\ref{boldL}), it is clear that the eom's obtained from the $m$-Loop $n$PI effective action, and the sd equations for $p=m$, both match the perturbative expansion and therefore each other (for $2\le i\le n$), to the same perturbative loop order.

In the rest of this paper we will choose  $m=n$ (see \ref{hierSection}). From Eqs. (\ref{calL}) and (\ref{boldL}) we find that, for $n=m=p$, the $n$PI  eom's and the sd equations both match the perturbative expansion, and therefore each other, to $L_{\rm pt} = {\cal L}[n,i]$. It is equivalent to say that Eq. (\ref{genericEOM}) can be rearranged in the form:
\bea
\label{eomRearrange}
{\cal V}_i = {\cal V}^0_i ~ +~f_i^{sd}[{\cal V}_j]\bigg|_{L\le {\cal L}[n,i]}~+~\underbrace{~~~{\rm extra}~~~}_{L_{\rm pt}\ge{\cal L}[n,i]+1}\,;~~j\in \{2,\dots{\rm min}[i+2,n]\}\,;~~~i\in\{2,\dots n\}\,.
\eea
The expression $f_i^{sd}[{\cal V}_j]$ on the right side of Eq. (\ref{eomRearrange}) represents a series of diagrams that have the same form as the diagrams in the sd equation for the vertex ${\cal V}^{sd}_i$ [Eq. (\ref{sdGeneral})], but taking only diagrams with $L\le {\cal L}[n,i]$ loops, and replacing the sd propagator and vertices ${\cal V}^{sd}_j$ with the variational propagator and vertices obtained from the $n$-Loop $n$PI effective action for $2\le j\le {\rm min}[i+2,n]$, and bare vertices for $j \ge {\rm min}[i+2,n]+1$. In Sec. \ref{PiUeqnSection} we verify Eq. (\ref{eomRearrange}) for $n\le 5$ and $i\in\{2,3\}$.
%In principle, the rearrangement in Eq. (\ref{eomRearrange}) is always possible, since we have not specified what terms are included in the piece marked extra; the point is to show that the extra terms have $L_{\rm pt}\ge ({\cal L}[m,i]+1)$ loops.

Before discussing the significance of Eq. (\ref{eomRearrange}), we look at an example to illustrate the notation. We consider the 5-Loop 5PI effective action and look at the eom for the 3-point function. We have min[$i+2,n$]=min[3+2,5]=5 and ${\cal L}[n,i]={\cal L}[5,3] =3$.  Since the sd equation contains only terms of $L\le 2< {\cal L}[5,3]=3$ loops (see Fig. \ref{sdUeqnLABEL}), we can drop the subscript and write  $f^{sd}_3[D,U,V,W]\big|_{L\le 3} = f^{sd}_3[D,U,V,W]$ so that Eq. (\ref{eomRearrange}) becomes
\bea
\label{rearrEx}
5{\rm -Loop} ~{\rm 5PI}:~~~U=U^0~+~f^{sd}_3[D,U,V,W]~+~\underbrace{\rm ~~~extra~~~}_{L_{\rm pt}\ge 4}\,.
\eea
Equation (\ref{rearrEx}) says that the variational eom for the 3-point function has the same form as the sd equation for the 3-point function, with the sd propagator and vertices replaced by the variational propagator and vertices, plus some terms that are 4-loop or higher in the perturbative expansion.

\ts
The term marked ``extra'' in Eq. (\ref{eomRearrange}) is of order ${\cal L}[n,i]+1 = {\cal L}[n+1,i]$ loops, and is the same order as terms that would be produced by $L=(n+1)$ loop terms in the effective action which were dropped when we performed the  truncation. An equivalent statement is that if we do a calculation using an $n$-Loop $n$PI effective theory, and replace the eom's for the variational propagators and vertices by the sd equations truncated with ${\cal V}_{n+1}={\cal V}_{n+1}^0$, the error we make is of the same order as terms that would come from contributions to the effective action that are beyond the truncation order. For a gauge theory, these terms  correspond to potentially gauge dependent contributions, and  have no physical interpretation.
Note that truncations of the sd equations also produce violations of underlying symmetries. This has been discussed extensively (see, for example, Refs.  \cite{BINOSIreview,alkofer,casal} and references therein).

\section{Integral Equations - Explicit Calculations}
\label{PiUeqnSection}

In this section we explicitly verify Eq. (\ref{eomRearrange}) for the 2- and 3-vertices to the level of the  5-Loop 5PI effective action.
The basic method is to rearrange the eom's by substituting them into themselves. We perform these substitutions by rewriting the variational eom's in Eq. (\ref{genericEOM}) for $3\le i\le 5$ in the form
\begin{subequations}
\begin{equation}
\label{genericSubU}
U = U^{0}+{\rm fcn}_3[{\cal V}_j]\,,
\end{equation}
\begin{equation}
\label{genericSubU0}
U^0 = U - {\rm fcn}_3[{\cal V}_j]\,,
\end{equation}
\begin{equation}
\label{genericSubV}
V = V^0+{\rm fcn}_4[{\cal V}_j]\,,
\end{equation}
\begin{equation}
\label{genericSubV0}
V^0 = V - {\rm fcn}_4[{\cal V}_j]\,, \end{equation}
\begin{equation}
\label{genericSubW}
W = {\rm fcn}_5[{\cal V}_j]\,.
\end{equation}
\end{subequations}
We compare the rearranged eom's with the sd equation, with propagators and vertices replaced by the corresponding variational ones. We show that the difference is a set of skeleton diagrams that corresponds to terms in the perturbative expansion with $L_{\rm pt}\ge {\cal L}[m,i]+1$ loops.
Throughout this section, when we refer to ``the sd equation,'' we mean the sd equation as a functional of variational propagators and vertices, as in Eq. (\ref{eomRearrange}).

\subsection{Integral Equation for the 2-point Vertex}
\label{PIall}
For the 2-point function, we will show that the structure of the calculation is particularly simple, and the extra terms in Eq. (\ref{eomRearrange}) cancel cleanly. Equivalently, the eom for the 2-point function has, formally, exactly the same structure as the sd equation.

\subsubsection{eom for $\Pi$ for 2-Loop 2PI}
\label{2PIpi}

We start with the 2-Loop 2PI effective theory. We have min[$i+2$,$n$]=min[2+2,2]=2 and ${\cal L}[m,i]={\cal L}[2,2]=1$. Equation (\ref{eomRearrange}) becomes
\bea
\label{rearrangePi2}
\Pi = -f_2^{sd}[D]\big|_{L\le 1}~~+~~\underbrace{~~~{\rm extra}~~~}_{L_{\rm pt}\ge 2}\,.
\eea
The origin of the minus sign in front of the first term on the rhs is discussed under Eq. (\ref{generaleom2}).

We start by extracting the eom for the 2-point function for 2-Loop 2PI from our previous results. The diagrams that contribute to $\Phi^0$ and $\Phi^{int}$ are the 2-Loop graphs in Figs. \ref{Phi0Fig} and \ref{Phi4Fig} with $U=U^0$. These are the diagrams labeled EIGHT, EGG, and EGG$_0$ (the EGG$_0$ and EGG diagrams combine when $U=U^0$).  The eom for $\Pi$ is obtained from Fig. \ref{PIintLABEL} by including the contributions from the 2-Loop diagrams listed above, and setting $U=U^0$. This set of diagrams is the lhs of Eq. (\ref{rearrangePi2}).
The rhs is the 1-loop terms in the sd equation, with the 3-vertex replaced by the bare one. From Figs. \ref{PIintLABEL} and \ref{sdPIeqnLABEL}, it is easy to see that Eq. (\ref{rearrangePi2}) is satisfied. We also find that for this example the extra terms in Eq. (\ref{rearrangePi2}) are identically zero. We will show that this is a general feature of the calculation for the 2-point function.

\subsubsection{eom for $\Pi$ for 3-Loop 3PI}
\label{3PIpi}

Now we consider the 3-Loop 3PI effective theory. We have min[$i+2$,$n$]=min[2+2,3]=3 and ${\cal L}[m,i]={\cal L}[3,2]=2$. Equation (\ref{eomRearrange}) becomes
\bea
\label{rearrangePi3}
\Pi = -f_2^{sd}[D,U]~~+~~\underbrace{~~~{\rm extra}~~~}_{L_{\rm pt}\ge 3}\,,
\eea
where we have dropped the subscript ${L\le 2}$ on the first term on the rhs, since all terms in the sd equation  have two or fewer loops.

We start by extracting the eom for the 2-point function for 3-Loop 3PI from our previous results. The diagrams that contribute to $\Phi^0$ and $\Phi^{int}$ are the 3-Loop graphs in Figs. \ref{Phi0Fig} and \ref{Phi4Fig} with $V=V^0$. These are the diagrams labeled EIGHT, EGG$_0$, EGG, BBALL$_0$, BBALL, HAIR, and MERCEDES (the BBALL$_0$ and BBALL diagrams combine when $V=V^0$, but the EGG$_0$ and EGG diagrams do not, because we have $U\ne U^0$).  The eom for $\Pi$ is obtained from Fig. \ref{PIintLABEL} by including the contributions from the 3-Loop diagrams listed above, and setting $V=V^0$.

We want to compare this eom with the corresponding sd equation, with the 4-vertex replaced by the bare one. Unlike the artificially simple 2-Loop 2PI example that we discussed above,  these two equations look very different. For example, in Fig. \ref{PIintLABEL}, the second diagram labeled $[\Pi_{{\rm HAIR}}]_2$ and the diagram labeled $\Pi_{{\rm MERCEDES}}$ do not appear in the sd equation at all. In addition, the 1-loop EGG-type topology contributes to both the eom and the sd equation, but in the sd equation it only appears with a bare vertex on the lhs, and in the eom there are also graphs with a bare vertex on the rhs, and with two corrected vertices.

\begin{figure}[H]
\begin{center}
\epsfig{file=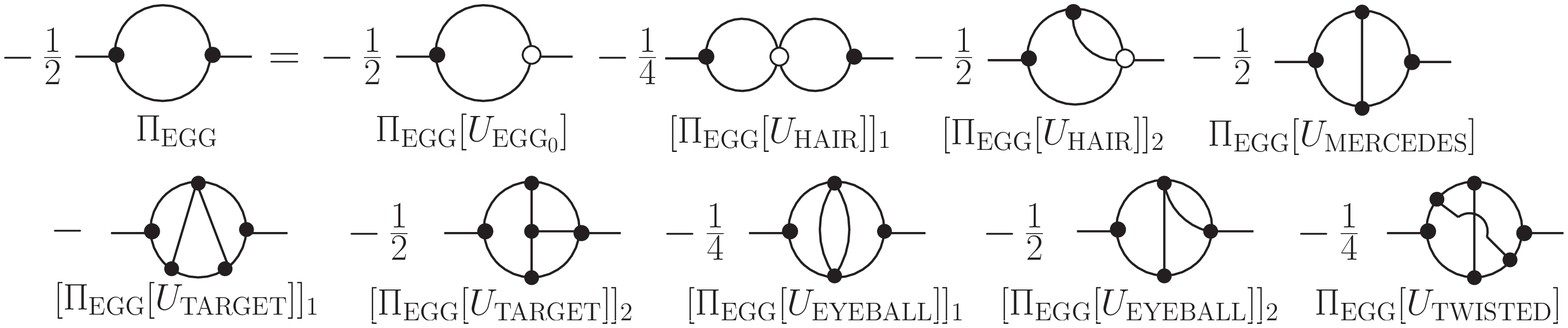,width=14cm}
\end{center}
\vspace*{-0.5cm} \caption{\small Result from replacing the 3-vertex on the rhs of the $\Pi_{{\rm EGG}}$ diagram
with the 3-vertex eom from the 4-Loop 4PI effective action (Fig. \ref{UintLABEL}). The graphs in the first line correspond to  3-Loop 3PI contributions in the effective action.} \label{figreplaceEGG}
\end{figure}

In spite of these apparent differences, we can show that the variational eom and the sd equation have the same form. We use Eq. (\ref{genericSubU}) to remove the variational vertex on the rhs of the $\Pi_{{\rm EGG}}$ diagram in the eom. The result is shown diagrammatically in the first line of Fig. \ref{figreplaceEGG} (the diagrams in the second line correspond to contributions from the 4-loop 4PI effective action which we will need later). We use labels of the form $\{U_{\eg_0},\,U_\ha,\,U_\mer\,\dots\}$ to indicate which term in the $U$ eom produced each graph. In addition, we label each graph $\Pi_{{\rm EGG}}$, to remind ourselves that the original graph from which they were produced is the $\Pi_{{\rm EGG}}$ graph. For example, there are two graphs that are produced by the substitution of the $U_\ta$ graphs in Fig. \ref{UintLABEL} into the $\Pi_\eg$ diagram, which are labeled $[\Pi_\eg[U_\ta]]_1$ and $[\Pi_\eg[U_\ta]]_2$ in Fig. \ref{figreplaceEGG}.
The final result is obtained by combining the diagrams in the first two lines in Fig. \ref{PIintLABEL}, with the $\Pi_{{\rm EGG}}$ diagram replaced by the graphs in the rhs of the first line of Fig. \ref{figreplaceEGG}. We extract the diagrams that correspond to the sd equation with $V=V^0$, and then show that the remaining diagrams cancel.

In order to explain how this procedure works, we must introduce some new notation. Throughout this paper, we combine permutations of external legs in equations and diagrams, whenever possible. In order to separate the sd contributions from the rearranged eom's, we need to separate contributions that correspond to permutations of external legs for some graphs.
When external leg permutations are separated, we indicate with a superscript (1p, 2p, $\dots$)  how many permutations are included in a specific term. For example we write:
\bea
&& \Pi_{\eg_0}=\Pi^{\rm 1p}_{\eg_0}+\Pi^{\rm 1p}_{\eg_0}\,, \\
&& U_\ta= U_\ta^{\rm 1p} + U_\ta^{\rm 2p}\,.\nonumber
\eea
We describe the content of these equations in words as follows:
The graph $\Pi_{\eg_0}$ in Fig. \ref{PIintLABEL}, which contains 2 permutations of external legs, is split into two pieces, each containing one permutation, and both of which are labeled $\Pi^{\rm 1p}_{\eg_0}$. The graph  $U_\ta$ in Fig. \ref{UintLABEL},  which contains 3 permutations, is split into two pieces $U_\ta^{\rm 1p}$ and $U_\ta^{\rm 2p}$ which contain 1 and 2 permutations, respectively. We note that the superscripts $\{$1p, 2p, $\dots \}$ indicate the number of  permutations in a given term, but do not tell us which ones. We use this abbreviated and incomplete notation in order to make the equations readable. Each time this notation is used, the specific permutation that is meant will either be explained in words or be clear from the corresponding figure.

Using this notation we can now extract the
terms that correspond to the sd equation from the rearranged eom. They are $\Pi_{\rm EIGHT}$, the permutation of the $\Pi_{\rm EGG_0}$ diagram with the bare vertex on the lhs, the permutation of the $[\Pi_{\rm HAIR}]_1$ diagram with the bare vertex on the lhs, and the sum of the $\Pi_\bb$ and $\Pi_{\bb_0}$ diagrams with $V=V^0$, which we write $\{ \Pi_{\rm EIGHT}$, $\Pi_{\rm EGG_0}^{\rm 1p}$, $[\Pi_{\rm HAIR}]^{\rm 1p}_1$, $\Pi_\bb$, $\Pi_{\bb_0} \}$. %From Fig. \ref{sdPIeqnLABEL}, it is clear that, for both graphs that carry the superscript 1p, the permutation we need for the sd equation is the one with the bare vertex on the lhs.

It is straightforward to see that all of the remaining terms cancel.
We list the sets of terms that cancel symbolically in Eq. (\ref{can3}). In this equation, the superscripts 1p indicate the permutations with bare vertices on the rhs, which were not included in the sd equation.
\bea
\label{can3}
&& \Pi_{{\rm EGG}_0}^{\rm 1p}+\Pi_{{\rm EGG}}[U_{{\rm EGG}_0}] =0\,,\\
&& \Pi_{{\rm MERCEDES}}+\Pi_{{\rm EGG}}[U_{\rm MERCEDES}] =0\,,\nonumber\\
&& [\Pi_{{\rm HAIR}}]_2+[\Pi_{{\rm EGG}}[U_{\rm HAIR}]]_1 =0\,,\nonumber\\
&& [\Pi_{{\rm HAIR}}]_1^{\rm 1p}+[\Pi_{{\rm EGG}}[U_{\rm HAIR}]]_2 =0\,.\nonumber
\eea
The content of this equation is described in words as follows.

\noindent 1) $\Pi^{\rm 1p}_{{\rm EGG}_0}$ is the permutation of the $\Pi_{{\rm EGG}_0}$ diagram in Fig. \ref{PIintLABEL} with the bare vertex on the rhs. It cancels with the diagram labeled $\Pi_{{\rm EGG}}[U_{{\rm EGG}_0}]$ in Fig. \ref{figreplaceEGG}.

\noindent 2) The diagram marked $\Pi_{{\rm MERCEDES}}$ in Fig. \ref{PIintLABEL} cancels with the diagram marked $\Pi_{{\rm EGG}}[U_{\rm MERCEDES}]$ in Fig. \ref{figreplaceEGG}.

\noindent 3) The graph marked $[\Pi_{{\rm HAIR}}]_2$ in Fig. \ref{PIintLABEL} cancels with the diagram marked $\Pi_{{\rm EGG}}[U_{\rm HAIR}]_1$ in Fig. \ref{figreplaceEGG}.

\noindent 4) $[\Pi_{{\rm HAIR}}]_1^{\rm 1p}$ is the permutation of the  $[\Pi_{{\rm HAIR}}]_1$ diagram in Fig. \ref{PIintLABEL} with the bare vertex on the rhs. It cancels with the diagram marked  $\Pi_{{\rm EGG}}[U_{\rm HAIR}]_2$ in Fig. \ref{figreplaceEGG}.

The result is that all graphs in the rhs of the first line of Fig. \ref{figreplaceEGG} cancel, and the surviving diagrams in Fig. \ref{PIintLABEL}
%are $\Pi_{{\rm EIGHT}}$, the permutation of the $\Pi_{{\rm EGG}_0}$ diagram with the bare vertex on the lhs, the permutation of the  $[\Pi_{{\rm HAIR}}]_1$ diagram with the bare vertex on the lhs, and a $\Pi_{{\rm BBALL}}$-type diagram with 2 bare vertices and symmetry factor 1/6 (from combining $\Pi_{{\rm BBALL}}$ and $\Pi_{{\rm BBALL}_0}$ and setting $V=V^0$).
have exactly the same form as the sd equation in Fig. \ref{sdPIeqnLABEL}, with $V=V^0$. Thus we have verified Eq. (\ref{rearrangePi3}), and found again that for the 2-point function the extra terms are exactly zero.

\ts

We point out an interesting feature of the cancellations listed above. If the contributions from the HAIR diagram are grouped together, the cancellations in Eq. (\ref{can3}) have the form
\bea
\label{master3}
\Pi_{\cal I} + \Pi_{{\rm EGG}}[U_{{\cal I}}] = 0\,;~~{\cal I}\in
\{\{3{\rm PI}\}\backslash {\rm super\!\!-\!\!BBALL}\}~~ {\rm and} ~~ \Pi \notin\Pi_{sd}\,,
\eea
where the notation ``${\cal I}\in\{\{3{\rm PI}\}\backslash {\rm super\!\!-\!\!BBALL}\}$ and $\Pi \notin\Pi_{sd}$'' means contributions to ${\cal I}$ from all diagrams in  the 3-Loop 3PI $\Phi$ except the super-BBALL diagrams (the EGG diagram), but removing $\Pi$ terms that are in the sd equation, which are not canceled.
We will show below that this pattern holds for the 4-Loop 4PI effective theory, and the 5-Loop 5PI effective theory.
Equation (\ref{master3}) can be represented symbolically by the first two terms in Fig. \ref{figpi5loop}.

\subsubsection{eom for $\Pi$ for 4-Loop 4PI}
\label{4PIpi}

For 4-Loop 4PI we have min[$i+2$,$n$]=min[2+2,4]=4 and ${\cal L}[m,i]={\cal L}[4,2]=3$. Equation (\ref{eomRearrange}) becomes
\bea
\label{rearrangePi4}
\Pi = -f_2^{sd}[D,U,V]~~+~~\underbrace{~~~{\rm extra}~~~}_{L_{\rm pt}\ge 4}\,,
\eea
where we have dropped the subscript ${L\le 3}$ on the first term on the rhs, since all terms in the sd equation  have two or fewer loops.

\begin{figure}[H]
\begin{center}
\epsfig{file=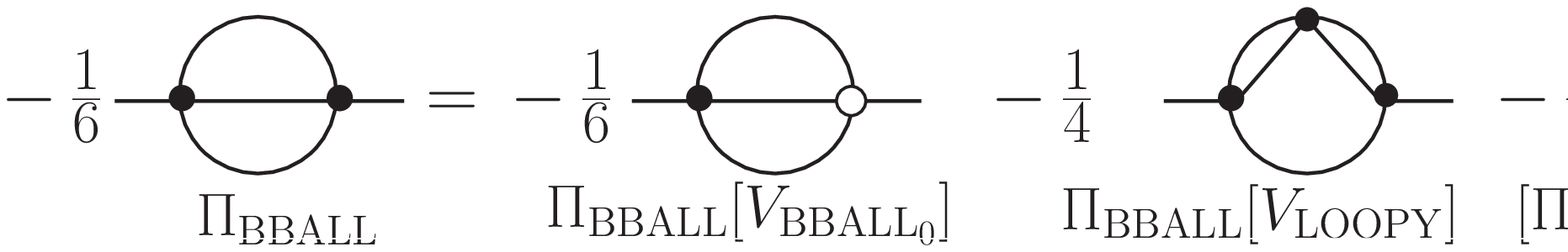,width=17cm}
\end{center}
\vspace*{-0.5cm} \caption{\small Result from replacing the 4-vertex on the rhs of the $\Pi_{{\rm BBALL}}$ diagram
with the corresponding eom from the 4-Loop 4PI effective action (Fig. \ref{VintLABEL}).}
\label{figreplaceBBALL}
\end{figure}

The eom for the 2-point function for 4-Loop 4PI is shown in Fig. \ref{PIintLABEL}. We use  Eq. (\ref{genericSubU}) to replace the variational proper 3-vertex on the rhs of the $\Pi_{{\rm EGG}}$ diagram, and Eq. (\ref{genericSubV}) to replace the variational proper 4-vertex on the rhs of the $\Pi_{{\rm BBALL}}$ diagram. The results for the $\Pi_{{\rm EGG}}$ and $\Pi_{{\rm BBALL}}$ diagrams are shown in Figs. \ref{figreplaceEGG} and \ref{figreplaceBBALL}. We use labels of the form $\{V_{\bb_0},\,V_\lo,\,V_\ey\,\dots\}$ to indicate which term in the $V$ eom produced each graph.

The final result is obtained by combining the diagrams in  Fig. \ref{PIintLABEL}, with the $\Pi_{{\rm EGG}}$ diagram replaced by the graphs on the rhs of Fig. \ref{figreplaceEGG}, and the $\Pi_{{\rm BBALL}}$ diagram replaced by the graphs on the rhs of Fig. \ref{figreplaceBBALL}.
The terms that correspond to the sd equation are $\{ \Pi_{\rm EIGHT}$, $\Pi_{\rm EGG_0}^{\rm 1p}$, $[\Pi_{\rm HAIR}]^{\rm 1p}_1$, $\Pi^{\rm 1p}_{\bb_0} \}$, where the superscripts 1p indicate the permutations with bare vertices on the lhs.
In Eq. (\ref{can4}) we list the sets of terms that cancel, in addition to those  in Eq. (\ref{can3}). In this equation the superscripts 1p refer to permutations with bare vertices on the rhs.
\bea
\label{can4}
&& \Pi_{{\rm BBALL}_0}^{\rm 1p}+\Pi_{{\rm BBALL}}[V_{\bb_0}] =0\,,\\
&& \Pi_{\tw} +\Pi_{\eg}[U_{\tw}]=0\,,\nonumber\\
&& \Pi_{\lo} +\Pi_{\bb}[V_{\lo}]=0\,,\nonumber\\
&& [\Pi_{\ta}]_1+ [\Pi_{\eg}[U_{\ta}]]_1=0\,,\nonumber\\
&& [\Pi_{\ta}]_2+\Pi_{\bb}[V_{\ta}]+[\Pi_{\eg}[U_{\ta}]]_2 =0\,,\nonumber\\
&& [\Pi_{\ey}]_1+ [\Pi_{\eg}[U_{\ey}]]_1 =0\,,\nonumber\\
&& [\Pi_{\ey}]_2+ [\Pi_{\bb}[V_{\ey}]]_1 =0\,,\nonumber\\
&& [\Pi_{\ey}]_3 + [\Pi_{\eg}[U_{\ey}]]_2+ [\Pi_{\bb}[V_{\ey}]]_2 =0\,.\nonumber
\eea

The result is that all graphs on the rhs of Figs. \ref{figreplaceEGG} and \ref{figreplaceBBALL} cancel, and the surviving diagrams in Fig. \ref{PIintLABEL} have exactly the same form as the sd equation in Fig. \ref{sdPIeqnLABEL}. Thus we have verified Eq. (\ref{rearrangePi4}), and found again that for the 2-point function the extra terms are exactly zero.

\ts

We note that Eq. (\ref{can4}) has the same structure as Eq. (\ref{can3}). If the contributions from the $\Pi_{{\rm HAIR}}$, $\Pi_{{\rm TARGET}}$ and $\Pi_{{\rm EYEBALL}}$ graphs are respectively grouped together, we can rewrite Eqs. (\ref{can3}) and (\ref{can4}) as
\bea
\label{master4}
\Pi_{\cal I} + \Pi_{{\rm EGG}}[U_{{\cal I}}] + \Pi_{{\rm BBALL}}[V_{{\cal I}}]= 0\,;~~{\cal I}\in\{\{4{\rm PI}\}\backslash {\rm super\!\!-\!\!BBALL}\}~~ {\rm and} ~~ \Pi \notin\Pi_{sd}\,,
\eea
where the notation ``${\cal I}\in\{\{4{\rm PI}\}\backslash {\rm super\!\!-\!\!BBALL}\}$ and $\Pi \notin\Pi_{sd}$'' means contributions to ${\cal I}$ from all diagrams in the 4-Loop 4PI $\Phi$ except the super-BBALL diagrams (EGG and BBALL),  but removing $\Pi$ terms that are in the sd equation, which are not canceled. Equation (\ref{master4}) can be represented symbolically by the first three terms in Fig. \ref{figpi5loop}.

\subsubsection{eom for $\Pi$ for 5-Loop 5PI}
\label{5PIpi}

For 5-Loop 5PI we have min[$i+2$,$n$]=min[2+2,5]=4 and ${\cal L}[m,i]={\cal L}[5,2]=4$. Equation (\ref{eomRearrange}) becomes
\bea
\label{rearrangePi5}
\Pi = - f_2^{sd}[D,U,V]~~+~~\underbrace{~~~{\rm extra}~~~}_{L_{\rm pt}\ge 5}\,,
\eea
where we have dropped the subscript ${L\le 4}$ on the first term on the rhs, since all terms in the sd equation  have two or fewer loops. We remark that although Eqs. (\ref{rearrangePi4}) and (\ref{rearrangePi5}) are identical in their structure, their content is different, because their functional arguments are variational propagators and vertices that are determined from different equations of motion.

\begin{figure}[H]
\begin{center}
\epsfig{file=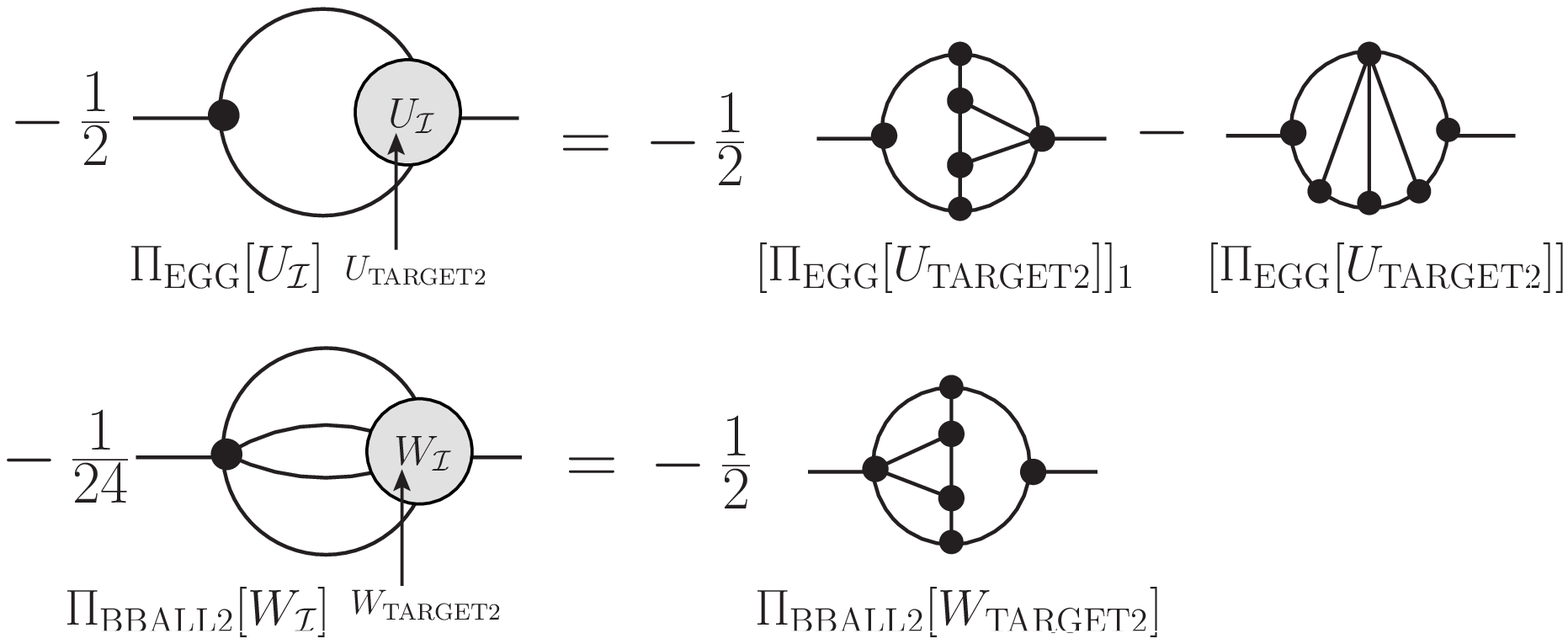,width=10.5cm}
\end{center}
\vspace*{-0.5cm} \caption{\small Contributions that cancel the graphs labeled $\Pi_{{\rm TARGET2}}$ in Fig. \ref{bigPIintLABEL}.}
\label{canNEWTAR}
\end{figure}

\begin{figure}[H]
\begin{center}
\epsfig{file=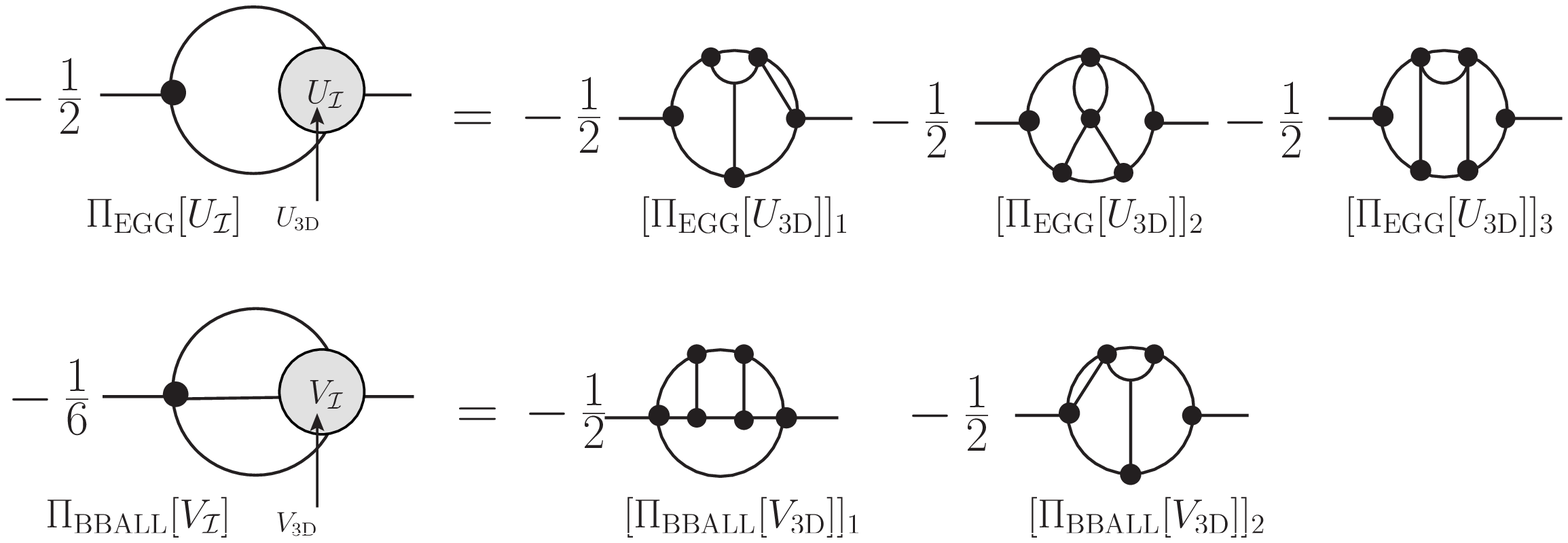,width=12cm}
\end{center}
\vspace*{-0.5cm} \caption{\small Contributions that cancel the graphs labeled $\Pi_{{\rm 3D}}$ in Fig. \ref{bigPIintLABEL}.}
\label{canNEW3D}
\end{figure}

In order to verify Eq. (\ref{rearrangePi5}), we follow the same strategy as before.
The new diagrams in the 5PI effective action are the 5-loop diagrams in Fig. \ref{fig:fig17} and the BBALL2 diagram in Fig. \ref{fig:fig16}.
It is straightforward to calculate the additional contributions to the eom's for the 2-, 3-, 4-, and 5-point functions that are produced by these diagrams. The results are shown in \ref{eomSection}.
%The BBALL2 diagram gives a contribution to the $\Pi$ eom that looks like the BBALL contribution in Fig. \ref{PIintLABEL}, except that there is an extra line between the vertices.
We use Eqs. (\ref{genericSubU}), (\ref{genericSubV}) and (\ref{genericSubW}) to replace the variational proper vertices on the rhs of the contributions to the $\Pi$ eom from the EGG, BBALL, and BBALL2 diagrams, respectively.
These substitutions produce 58 different 4-loop topologies which cancel exactly with the 58 topologies in Fig. \ref{bigPIintLABEL}. The final result is that all contributions cancel, except for the graphs that correspond to the sd equation for the 2-point function, which verifies Eq. (\ref{rearrangePi5}). Once again, the extra terms cancel identically.
%In order to prove that all 4-loop terms cancel, one must separate all contributions to each of the 58 different topologies by writing explicitly all external indices. We have done this.
%
As examples, in Eqs. (\ref{TARGET2}) and (\ref{3D}) we list the sets of graphs that cancel the diagrams labeled $\Pi_{{\rm TARGET2}}$ and $\Pi_{{\rm 3D}}$ in Fig. \ref{bigPIintLABEL}:
\bea
\label{TARGET2}
&& [\Pi_{{\rm TARGET2}}]_1+ [\Pi_{\eg}[U_{\rm TARGET2}]]_1+ \Pi_{{\rm BBALL2}}[W_{\rm TARGET2}]=0\,,\\
&& [\Pi_{{\rm TARGET2}}]_2+ [\Pi_{\eg}[U_{\rm TARGET2}]]_2=0 \,,\nonumber
\eea
\bea
\label{3D}
&& [\Pi_{{\rm 3D}}]_1+ [\Pi_{\eg}[U_{\rm 3D}]]_3=0\,,\\
&& [\Pi_{{\rm 3D}}]_2+ [\Pi_{\eg}[U_{\rm 3D}]]_2=0\,,\nonumber\\
&& [\Pi_{{\rm 3D}}]_3+ [\Pi_{\eg}[U_{\rm 3D}]]_1 + [[\Pi_{{\rm BBALL}}[V_{\rm 3D}]]]_2=0\,, \nonumber\\
&& [\Pi_{{\rm 3D}}]_4+ [\Pi_{{\rm BBALL}}[V_{\rm 3D}]]_1=0\,.\nonumber
\eea
The substitutions that produce the terms on the rhs of Eqs. (\ref{TARGET2}) and (\ref{3D}) are shown in Figs. \ref{canNEWTAR} and \ref{canNEW3D}, respectively.

\begin{figure}[H]
\begin{center}
\epsfig{file=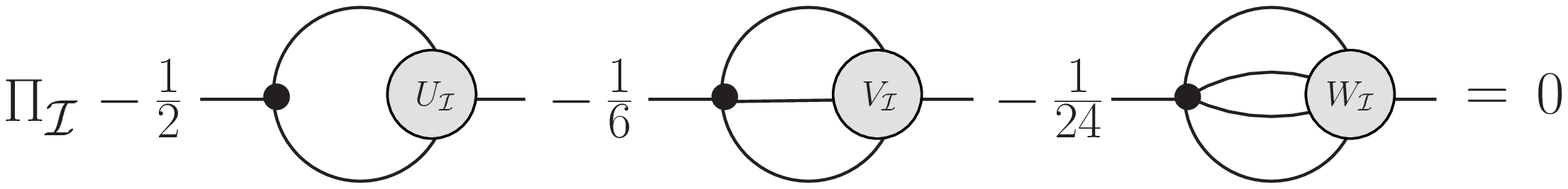,width=12.5cm}
\end{center}
\vspace*{-0.5cm} \caption{\small Contributions to $\Pi$ from the 5-Loop 5PI effective action that cancel.} \label{figpi5loop}
\end{figure}

All of the cancellations have the same structure as the examples given in Eqs. (\ref{TARGET2}) and (\ref{3D}),  and  we have the same pattern of cancellation as in Eqs. (\ref{master3}) and (\ref{master4}):
\bea
\label{master5}
&& \Pi_{\cal I} + \Pi_{{\rm EGG}}[U_{{\cal I}}] + \Pi_{{\rm BBALL}}[V_{{\cal I}}]+ \Pi_{{\rm BBALL2}}[W_{{\cal I}}]= 0\,;\\
&& \hspace*{4cm} {\cal I}\in\{\{5{\rm PI}\}\backslash  {\rm super\!\!-\!\!BBALL}\}~~ {\rm and} ~~ \Pi \notin\Pi_{sd}\,,\nonumber
\eea
where the notation ``${\cal I}\in\{\{5{\rm PI}\}\backslash  {\rm super\!\!-\!\!BBALL}\}$ and $\Pi \notin\Pi_{sd}$'' means contributions to ${\cal I}$ from all diagrams in the 5-Loop 5PI $\Phi$ except the super-BBALL  diagrams (EGG, BBALL and BBALL2), but removing $\Pi$ terms that are in the sd equation, which are not canceled.
Equation (\ref{master5}) is represented symbolically in Fig. \ref{figpi5loop}.

\subsubsection{eom for $\Pi$ for $n$-Loop $n$PI}
\label{nPIpi}

We describe below a general method to rearrange the eom for the 2-point function to show that it satisfies Eq. (\ref{eomRearrange}). In the subsections above, we have demonstrated that the method works for the $n$-Loop $n$PI effective theory, for $n\le 5$. We expect that it will work at higher orders.
%All of the terms in the sd equation are contained in the eom for $\Pi$ obtained from the 4-Loop 4PI effective action (compare Figs.  \ref{PIintLABEL} and \ref{sdPIeqnLABEL}).
The strategy is to perform substitutions on the diagrams in the eom that were produced by the super-BBALLs. We rearrange these terms by using Eqs. (\ref{genericSubU}), (\ref{genericSubV}), $\dots$ to replace the variational proper vertex on the rhs.
These substitutions produce graphs that cancel the diagrams that appear in the original eom, and not in the sd equation.
We have found that each diagram in the effective action, except the super-BBALLs, produces contributions that cancel among themselves, excluding the diagrams in the eom that appear in the sd equation. This pattern of cancellation is depicted in Fig. \ref{figpi5loop} for 5-Loop 5PI. An equivalent statement is that if we remove any set of diagrams from $\Phi^{int}$ (not including super-BBALL diagrams),  the pattern of cancellation would not be destroyed:  we would still find that the eom for the 2-point function has the same form as the sd equation.
The corresponding calculation for the 3-point function is much more complicated. We discuss this in the next section.

\subsection{Integral Equation for the 3-point Vertex}

\subsubsection{eom for $U$ for 3-Loop 3PI}
\label{3PIU}

We start with the 3-Loop 3PI effective action.
We have min[$i+2,n$] = min[3+2,3]=3 and ${\cal L}[n,i]={\cal L}[3,3] = 1$. Equation (\ref{eomRearrange}) becomes
\bea
\label{exA}
3{\rm -Loop} ~3{\rm PI}:~~~U=U^0+f^{sd}_3[D,U]\big|_{L\le 1}~+~\underbrace{~~~\rm extra~~~}_{L_{\rm pt}\ge 2}\,.
\eea
This equation says that the variational eom for the 3-point function has the same form as the 1-loop terms in the sd equation for the 3-point function with $V=V^0$, plus some terms that are 2-loop or higher in the perturbative expansion.

This result is easy to see directly for the simple example we are discussing in this subsection.  We start by extracting the eom for the 3-point function for 3-Loop 3PI from Fig. \ref{UintLABEL} by including the contributions from the EGG$_0$, EGG, EIGHT, BBALL$_0$, BBALL, HAIR, and MERCEDES diagrams, and setting $V=V^0$. The first three diagrams on the rhs of Fig. \ref{UintLABEL} give the lhs of Eq. (\ref{exA}). Note that these diagrams are all $L\le 1$ loop.

Now we look at the  rhs of Eq. (\ref{exA}). The subscript $L\le 1$ indicates that we can ignore all 2-loop diagrams in the sd equation. In addition, we can freely interchange bare and variational proper vertices in 1-loop diagrams of the sd equation (since the difference is always 2-loop or higher in the perturbative expansion).
%For example, the MERCEDES graph in Fig. \ref{UintLABEL} and  graph (2) in Fig. \ref{sdUeqnLABEL} are equivalent,  since replacing the bare 3-vertex in the sd graph with the variational proper 3-vertex would only contribute an error at 2-loops.
In order to show that the ``extra'' term is 2-loop or higher, we only have to observe that the symmetry factors for the 1-loop diagrams in Figs. \ref{UintLABEL} and \ref{sdUeqnLABEL} are the same.

\subsubsection{eom for $U$ for 4-Loop 4PI}
\label{4PIU}
Now we consider the 4-Loop 4PI effective theory. We have min[$i+2$,n]=min[3+2,4]=4 and ${\cal L}[m,i]={\cal L}[4,3]=2$. Equation (\ref{eomRearrange}) becomes
\bea
\label{rearrangeU3}
U = U^0 + f_3^{sd}[D,U,V]~~+~~\underbrace{~~{\rm extra}~~}_{L_{\rm pt}\ge 3}\,.
\eea
We drop the subscript $L \le 2$ on the second term on the rhs, since all terms in the sd equation have two or fewer loops.

\begin{figure}[H]
\begin{center}
\includegraphics[width=16cm]{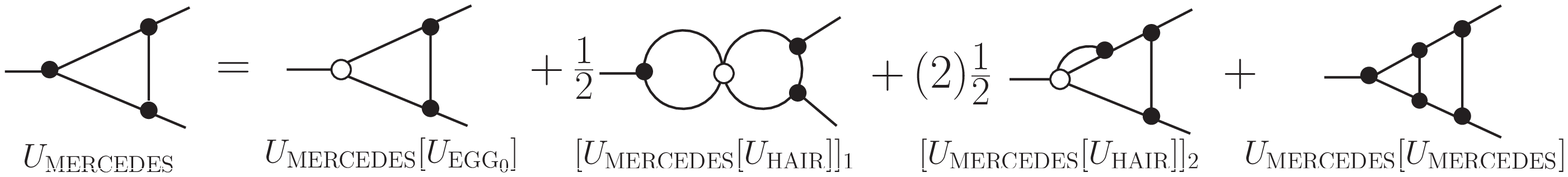}
\end{center}
\caption{\label{ureplace1}Rearrangement of the diagram labeled $U_{{\rm MERCEDES}}$ in the eom for 3-vertex. }
\end{figure}

\begin{figure}[H]
\begin{center}
\includegraphics[width=14cm]{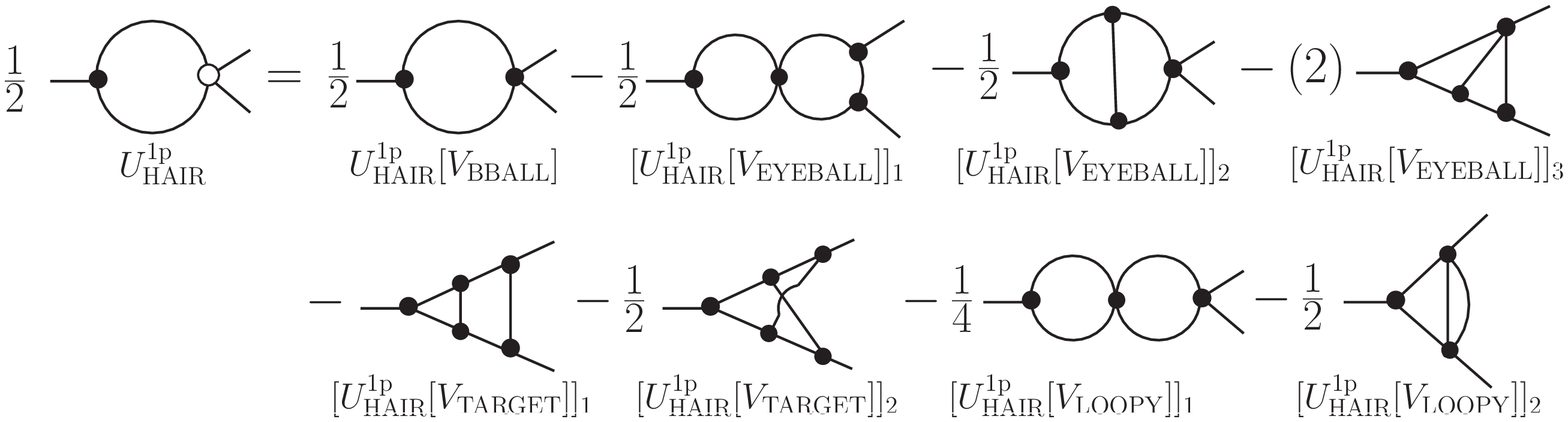}
\end{center}
\caption{\label{ureplace2}Rearrangement of the $U^{\rm 1p}_{{\rm HAIR}}$ diagram in the eom for the 3-vertex. }
\end{figure}

The eom for the 4-Loop 4PI 3-point function is given in Fig. \ref{UintLABEL}. In order to prove Eq. (\ref{rearrangeU3}), we need to rearrange the eom for the 3-point function as follows:

1) The diagram marked $U_{{\rm MERCEDES}}$ in Fig. \ref{UintLABEL} is rearranged by replacing the 3-vertex on the lhs using Eq. (\ref{genericSubU}). We represent this rearrangement in equation form as
\bea
\label{genmer}
U_{{\rm MERCEDES}}=U_{{\rm MERCEDES}}[U_{{\rm EGG}_0}]+\sum_{k=1}^{n-2}U_{{\rm MERCEDES}}[{\rm fcn}_3^{(k-{\rm loop})}]\,.
\eea
In this subsection we are only interested in contributions with $L\le 2$ (all terms with $L\ge 3$ loops can be lumped into the extra term). Since the $U_{{\rm MERCEDES}}$ diagram is itself 1-loop, we only need to consider 1-loop terms in ${\rm fcn}_3$, which means we can take $k=1$ in
Eq. (\ref{genmer}). The result is  shown in Fig. \ref{ureplace1}\footnote{In Figs. \ref{ureplace1}, \ref{ureplace2} \ref{newfig3} and  \ref{UcanF}, the bracketed numerical factors indicate permutations of the external legs on the rhs of the graph only.}.
\par

%Note that at 4-loop 4PI level, we should have 3-loop terms in the figure which we don't show explicitly for the moment.
%Note that the three 3-vertices in diagram MERCEDES are symmetric, we can replace any one of the three 3-vertices. However, the 2 terms on the rhs of Eq. (\ref{genmer}) correspond to the replacement for the same 3-vertex in MERCEDES. The same applies for the following equations.

2) There are 3 permutations of the diagram marked $U_{{\rm HAIR}}$ in Fig. \ref{UintLABEL}. We separate them by writing:
\bea
U_{{\rm HAIR}}=U^{\rm 2p}_{{\rm HAIR}}+U^{\rm 1p}_{{\rm HAIR}}\,.
\eea
The two permutations a bare vertex on the lhs are written $U^{\rm 2p}_{{\rm HAIR}}$, and
correspond to the diagrams labeled (4,5) in the sd equation in Fig. \ref{sdUeqnLABEL}. The third permutation is shown on the lhs of Fig. \ref{ureplace2} and labeled $U^{\rm 1p}_{{\rm HAIR}}$.
This graph has the same form as diagram (3) in the sd equation, but with corrected and bare vertices in the wrong places. We rearrange the $U^{\rm 1p}_{{\rm HAIR}}$ diagram as follows. The first step is to remove the bare 4-vertex using Eq. (\ref{genericSubV0}). This substitution is written as an equation:
\bea
\label{genhair}
U^{\rm 1p}_{{\rm HAIR}}=U^{\rm 1p}_{{\rm HAIR}}[V_{{\rm BBALL}}]-\sum_{k=1}^{n-3}U^{\rm 1p}_{{\rm HAIR}}[{\rm fcn}_4^{(k-{\rm loop})}]\,.
\eea
As argued above, we can take $k$=1 since we are working to 2-loop level, which gives the diagrams shown in Fig. \ref{ureplace2}.
\par

\begin{figure}[H]
\begin{center}
\includegraphics[width=16cm]{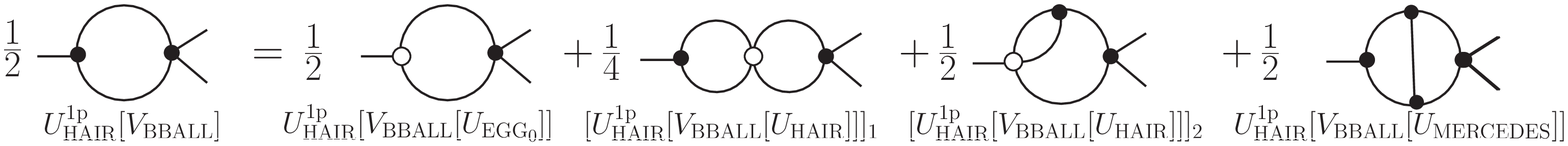}
\end{center}
\caption{\label{ureplace3}Rearrangement of the diagram labeled $U^{\rm 1p}_{{\rm HAIR}}[V_{{\rm BBALL}}]$ in Fig. \ref{ureplace2}. }
\end{figure}

The second step is to extract the graph in Fig. \ref{ureplace2} labeled $U^{\rm 1p}_{{\rm HAIR}}[V_{{\rm BBALL}}]$ and use Eq. (\ref{genericSubU}) to obtain the bare 3-vertex that we need to reproduce the graph in the sd equation that is labeled (3) in Fig. \ref{sdUeqnLABEL}.  This substitution is represented in the equation
\bea
\label{gen}
U^{\rm 1p}_{{\rm HAIR}}[V_{{\rm BBALL}}]=U^{\rm 1p}_{{\rm HAIR}}[V_{{\rm BBALL}}[U_{{\rm EGG}_0}]]+\sum_{k=1}^{n-2}U^{\rm 1p}_{{\rm HAIR}}[V_{{\rm BBALL}}[{\rm fcn}_3^{(k-{\rm loop})}]]\,.
\eea
To 2-loop level, we only need the $k=1$ term in Eq. (\ref{gen}). The result is shown diagrammatically in Fig. \ref{ureplace3}.
\par

Now we discuss how to combine all of these terms. The graphs that have to be added together are as follows. From the original eom (Fig. \ref{UintLABEL}), we need the graphs that have not been involved in the rearrangements above: the $U_{{\rm EGG}_0}$, $U^{\rm 2p}_{{\rm HAIR}}$, $U_{{\rm EYEBALL}}$, $U_{{\rm TARGET}}$, and $U_{{\rm TWISTED}}$ graphs. The $U_{{\rm MERCEDES}}$ graph is replaced by the set of graphs shown in Fig. \ref{ureplace1}. The $U^{\rm 1p}_{{\rm HAIR}}$ graph is replaced by the set of graphs shown in Fig. \ref{ureplace2}, with the first graph in this figure replaced in turn by the graphs in Fig. \ref{ureplace3}.

First we extract from this set of graphs the diagrams that correspond to the sd equation.

\begin{itemize}

\item In Fig. \ref{UintLABEL} the $U_{{\rm EGG}_0}$ and $U^{\rm 2p}_{{\rm HAIR}}$ graphs correspond, respectively, to the diagrams labeled (1), (4,5) in the sd equation (Fig. \ref{sdUeqnLABEL}).

\item From Fig. \ref{UintLABEL} we take the permutations of the $U_{{\rm TARGET}}$ and $U_{{\rm EYEBALL}}$ graphs which have the same structure as
(8) and (9,10) in the sd equation (Fig. \ref{sdUeqnLABEL}). We label these pieces $U^{\rm 1p}_{{\rm TARGET}}$ and $U^{\rm 2p}_{{\rm EYEBALL}}$  (the permutations that have not been accounted for are labeled $U^{\rm 2p}_{{\rm TARGET}}$ and $U^{\rm 1p}_{{\rm EYEBALL}}$, and will be considered later).
Since the vertex on the lhs of the $U^{\rm 1p}_{{\rm TARGET}}$ and $U^{\rm 2p}_{{\rm EYEBALL}}$ diagrams can be replaced with bare vertices (because the difference is 3-loop in the perturbative expansion), these graphs correspond to the terms labeled (8) and (9,10) in the sd equation.

\item The first and third graphs on the rhs of Fig. \ref{ureplace1} %(($U_{\mer}[U_{\eg_0}]$ and $[U_{\mer}[U_\ha]]_2$)
correspond, respectively, to the graphs labeled (2) and (6,7) in the sd equation.

\item The first  and third graphs on the rhs of  Fig. \ref{ureplace3} %($U_{\ha}[V_{\bb}[U_{\eg_0}]]$ and $[U_{\ha}[V_\bb[U_\ha]]]_2$)
correspond, respectively, to the graphs labeled (3) and (11) in the sd equation.

\item The graph labeled (12) in the sd equation is identically zero at the level of the 4PI effective theory, because the 5-point vertex on the rhs reduces to $W=W^0=0$.
\end{itemize}
We have produced all of the terms in the sd equation, and thus verified the first two terms on the rhs in Eq. (\ref{rearrangeU3}).

\ts

It is straightforward to see that the remaining 2-loop diagrams cancel, which verifies that the extra term in Eq. (\ref{rearrangeU3}) contains only terms of $L_{\rm pt}\ge 3$ loops. We list the contributions that cancel.

First, we give the sets of diagrams that cancel identically:
\bea
\label{canU3}
&&U_{{\rm MERCEDES}}[U_{\rm {MERCEDES}}]+[U^{\rm 1p}_{{\rm HAIR}}[V_{\rm {TARGET}}]]_1=0\,, \\
&& U_{{\rm TWISTED}}+[U^{\rm 1p}_{{\rm HAIR}}[V_{\rm {TARGET}}]]_2=0\,,\nonumber \\
&& U^{\rm 1p}_{{\rm HAIR}}[V_{\rm {BBALL}}[U_{\rm {MERCEDES}}]]+[U^{\rm 1p}_{{\rm HAIR}}[V_{\rm {EYEBALL}}]]_2=0 \,,\nonumber\\
&& U^{\rm 1p}_{{\rm EYEBALL}}+[U^{\rm 1p}_{{\rm HAIR}}[V_{\rm {LOOPY}}]]_2=0\,,\nonumber \\
&& U^{\rm 2p}_{{\rm TARGET}}+ [U^{\rm 1p}_{{\rm HAIR}}[V_{\rm {EYEBALL}}]]_3=0 \,.\nonumber
\eea
%In the last two lines in Eq. (\ref{canU3}), the numerical factors in brackets indicate the number of permutations of external legs that are included. For example, in the fourth line of Eq. (\ref{canU3}), the factor (1) indicates that one of the three permutations of the $U_{{\rm EYEBALL}}$ diagram in Fig. \ref{UintLABEL} cancels with the graph labelled $U_{{\rm HAIR}}[V_{\rm {LOOPY}}]_2$ in Fig. \ref{ureplace2}.

In addition, there are two pairs of diagrams which have the same form, except that one contains a bare vertex $V^0$ and the partner diagram contains a variational vertex $V$. If we use Eq. (\ref{genericSubV0}) to remove the bare vertex, the surviving terms in each pair are at least 3-loop in the perturbative expansion. Therefore, at the 2-Loop level we have the additional cancellations:
\bea
\label{V0can}
&& [U_{\mer}[U_{\ha}]]_1+ [U^{\rm 1p}_{\ha}[V_{\ey}]]_1=0\,,\\
&& [U^{\rm 1p}_{\ha}[V_{\bb}[U_{\ha}]]]_1+[U^{\rm 1p}_{\ha}[V_{{\rm {LOOPY}}}]]_1=0 \,.\nonumber
\eea

\subsubsection{eom for $U$ for 5-Loop 5PI}
\label{5PIU}

Now we consider the 5-Loop 5PI effective theory. We have min[$i+2$,n]=min[3+2,5]=5 and ${\cal L}[m,i]={\cal L}[5,3]=3$. Equation (\ref{eomRearrange}) becomes
\bea
\label{rearrangeU4}
U = U^0 + f_3^{sd}[D,U,V,W]~~+~~\underbrace{~~~{\rm extra}~~~}_{L_{\rm pt}\ge 4}\,,
\eea
and we drop the subscript $L \le 2$ on the second term on the rhs, since all terms in the sd equation have two or fewer loops.

\begin{figure}
\begin{center}
\includegraphics[width=15cm]{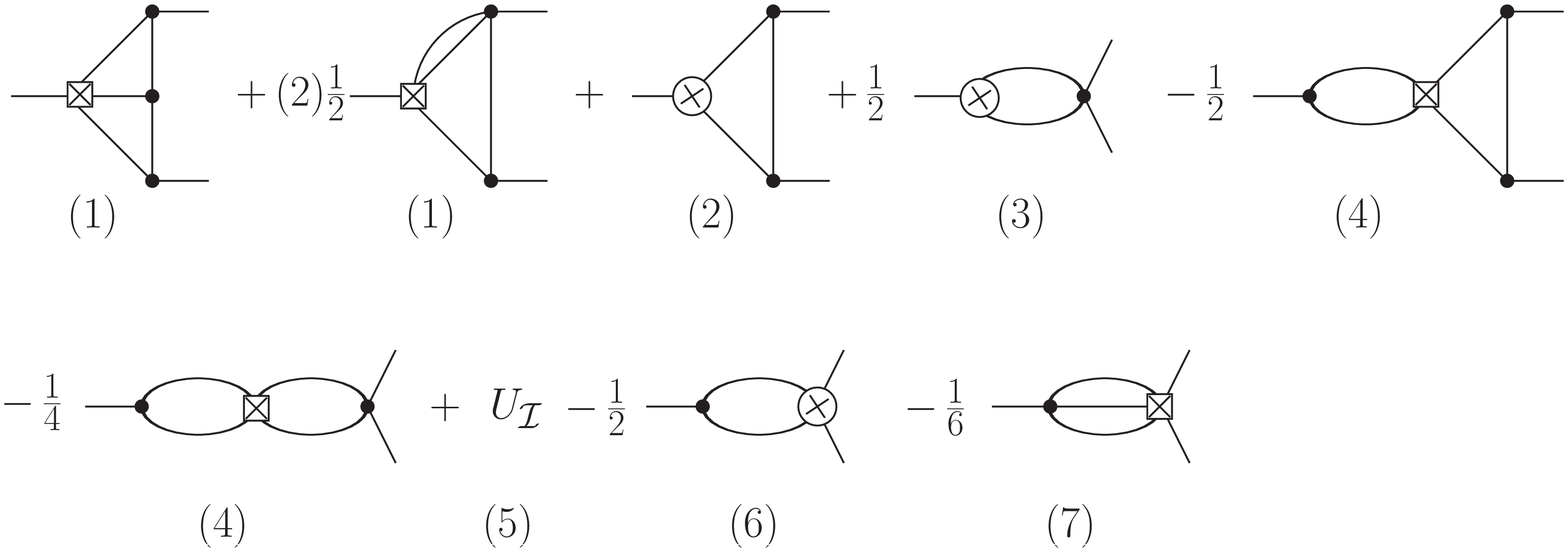}
\end{center}
\caption{\label{yunFig}The 3-loop terms in the $U$ eom for $5$-Loop $5$PI. The circles represent 2-loop insertions and the squares are 1-loop insertions. The symbol $U_{\cal I}$ means contributions to the $U$ eom from differentiating the 5-loop diagrams in the 5PI effective action. The numbers under the diagrams correspond to the numbers in the list under Eq. (\ref{rearrangeU4}). }
\end{figure}

In order to prove Eq. (\ref{rearrangeU4}), we must show that the diagrams in the extra term cancel to $L_{\rm pt}=3$ loop order. We start by listing the 3-loop diagrams that were dropped in the discussion in the previous subsection.
\begin{enumerate}

\item In the second item of the list under Fig. \ref{ureplace3}, we said that the $U^{\rm 2p}_{{\rm EYEBALL}}$ graphs and the $U^{\rm 1p}_{{\rm TARGET}}$ graph in the eom produce the diagrams marked (9,10) in the sd equation, if we use Eq. (\ref{genericSubV}) to replace the variational vertices on the lhs by bare vertices, and drop 3-loop terms. Now we need to keep these 3-loop terms, which we will write

$U^{\rm 2p}_{{\rm EYEBALL}}[{\rm fcn}_4^{({\rm 1-loop})}]$

$U^{\rm 1p}_{{\rm TARGET}}[{\rm fcn}_4^{({\rm 1-loop})}]$
\item There are 3-loop contributions that we dropped in Fig. \ref{ureplace1} which we write

$U_{{\rm MERCEDES}}[{\rm fcn}_3^{({\rm 2-loop})}]$

\item There are 3-loop contributions that we dropped in Fig. \ref{ureplace3} which we write

$U^{\rm 1p}_{\ha}[V_{\rm {BBALL}}[{\rm fcn}_3^{({\rm 2-loop})}]]$

\item In Eq. (\ref{V0can}), we had cancellation at the 2-Loop level when we used Eq. (\ref{genericSubV0}) to replace the $V^0$ vertex. The 3-loop terms have the form

$U_{\mer}[U_{\ha_1}[{\rm fcn}_4^{({\rm 1-loop})}]]$

$U^{\rm 1p}_{\ha}[V_{\bb}[U_{\ha_1}[{\rm fcn}_4^{({\rm 1-loop})}]]]$

\end{enumerate}
We must show that these diagrams cancel with the new 3-loop diagrams that are introduced at the level of 5-Loop 5PI. Using Eq. (\ref{calL}) we can see where these diagrams will appear:

\begin{enumerate}
\setcounter{enumi}{4}
\item Since ${\cal L}[5,3]=3$, there will be new 3-loop contributions to the eom for $U$ (Fig. \ref{bigUintLABEL}) from the 5-loop diagrams in the effective action\footnote{There is one diagram that is 4-loop 5PI which we call BBALL2 (see Fig. \ref{fig:fig16}), but this diagram does not contribute to the $U$ eom because it does not depend on $U$ and thus goes to zero when the functional derivative acts on it.}.

\item Since ${\cal L}[5,4]=2$ there will be 2-loop contributions to the eom for $V$, which will contribute at the 3-loop level when Eq. (\ref{genericSubV}) is used to rearrange the $U_{{\rm HAIR}}$ diagram (see Fig. \ref{ureplace2}). We write these terms:

$U^{\rm 1p}_{\ha}[{\rm fcn}_4^{({\rm 2-loop})}]$

\item There is no graph in the original (unrearranged) eom for the 3-point function with the same structure as the graph labeled (12) in the sd equation (Fig. \ref{sdUeqnLABEL}), and it cannot be produced by performing substitutions on lower loop graphs. We insert this graph into the eom by hand and group it with the other terms that make up the sd equation. We then subtract the same graph from the extra terms. Since the graph contains a 5-point vertex, and since the bare 5-vertex is identically zero, this subtracted counterterm contributes to the list of extra terms at $L_{\rm pt}= 3$ loop order. In addition, we can replace the bare 4-vertex with the proper variational 4-vertex, up to corrections at the  $L_{\rm pt}=4$ loop level that can be grouped with the extra term. We name the subtracted graph $U_{{\rm SD12}}$. The 3-loop terms that are produced by the replacements of the 5-point vertex using Eq. (\ref{genericSubW}), and the bare 4-point vertex using Eq. (\ref{genericSubV0}), are written:

$U_{{\rm SD12}}[V_{\rm BBALL}[{\rm fcn}_5^{({\rm 1-loop})}]]$
\end{enumerate}
\ts
The 3-loop terms listed above are shown diagrammatically in Fig. \ref{yunFig}. Expanding the vertex insertions produces 61 different topologies. We have verified that they all cancel. To do this calculation one must write explicitly all external indices to  separate the contributions to each topology. We give several examples below.
\par

\begin{figure}[H]
\begin{center}
\includegraphics[width=17cm]{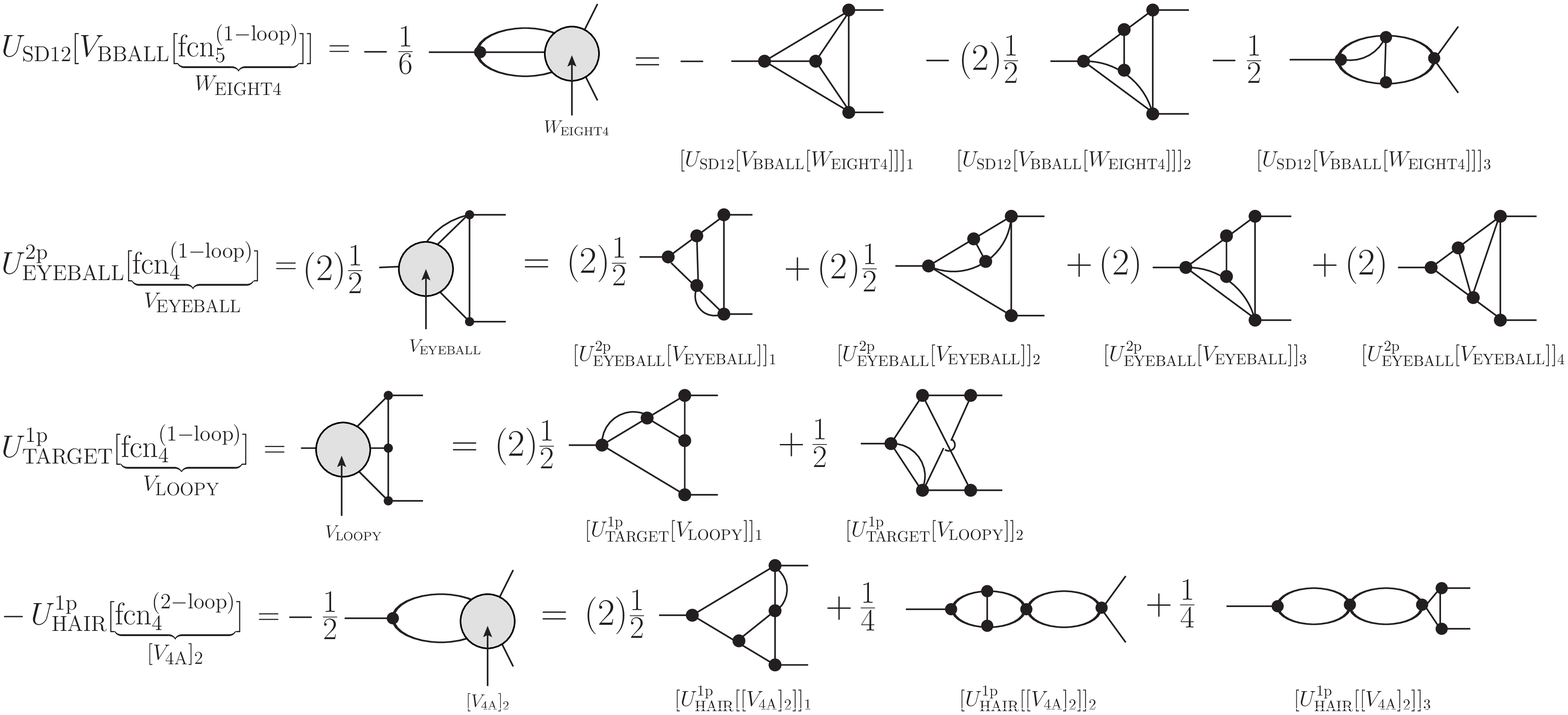}
\end{center}
\caption{\label{newfig3}Some examples of 3-loop diagrams in the eom of $U$ that cancel.}
\end{figure}

The PEA diagram in the effective action produces only one contribution to the $U$ eom which is labeled $U_{\rm PEA}$ and shown in Fig. \ref{bigUintLABEL}.  It is canceled by $U_{\rm SD12}[V_{\rm BBALL}[W_{\rm EIGHT4}]]_1$ as shown in Fig. \ref{newfig3}.

\begin{figure}[H]
\begin{center}
\includegraphics[width=15cm]{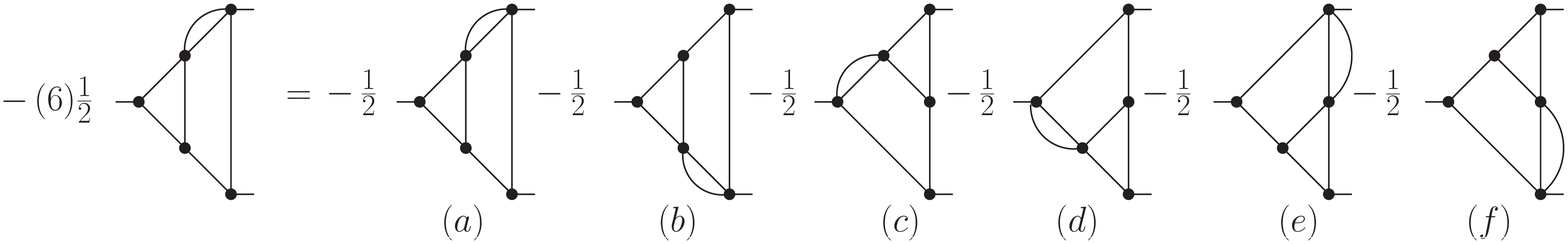}
\end{center}
\caption{\label{UcanF}The contributions to the $U$ eom from the diagram 3D in the effective action. }
\end{figure}

The diagram 3D in the effective action produces six different permutations in eom for $U$. They are labeled $U_{\rm 3D}$ and shown in Fig. \ref{bigUintLABEL}. In order to see how they are canceled, we draw the six permutations separately in parts (a)-(f) of Fig. \ref{UcanF}. Comparing with Fig. \ref{newfig3}, we can see that all six different permutations of $U_{\rm 3D}$ are canceled exactly. The first two permutations are canceled by the graph labeled $[U^{\rm 1p}_{\rm EYEBALL}[V_{\rm EYEBALL}]]_1$, the second two are canceled by $[U^{\rm 2p}_{\rm TARGET}[V_{\rm LOOPY}]]_1$, and the last two are canceled by $[U^{\rm 1p}_{\rm HAIR}[V_{\rm 4A}]_2]_1$.
\par
\par

Note that some of the topologies that are produced by the substitutions in Fig. \ref{newfig3} do not have the same form as any of the graphs in Fig. \ref{bigUintLABEL}. For example, the diagrams labeled by $[U^{\rm 1p}_{\rm HAIR}[[V_{\rm 4A}]_2]]_2$ and $[U^{\rm 1p}_{\rm HAIR}[[V_{\rm 4A}]_2]]_3$ in Fig. \ref{newfig3} are canceled by contributions obtained from $[U^{\rm 1p}_{\rm HAIR}[V_{\rm BBALL}[U_{\rm HAIR}]]]_1$ and $[U_{\rm MERCEDES}[U_{\rm HAIR}]]_1$ by removing the bare 4-vertex using Eq. (\ref{genericSubV0}), and taking the pieces that correspond to $V_\ey$ and $V_\lo$, respectively.

\subsection{Comparison of the structure of the $\Pi$ and $U$ eom's}
\label{nPIU}

There are three special features of the cancellations for the 2-point function (see Sec. \ref{nPIpi}) that do not apply to the cancellations for the 3-point function:
\begin{enumerate}
\item For the 2-point function, all of the graphs in the sd equation appear in the eom at the level of 4-Loop 4PI.
For the 3-point function this is not true: there is no term in the eom for $U$ with the form of the last term in Fig. \ref{sdUeqnLABEL}, for any $n$PI effective theory, and it cannot be produced by rearranging a lower loop graph.
%We have shown above for the 5-Loop 5PI effective theory that one can insert this missing term in the eom (by adding and subtracting it), and verify that the extra graphs cancel, to the order of the truncation. However, these cancellations are much more delicate than the corresponding cancellations for the 2-point function.

%In general, Eq. (\ref{eomRearrange}) tells us that for the $n$-Loop $n$PI effective theory, extra terms in the $U$ eom with $L_{\rm pt}\le (n-2)$ loops must cancel.
%The basic structure the cancellation shown in Fig. \ref{yunFig}.
%It is clear that the basic structure of Fig. \ref{figpi5loop} is much simpler than Fig. \ref{yunFig}.
%We expect that these observations will hold for $n$-Loop $n$PI effective actions for $n\ge 5$.
\item The 3-point function cancellations depend on a complicated level of cooperation between contributions from different diagrams (see Figs. \ref{newfig3} and \ref{UcanF}): unlike the case of the 2-point function, it is not true that we can add diagrams one at a time to the effective action and find that Eq. (\ref{eomRearrange}) is satisfied diagram by diagram (see Fig. \ref{figpi5loop}).
\item For the 2-point function, every topology that is canceled gets a contribution from the term labeled $\Pi_{\cal I}$ in Fig. \ref{figpi5loop}, while it is not true that all canceled 3-point topologies have a contribution of the form $U_{\cal I}$.
\end{enumerate}

\section{Conclusions}
\label{concSection}

In this paper we have calculated the 5-Loop 5PI effective action for a scalar theory with cubic and quartic interactions. The result has some surprising features.

The effective action does not contain only 5-particle irreducible diagrams,  even when a 5PR diagram is defined in the strictest possible sense, as a diagram that cannot be divided into two pieces by cutting five or fewer lines such that each piece contains at least one closed loop.

It is not true that all diagrams (except the super-BBALL and super-BBALL$_0$ diagrams) carry symmetry factors that are produced by the usual combinatoric rules.

Neither of these features has been seen previously, since they do not appear at the level of the 4-Loop 4PI effective action, which is the highest Legendre transform that has appeared in the literature to date.

We have shown that the skeleton diagrams in the $m$-Loop $n$PI effective action correspond to an infinite resummation of perturbative diagrams that is void of double counting at the $m$-Loop level.

From our calculation of the 5-Loop 5PI effective action we are able to obtain results for the 3PI and 4PI effective action up to 5 loops. The result is that
the 3PI effective action contains only 3-particle irreducible diagrams up to 5 loops. However, although the 4PI effective action contains only 4-particle irreducible diagrams up to 4 loops, there are 4-particle reducible diagrams at the 5-loop level.
The conclusion is that the standard idea that the $n$PI effective action contains only $n$-particle irreducible diagrams is not applicable at arbitrary loop order.

We stress that the absence of double counting is a property of the Legendre transforms, and the cancellation of $n$PR diagrams is not.
At lower levels in the loop expansion, the effective action contains only $n$-particle irreducible graphs in the skeleton loop expansion, but at higher orders, the counting becomes more complicated and a combination of irreducible and irreducible graphs is needed.

We have worked with a toy model which has the same basic diagrammatic structure as QED or QCD. We expect that it would be straightforward to use the same
method to show that the $m$-Loop $n$PI effective action for these gauge theories matches the corresponding perturbative expansion to $m$ loops. This would prove that the $m$-Loop effective action respects gauge invariance to order $\sim g^{2m-2}$.
%It has been shown previously that the gauge dependence of the effective action always appears at higher order than the truncation order. If the effective action is truncated at $n$-Loops ($\sim g^{2n-2}$) in the skeleton expansion, the non-perturbative result for the resummed effective action that is obtained by substituting the variational solutions can be gauge dependent only at order $g^{2n}$.
This result has been obtained previously in Refs. \cite{smit,HZ} using a completely different method.

We have also shown that the variational equations of motion produced by the $n$-Loop $n$PI effective theory are equivalent to the Schwinger-Dyson equations, up to the truncation order.
The equation of motion for the 2-point function has {\it exactly} the same structure as the corresponding Schwinger-Dyson equation.
An equivalent statement is that if we did a calculation using an $n$-Loop $n$PI effective theory, and replaced the equations of motion for the variational propagators and vertices by the Schwinger-Dyson equations truncated by setting the ($n+1$)-vertex to the bare one, the error we would make is of the same order as terms that would come from contributions to the effective action that are beyond the truncation order.

\appendix

\section{Eom's from the 5-Loop 5PI effective action}
\label{eomSection}

In this appendix we give the eom's for the 5-Loop 5PI effective action in diagrammatic form.

\par\begin{figure}[H]
\begin{center}
\includegraphics[width=18cm]{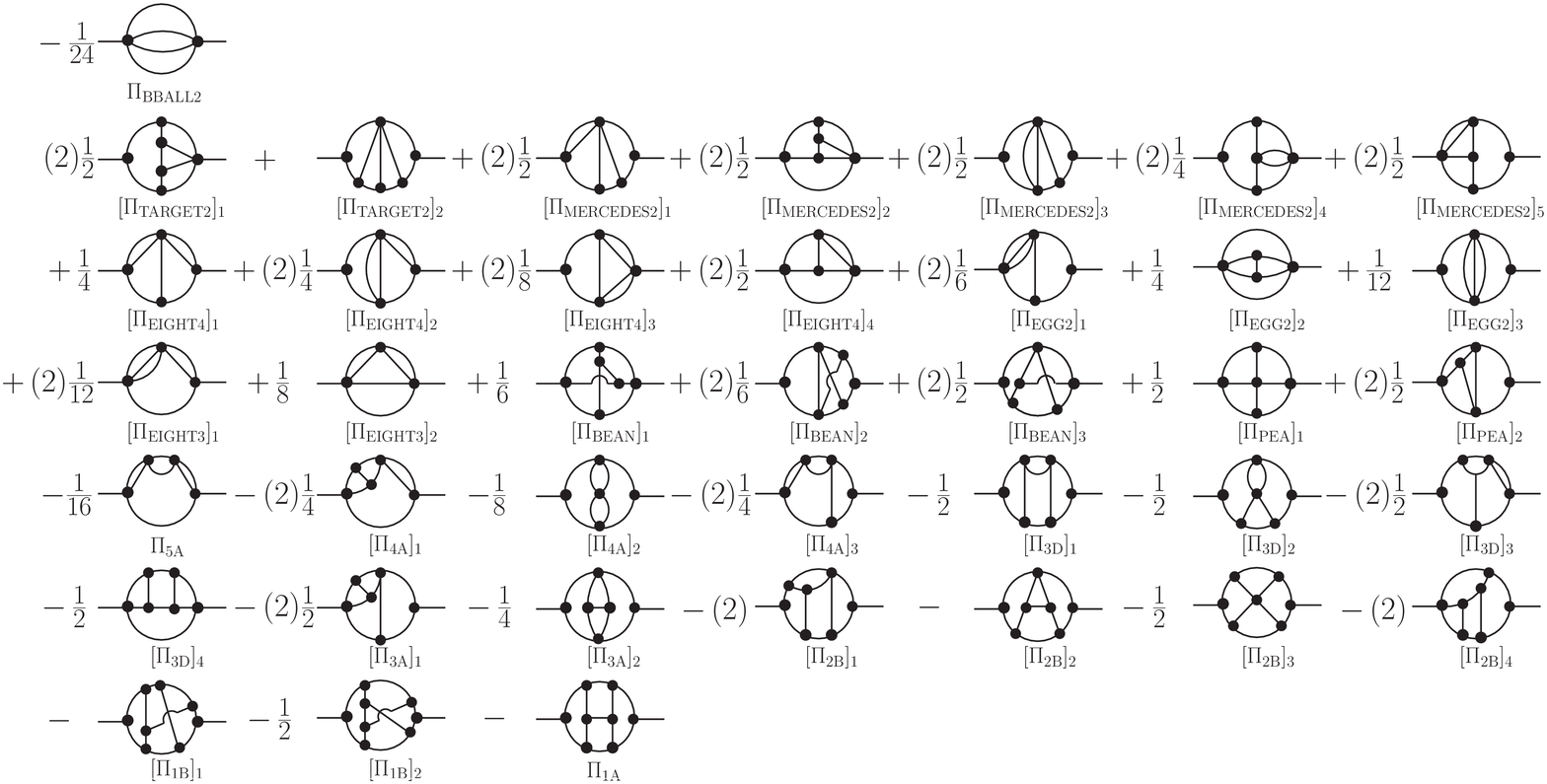}
\end{center}
\caption{\label{bigPIintLABEL}Contributions to the $\Pi$-integral equation from the 5-loop 5PI effective action, and the BBALL2 diagram from Fig. \ref{fig:fig16}.}
\end{figure}

\par\begin{figure}[H]
\begin{center}
\includegraphics[width=14cm]{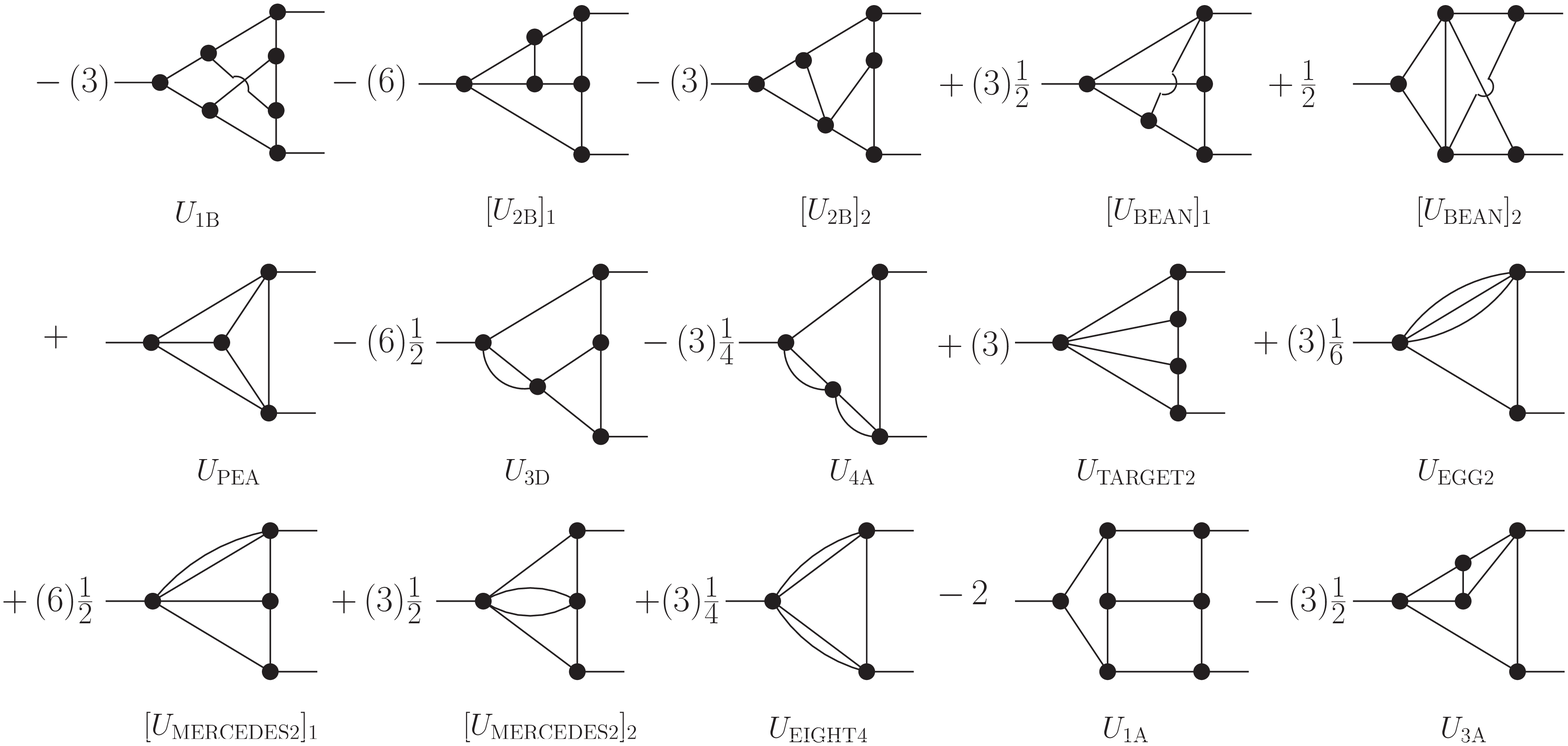}
\end{center}
\caption{\label{bigUintLABEL}3-loop contributions to the $U$-integral equation from the 5-loop 5PI effective action. }
\end{figure}
\par\begin{figure}[H]
\begin{center}
\includegraphics[width=16cm]{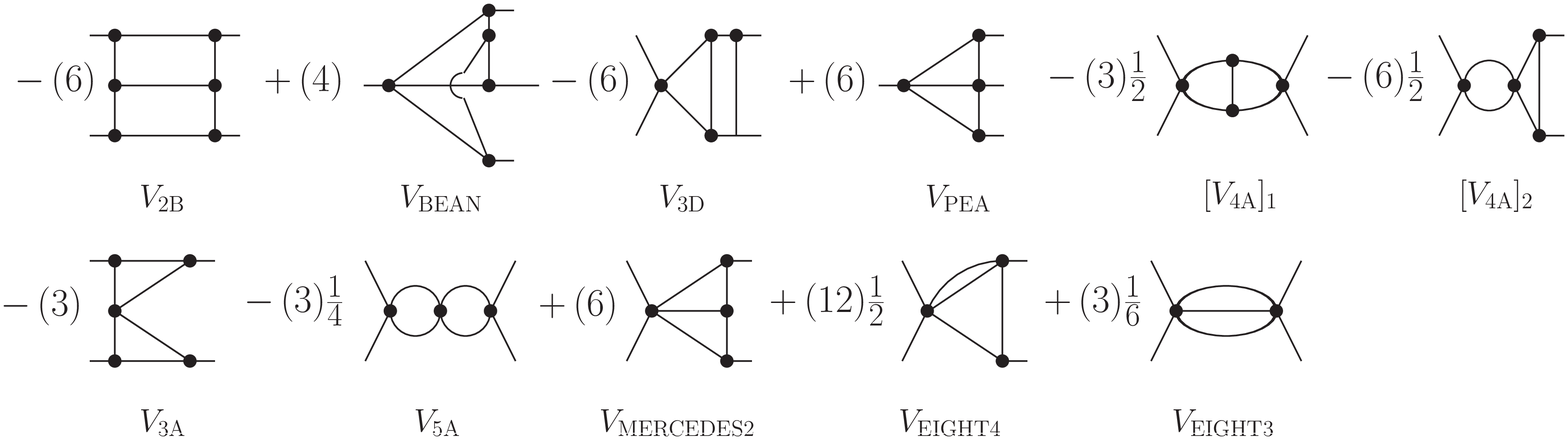}
\end{center}
\caption{\label{bigVintLABEL}2-loop contributions to the $V$-integral equation from the 5-loop 5PI effective action. }
\end{figure}
\par\begin{figure}[H]
\begin{center}
\includegraphics[width=13cm]{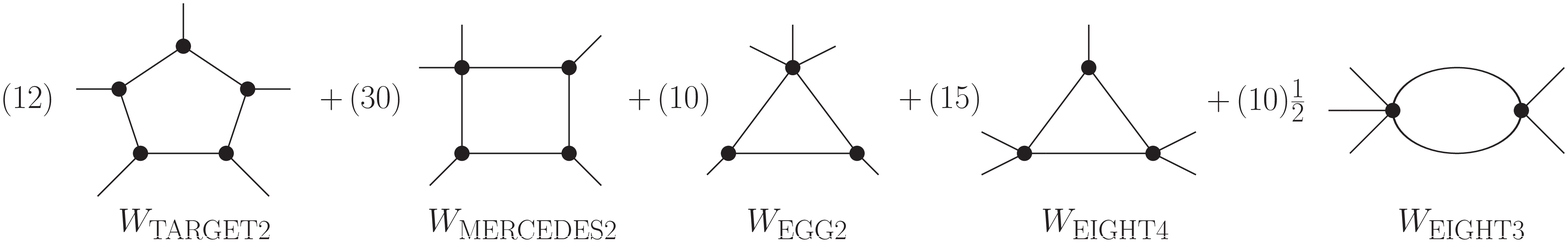}
\end{center}
\caption{\label{bigWintLABEL}1-loop contributions to the $W$-integral equation from the 5-loop 5PI effective action. }
\end{figure}

\section{Equivalence Hierarchy}
\label{hierSection}

In Ref. \cite{bergesReview} it is argued that at $m$-Loop order, the $n$PI effective action for $n\ge m$ depends only on ${\cal V}_i,~i\le m$. An equivalent statement is that the $n$-Loop $n$PI effective action provides a complete self-consistent description. We show below that the arguments we have used in Sec. \ref{accuracySection} about the structure of the diagrams in the effective action are consistent with the equivalence hierarchy proposed in \cite{bergesReview}.

Consider the $m$-Loop effective action. We have argued that vertices  ${\cal V}_{i}$ for $i>m+2$ do not appear in the $m$-Loop effective action, since these vertices would appear in graphs that would produce disconnected contributions to the eom's.  It is straightforward to calculate the eom for the vertex ${\cal V}_{m+1}$. From Sec. \ref{accuracySection}, the only terms in the effective action that contain this vertex are the $m$-loop super-BBALL and super-BBALL$_0$ diagrams. The eom produced by these diagrams is
\bea
&& {\cal V}_{m+1} = {\cal V}^0_{m+1}~~{\rm for}~~m\le 3\,, \\
&& {\cal V}_{m+1} = 0\, ~~~~~~~{\rm for}~~m\ge 4 \,.\nonumber
\eea
The conclusion is that the $m$-Loop effective action depends on the vertices ${\cal V}_i,~i\le m$:
\bea
\label{jbH}
\Gamma^{(m)}_{n\ge m}=\Gamma^{(m)}_m .
\eea
Our result for the 5-Loop 5PI effective action verifies Eq. (\ref{jbH}) for  $m=4$.

\section{The sd equation for the 4-point vertex}
\label{sdAppendix}

\begin{figure}[H]
\begin{center}
\includegraphics[width=15cm]{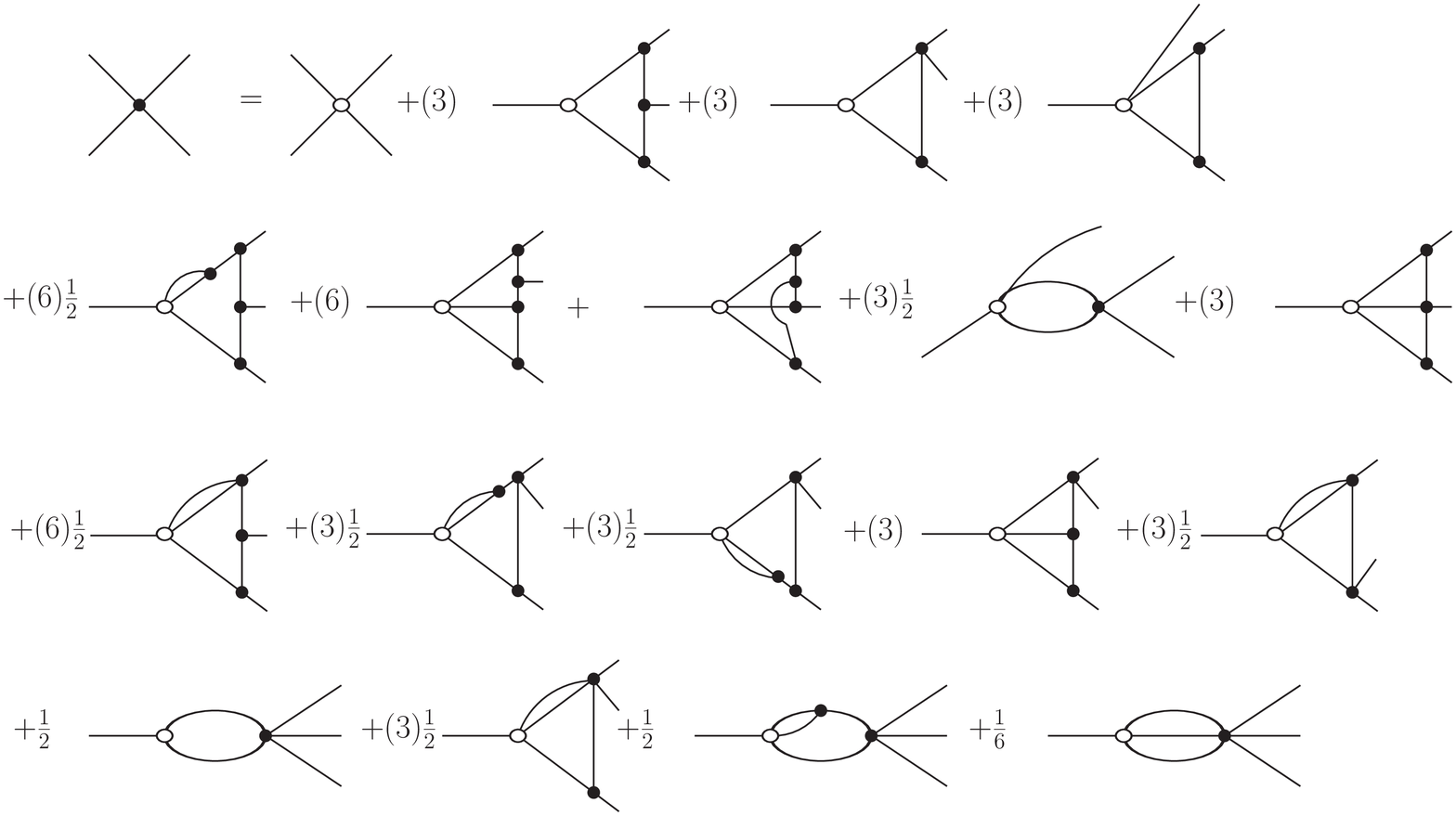}
\end{center}
\caption{\label{SD4}Schwinger-Dyson equation for the 4-point vertex.}
\end{figure}

In Fig. \ref{SD4} we give the sd equation for the 4-point vertex in diagrammatic form.
\par

%%%%%%%%%%%%%%%%%%%%
%%%%%%%%%%%%%%%%%%%%

\end{document}